\documentstyle[12pt]{report}
\def\N{I\!\!N}
\def\bbbn{I\!\!N}
\def\bbbr{\rm I\!R}
\def\bbbc{C\!\!\!\!I}
\def\bbbq{Q\!\!\!\!I}
\def\bbbz{Z\!\!\!Z}

\def\U{\hbox{U}}

\def\Fix{\rm {\hbox{Fix}}}
\def\Gb{{\overline G}}
\def\phib{{\overline \phi}}

\def\cR{{\cal R}}
\def\cS{{\cal S}}
\newtheorem{theorem}{Theorem}
\newtheorem{lemma}{Lemma}
\newtheorem{corollary}{Corollary}
\newtheorem{proposition}{Proposition}
\newtheorem{definition}{Definition}
\newtheorem{example}{Example}
\newtheorem{remark}{Remark}

\newcommand{\tr}{{\rm Tr\ }}
\newcommand{\coker}{{\rm Coker\ }}
\newcommand{\fix}{{\rm Fix\ }}
\newcommand{\rank}{{\rm Rank\ }}
\newcommand{\spec}{{\rm Spec\ }}
\newcommand{\im}{{\rm Im\ }}
\newcommand{\ind}{{\rm Index\ }}
\newcommand{\mod}{{\rm \; mod\ }}

\begin{document}

\begin{titlepage}
\centerline{\Huge\bf Dynamical Zeta Functions, }
\bigskip
\centerline{\Huge\bf  Nielsen Theory}
\bigskip

\centerline{\Huge\bf and Reidemeister Torsion}
\vskip 3cm

\bigskip
\centerline{\Huge\bf Alexander  Fel'shtyn}
\bigskip

\end{titlepage}

\tableofcontents

\addcontentsline{toc}{chapter}{Introduction}
\chapter*{Introduction}
\markboth{\sc Introduction}{\sc Introduction}

  \section{From Riemann zeta function to dyna\-mical zeta functions}

 In this subsection we shall try to explain where dynamical zeta functions come from . 
In a sense the study of dynamical zeta functions is part of the theory of dynamical systems, but it is also intimately related to algebraic geometry,  number theory, topology  and statistical mechanics.

\subsection{Riemann zeta function}
  The theory of the Riemann zeta function and its generalisations represent one of the most beatiful developments in mathematics. The Riemann zeta function is that function defined on $\{ s\in \bbbc: Re(s) >1\}$ by the series
 
 $$
 \zeta(s)\equiv \sum_{n=0}^\infty \frac{1}{n^s}.
 $$
 There is a second representation of $\zeta$ which was discovered by Euler in 1749 and which is the reason for the significance of the Riemann zeta function in arithmetic. This is Euler product formula:
 $$
 \zeta(s)= \prod_{ p \> prime}(1-p^{-s})^{-1}.
 $$
 Riemann's significant contribution here was his consideration in 1858 of the zeta function as an analytic function.He first showed that the  zeta function has an analytic continuation to the complex plane  as a meromorphic function with a single pole at $s=1$ whose properties can
 on the one hand be investigated by the techniques of complex analysis, and  on the other
 yield difficult theorems concerning the integers\cite{patti}. It is this connection between the continuous
 and the discrete that is so wonderful.
 Riemann also showed that the zeta function satisfied a functional equation of the form
 $$
 \zeta(1-s)=\gamma(s)\cdot \zeta(s)
 $$
 where 
 $$
 \gamma(s)=\pi^{1/2 -s}\cdot \Gamma(s/2)/\Gamma((1-s)/2) 
 $$
 and $\Gamma$ is the Euler gamma-function.
 In the course of his investigations Riemann was led to suspect that all  nontrivial zeros are 
 on the line $ Re(s)=1/2$, this is the Riemann Hypothesis which has been the central goal of research to the present day. Although no proof has yet appeared various weak forms of this conjecture, and in other contexts, analogues of it have been of considerable  significance.

  \subsection{Problems concerning zeta functions}
         Since the 19 th century, many special functions called zeta functions have been defined
  and investigated. The  main problems concerning zeta functions are:
  
  (I) Creation of new zeta functions.
  
  (II) Investigation of the properties of zeta functions. Generally , zeta functions have 
the following  properties in common: 1) They are meromorphic on the whole complex plane ;
2) they have Dirichlet series expansions ; 3) they have Euler product expansions ; 4) they satisfy certain functional equations; 5) their special values play important role.
 Also, it is an important problem  to find the poles, residues,
and zeros  of zeta functions .
  
  (III) Application to number theory, geometry, dynamical systems.
  
  (IV) Study of the relations between different zeta functions.
  
Most of the functions called zeta functions or $L$-functions have the properties of problem (II).

 \subsection{Important types of  zeta functions}

For a general discussion of a zeta functions see article  " Zeta functions " in the Encyclopedic Dictionary of Mathematics \cite{encycloped}. The following is a classification of the important types of zeta functions that are already known:

1) The zeta and $L$-functions of algebraic numbers fields: the Rie\-mann zeta function,
Dirichlet $L$-functions, Dedekind zeta functions, Hecke $L$-functions, Artin $L$-functions.

2) The p-adic $L$-functions of Leopoldt and Kubota.

3) The zeta functions of quadratic forms: Epstein zeta functions, Siegel zeta functions.

4) The zeta functions associated with Hecke operators.
 
 5)The zeta and $L$-functions attached to algebraic varieties defined over finite fields: Artin zeta function, Hasse-Weil zeta functions.
 
 6) The zeta functions attached to discontinuous groups :  Selberg zeta functions.
 
 7) The dynamical zeta functions: Artin-Mazur zeta function, Lefschetz zeta function,
 Ruelle zeta function for discrete dynamical systems, Ruelle zeta function for flows.

 \subsection{ Hasse-Weil zeta function}

   Let $V$ be a nonsingular projective algebraic variety of dimension $n$ over a finite
 field $k$ with $q$ elements. The variety $V$ is thus defined by homogenous polynomial equations with coefficients in the field $k$ for $m+1$ variables $x_0, x_1, ... , x_m $.
 These variables are in the algebraic closure $\bar k$ of the field $k$, and constitute the homogeneous coordinates of a point of $V$.The variety $V$ is invariant under the Frobenius map $F: (x_0, x_1, ... , x_m ) \rightarrow (x_0^q, x_1^q, ... , x_m^q) $. Arithmetic considerations lead Hasse and Weil to introduce a zeta function which counts the points of $V$ with coordinates in the different finite extensions of the field $k$, or equivalently points of $V$ which are fixed under
 $F^n$ for some $n\geq 1$:
  $$ 
 \zeta(z,V)  :=  \exp\left(\sum_{n=1}^\infty \frac{\# \fix(F^n)}{n} z^n \right)
 $$
 Note that  $\zeta(z,V)$  can be written as a  Euler product 
 $$
  \zeta(z,V) =\prod_{\gamma}\frac{1}{1-z^{\#\gamma}},
  $$
  over  all primitive  periodic orbits $\gamma $ of $F$ on $V$.
  For comparison with Riemann's zeta function one has to put $z=q^{-s}$.  
  Certain conjectures  proposed by Weil \cite{weil} on the properties 
  of $\zeta(z,V)$  led to lot of work by Weil, Dwork, Grothendieck, and complete
  proof was finally obtained by Deligne \cite{del}.
  The story of the Weil conjectures is  one of the most striking instances exhibiting the fundamental unity of mathematics.
  It is found that $\zeta(z,V)$ is a rational function of $z$ with a functional equation :
  $$
  \zeta(z,V)=\prod_{i=0}^{2m} P_l(z)^{(-1)^{l+1}}
  $$
  where the zeros of the polynomial $P_l$ have absolute value $q^{-l/2}$ and  the $P_l(z)$
  have a cohomological interpretation: the polynomial $P_l$ is roughly the characteristic polynomial associated with the induced action of the Frobenius morphism on the etale cohomology: $ P_l(z)= \det(1-z\cdot F^*| H^l(V))$.

  \subsection{Dynamical zeta functions}
  
  Inspired by the Hasse-Weil  zeta function of an algebraic variety over a finite field,
  Artin and Mazur \cite{am} defined  the Artin - Mazur zeta function for an arbitrary map $f: X \rightarrow X $
  of a topological space $X$:
 $$ 
 F_f(z)  :=  \exp\left(\sum_{n=1}^\infty \frac{F(f^n)}{n} z^n \right)
 $$
  where $F(f^n)$ is the number of isolated fixed points of $f^n$.
Artin and Mazur showed that for a dense set of the space of smooth maps of a compact smooth manifold into itself the  Artin-Mazur zeta function $ F_f(z) $ has a positive radius of convergence.Later   Manning \cite{m} proved the rationality of the Artin - Mazur zeta function for diffeomorphisms of a smooth compact manifold satisfying Smale  axiom A, after partial results were obtained by Williams and Guckenheimer. On the other hand there exist maps for which Artin-Mazur zeta function is transcendental \cite{bl}.   
 
  The Artin-Mazur zeta function  was adopted later by Milnor and Thurston \cite{mt}
  to count periodic points for a piecewise monotone map of the interval.
  
   The Artin-Mazur zeta function was  historically the first dynamical zeta function 
   for $discrete$ dynamical system. The next dynamical zeta function was defined by Smale \cite{s} .This is  the Lefschetz zeta function of discrete dynamical system:
   $$
  L_f(z) := \exp\left(\sum_{n=1}^\infty \frac{L(f^n)}{n} z^n \right)
 $$
  , where
 $$
   L(f^n) := \sum_{k=0}^{\dim X} (-1)^k \tr\Big[f_{*k}^n:H_k(X;\bbbq)\to H_k(X;\bbbq)\Big]
 $$
 is the Lefschetz number of $f^n$. Smale considered   $ L_f(z) $ in the case when $f$ is diffeomorphism of a compact manifold, but it is well defined for any continuous map $f$ of compact polyhedron $X$.
 The Lefschetz zeta function is a rational function of $z$ and
is given by the formula:
$$
 L_f(z) = \prod_{k=0}^{\dim X}
          \det\big(I-f_{*k}\cdot z\big)^{(-1)^{k+1}}.
$$

 Afterwards, J.Franks \cite{fran}  defined reduced mod 2 Artin-Mazur and Lefschetz
 zeta functions, and D. Fried \cite{fri} defined twisted Artin-Mazur and Lefschetz
 zeta functions, which have coefficients in the group rings $\bbbz H$ or $\bbbz_2 H$ 
 of an abelian group $H$.
  The above zeta functions are directly analogous to the Hasse-Weil zeta function.

   Ruelle has found another generalization  of the Artin-Mazur zeta function.
He was motivated by ideas from equilibrium statistical mechanics and has replaced in  
the Artin-Mazur zeta function \cite{ruel} simple counting of the periodic points by counting with weights. He defined the Ruelle zeta function as
$$
 {F_f}^g(z) := \exp\left(\sum_{n=1}^\infty \frac{ z^n }{n}\sum_{x\in \fix (f^n)} \prod_{k=0}^{n-1}g(f^k(x)) \right),
$$
where $g:X\to \bbbc $ is a weight function(if $g=1$ we recover $F_f(z) $).

 Dynamical zeta functions have relations with statistical mechanics( entropy, pressure, Gibbs states, equilibrium states).  Manning used  Markov partitions and corresponding symbolic dynamics in his proof of the rationality of the Artin-Mazur zeta function. This symbolic dynamics is reminiscent of the statistical mechanics of one-dimensional lattice spin system.

 \section{ Dynamical zeta functions and Nielsen fixed point theory}

 In contrast with the Artin- Mazur zeta function which counts the periodic points of the map geometrically, the  Lefschetz zeta function does this algebraically.There is another way of counting the fixed points of  $f^n$  - according to Nielsen  \cite{j}.
 
 Let $p:\tilde{X}\rightarrow X$ be the universal covering of $X$
and $\tilde{f}:\tilde{X}\rightarrow \tilde{X}$ a lifting
of $f$, ie. $p\circ\tilde{f}=f\circ p$.
Two liftings $\tilde{f}$ and $\tilde{f}^\prime$ are called
{\it conjugate} if there is a $\gamma\in\Gamma\cong\pi_1(X)$
such that $\tilde{f}^\prime = \gamma\circ\tilde{f}\circ\gamma^{-1}$.
The subset $p(Fix(\tilde{f}))\subset Fix(f)$ is called
{\it the fixed point class of $f$ determined by the lifting class $[\tilde{f}]$}.
A fixed point class is called $essential$ if its index is nonzero.
The number of lifting classes of $f$ (and hence the number
of fixed point classes, empty or not) is called the {\it Reidemeister Number} of $f$,
denoted $R(f)$. This is a positive integer or infinity.
The number of essential fixed point classes is called the $Nielsen$ $number$
of $f$, denoted by $N(f)$.
The Nielsen number is always finite. $R(f)$ and $N(f)$ are homotopy
type invariants.
In the category of compact, connected polyhedra the Nielsen number
of a map is equal to the least number of fixed points
of maps with the same homotopy type as $f$.In Nielsen fixed point  theory the main objects for investigation are the Nielsen and Reidemeister numbers and their modifications \cite {j}.

Let $G$ be a group and $\phi:G\rightarrow G$ an endomorphism.
Two elements $\alpha,\alpha^\prime\in G$ are said to be
$\phi-conjugate$ iff there exists $\gamma \in G$ such that
$\alpha^\prime=\gamma . \alpha . \phi(\gamma)^{-1}$.
The number of $\phi$-conjugacy classes is called the $Reidemeister$
$number$ of $\phi$, denoted by $R(\phi)$.

 In papers \cite {f1,f2,f3,f6}  we have introduced new dynamical zeta functions connected with Nielsen fixed point theory.We defined the Nielsen zeta function of $f$  and Reidemeister zeta functions of $f$ and $\phi$ as power series:
 \begin{eqnarray*}
 R_\phi(z) & := & \exp\left(\sum_{n=1}^\infty \frac{R(\phi^n)}{n} z^n \right), \\
 R_f(z) & := & \exp\left(\sum_{n=1}^\infty \frac{R(f^n)}{n} z^n \right), \\
 N_f(z) & := & \exp\left(\sum_{n=1}^\infty \frac{N(f^n)}{n} z^n \right).
\end{eqnarray*}

We assume  that $R(f^n)<\infty$ and
$R(\phi^n)<\infty$ for all $n>0$.

We have investigated the following problem: for which spaces and maps and for which groups and endomorphisms are the Nielsen and Reidemeister zeta functions a rational functions?When they have a functional equation?Are these functions algebraic functions?

In \cite {f1}  we proved that the Nielsen zeta function has a positive radius of convergence which admits a sharp estimate in terms of the topological entropy of the map. Later, in \cite{fh4} we propose another prove of positivity of radius and proved an exact algebraic lower estimation for it. With the help of Nielsen - Thurston theory \cite {ht} of surface homeomorphisms ,  in \cite  {pf} we proved that for an orientation-preserving homeomorphism of a compact surface the Nielsen zeta function is either a rational function or the radical of rational function. For  a periodic map of any compact polyhedron in  \cite {pf} we proved  a product formula for Nielsen zeta function which implies that Nielsen zeta function is a radical of a rational function.

The investigation and computation of the Reidemeister zeta function  $R_\phi(z) $of a group endomorphism  $\phi $ is an algebraic ground of the computation and investigation of zeta functions $R_f(z)$ and $N_f(z)$. In \cite {f3} we investigated the behavior of  $R_\phi(z) $under the extension of a group and proved rationality and a functional equation of $R_\phi(z) $ and a trace formula for the $R(\phi^n)$ for the endomorphism of a finitely generated free Abelian group and group $ \bbbz/p\bbbz$.  An endomorphism $\phi : G\rightarrow G $ is said to be eventually commutative if there exists a natural number $n$ such that the subgroup $\phi^n(G) $ is commutative. A map $f:X\rightarrow X $ is said to be eventually commutative if the induced endomorphism on fundamental group is eventually commutative. In \cite {fh,fh1,fh2}  we proved that $R_\phi(z) $ is a rational function with functional equation in the case of any endomorphism $\phi$ of  any finite group $G$ and in the case that $\phi$ is eventually commutativ
e and $G$ is finitely generated. As a consequence we obtained rationality and a functional equation for $ R_f(z) $ where either the fundamental group of $X$ is finite, or the map $f$ is eventually commutative.We obtained sufficient conditions under which the Nielsen zeta function coincides with the Reidemeister zeta function and is a rational function with functional equation.
As an application we calculate the Reidemeister and Nielsen zeta functions of
all self-maps of lens spaces, nilmanifolds and tori.

In \cite {f2,f3,f5} we found a connection between the rationality of the Nielsen and Reidemeister zeta functions for the maps of fiber,base and total space of a fiber map of a Serre bundle using results of Brown, Fadell, and You (see \cite {j} ) about Nielsen and Reidemeister numbers of a fiber map.
 
 In \cite{fh3} we proved the rationality of the Reidemeister zeta function and the trace formulas for the Reidemeister numbers of group endomorphisms  in the following cases: the group is finitely generated and an endomorphism is eventually commutative; the group is finite ; the group is a direct sum of a finite group and a finitely generated free Abelian group; the group is finitely generated, nilpotent and torsion free .
 
 \section{ Congruences for Reidemeister numbers}
 
 In his article \cite{d}, Dold found a remarkable
arithmetical property of the Lefschetz numbers for the iterations of
a map $f$. He proved the following formula
$$
 \sum_{d\mid n} \mu(d)\cdot L(f^{n/d}) \equiv 0 \mod n 
 $$
 where $n$ is any natural number and $\mu$ is the M\"obius function.  
This result had previously been obtained for prime $n$
by Zabreiko, Krasnosel'skii \cite{zk} and Steinlein \cite{stein}.

The congruences for Lefschetz numbers are directly connected with the  rationality
of the Lefschetz zeta function \cite{d}.

In \cite {fh}, \cite{fh4} we proved, under additional conditions, similar congruences :
$$
 \sum_{d\mid n} \mu(d)\cdot R(\phi^{n/d}) \equiv 0 \mod n,
$$
$$
 \sum_{d\mid n} \mu(d)\cdot R(f^{n/d}) \equiv 0 \mod n
$$
 for the Reidemeister numbers of the iterations of a group endomorphism  $\phi$ and a map $f$.
 
 This result implies , in special cases ,  the corresponding congruences for
the Nielsen numbers. For  $n$-toral maps in the case when Jiang subgroup
coincide with fundamental group this congruences for Nielsen numbers
were proved by
Heath, Piccinini, and You \cite{hp} .

In \cite{fhw}, we generalize the arithmetic congruence relations among the
Reidemeister numbers of iterates of maps to similar congruences for Reidemeister numbers of
equivariant group endomorphisms and maps.

In the  article \cite {fh1} we conjected a general
 connection between the Reidemeister number of a group endomorphism
 and the number of fixed points of the induced map on the space
 of irreducible unitary representations.
This can be reformulated in terms of self-maps and pullbacks of
 vector bundles.
The results \cite {fh1,fh2} are essentially proofs of these
 conjectures under the condition that $\phi$ is eventually
 commutative or $G$ is finite.

\section{Reidemeister torsion.}

    Dynamical zeta functions in the Nielsen theory are closely connected with 
the Reidemeister torsion.
     
     Reidemeister torsion is a very important topological invariant which has useful applications in knots theory,quantum field theory and dynamical systems.In 1935 Reidemeister \cite{re} classified up to $PL$ equivalence the lens spaces $S^3/\Gamma$
where $\Gamma$ is a finite cyclic group of fixed point free orthogonal transformations.
He used a certain new invariant which was quickly extended by
Franz , who used it to classify the
generalized lens spaces $S^{2n+1}/\Gamma$.
This invariant is a ratio of determinants concocted from a $\Gamma$-equivariant
chain complex of $S^{2n+1}$ and a nontrivial character
$\rho:\Gamma\rightarrow U(1)$ of $\Gamma$.
Such a $\rho$ determines a flat bundle $E$ over $S^{2n+1}/\Gamma$
such that $E$ has holonomy $\rho$.
The new invariant is now called the {\it Reidemeister torsion},
or {\it R-torsion} of $E$.

The results of Reidemeister and Franz were extended by de Rham 
to spaces of constant curvature +1.
 
 Whitehead refined and generalized Reidemeister torsion  in defining the torsion of a
 homotopy equivalence in 1950, and his work was to play a crucial role in the development
 of geometric topology and algebraic K-theory in the 60's.
The Reidemeister torsion is closely related to the $ K_1 $ groups of algebraic  $ K $-theory.

Later Milnor identified the Reidemeister torsion
with the Alexander polynomial, which plays a fundamental role
in the theory of knots and links.

In 1971, Ray and Singer \cite{rs} introduced an analytic torsion associated with
the de Rham complex of forms with coefficients in a flat bundle over a compact Riemannian manifold, and conjectured it was the same as the Reidemeister torsion associated with the action
of the fundamental group on the covering space, and the representation associated with the flat bundle. The Ray- Singer conjecture was established independently by Cheeger \cite{c}
and M\"uller \cite{mu}
a few years later.

In 1978  A.Schwartz \cite{schw} showed how to construct  a quantum field theory on a manifold 
$M$ whose partition function is a power of the analytical torsion of $M$.
 Witten \cite{wit2} have used analytical torsion to study non-Abelian Chern-Simons gauge field theory. Its partition function is the Witten-Reshetikhin-Turaev invariant\cite{tu1} for the three manifold $M$ and the analytic torsion appears naturally in the asymptotic formula for the partition function obtained by the method of stationary phase approximation  \cite{wit2}.

Recently, the Reidemeister torsion has found interesting applications in dynamical systems theory. A connection between the Lefschetz type dynamical zeta functions
and the Reidemeister torsion was established by D. Fried \cite{fri}.
The work of Milnor \cite{mi} was the first indication that such a connection exists.
Fried also has shown  that, for some  flows, the value at 0 of the Ruelle zeta function coincides with the Reidemeister torsion.

In \cite{f4}, \cite {fh1,fh2,fh3} we established a connection between the Reidemeister torsion and Reidemeister zeta function. We obtained an expression for the Reidemeister torsion
of the mapping torus of the dual map of a group
endomorphism, in terms of the Reidemeister zeta function of
the endomorphism.
The result is
obtained by expressing the Reidemeister zeta function in terms
of the Lefschetz zeta function of the dual map, and then
applying the theorem of D. Fried.
What this means is that the Reidemeister torsion counts the fixed point classes of all iterates of map $f$ i.e. periodic point classes of $ f $.

 In \cite{f7} we   established a connection between the Reidemeister torsion of a mapping torus, the eta-invariant, the Rochlin invariant and the multipliers of the
theta function.
The formula is obtained via the Lefschetz zeta function
 and the results on the holonomy of determinant line bundles due to
 Witten \cite{wi}, Bismut-Freed \cite{bf},
 and Lee, Miller and Weintraub \cite{lm}.

Note, that the work of Turaev \cite{tu} was the first indication that the Rochlin invariant is connected with the Reidemeister torsion for three-dimensional rational homology spheres.

  In \cite{gf}, we proved an analogue of the Morse inequalities for the attraction domain of an attractor, and the level surface of the Lyapunov function.These inequalities describe the connection between the topology of the attraction domain and dynamic of a Morse-Smale flow on the attractor.

  In \cite{f4,f8} we  described with the help of the Reidemeister torsion the connection between the topology of the attraction domain of an attractor and the dynamic of flow with circular chain-recurrent set on the attractor. We showed that for flow with circular chain-recurrent set, the Reidemeister torsion of the attraction domain of an attractor and Reidemeister torsion of the level surface of the Lyapunov function is a special value of the twisted Lefschetz zeta function building via closed orbits in the attractor. 
   
   In \cite{f8} we  found that for the integrable Hamiltonian system on the four-dimensional symplectic manifold, the Reidemeister torsion of the isoenergetic surface counts the critical circles (which are  the closed trajectories of the system)  of the second independent Bott integral on this surface.

\section{ Table of contents}

The monograph consists of four  parts. Part I( Chapter 1) presents a brief account of the Nielsen
fixed point theory. Part II( Chapters 2 - 4 )  deals with dynamical zeta functions connected
with Nielsen fixed point theory. Part III ( Chapter 5)  is concerned  with congruences for the
Reidemeister and Nielsen numbers. Part IV (Chapter 6) deals with the Reidemeister torsion .

The content of the chapters should be clear from the headings. The following remarks give more directions to the reader.  
  
In Chapter 1  we define the lifting and fixed point classes, fixed point index,  Reidemeister and Nielsen numbers. The relevant definitions and results will be used throughout
the book.

In Chapter 2 - 4 we introduce  the Reidemeister zeta functions of a group endomorphism and of a map and the Nielsen zeta function of a map which are the main objects of the monograph.
 
 In Chapter 2 we prove that the Reidemeister zeta function of a group endomorphism is a rational function with functional equation in the following cases: the group is finitely generated and an endomorphism is eventually commutative;  the group is finite ;  the group is a direct sum of a finite group and a finitely generated free abelian group;  the group is finitely generated, nilpotent and torsion free. As a consequence we obtained rationality and a functional equation for the Reidemeister zeta function of a continuous map where  the fundamental group of $X$ is as above.

In Chapter 3 we show  that the Nielsen zeta function has a positive radius of convergence which admits a sharp estimate in terms of the topological entropy of the map. We also give
an exact algebraic lower estimation for the radius.With the help of Nielsen - Thurston theory of surface homeomorphisms  we prove that for an orientation-preserving homeomorphism of a compact surface the Nielsen zeta function is either a rational function or the radical of rational function.  For  a   periodic map of a compact polyhedron  we prove  a product formula for Nielsen zeta function which implies that Nielsen zeta function is a radical of a rational function.
In section 3.4 and 3.5 we give sufficient conditions under which the Nielsen zeta function coincides with the Reidemeister zeta function and is a rational function with functional equation.
In section 3.6 we describe  connection between the rationality of the Nielsen zeta functions for the maps of fiber, base and total space of a fiber map of a Serre bundle. We would like to mention that in all known cases the Nielsen zeta function is a
nice function. By this we mean that it is a product of an exponential of a polynomial with a function some power of which is rational. May be this is a
general pattern.

In Chapter 4  we generalize the results of Chapter 2-3 to the Nielsen and Reidemeister zeta functions modulo normal subgroup of the fundamental group. 

In Chapter 5 we prove analog of Dold congruences for Reidemeister and Nielsen numbers.

In Chapter 6 we explain how dynamical zeta functions give rise to the Reidemeister torsion, a very important topological invariant  .
In section 6.2 we establish a connection between the Reidemeister torsion and Reidemeister zeta function. We obtain an expression for the Reidemeister torsion of the mapping torus of the dual map of a group
endomorphism, in terms of the Reidemeister zeta function of
the endomorphism.This means  that the Reidemeister torsion counts the fixed point classes of all iterates of map $f$ i.e. periodic point classes of $ f $.

In section 6.3  we establish a connection between the Reidemeister torsion of a mapping torus,  the eta-invariant,  the Rochlin invariant and the multipliers of the theta function.

In section 6.4 we describe with the help of the Reidemeister torsion and of an analog of Morse inequalities the connection between the topology of the attraction domain of an attractor and the dynamic of the system on the attractor.
 
 In section 6.5  we show  that for the integrable Hamiltonian system on the four-dimensional symplectic manifold, the Reidemeister torsion of the isoenergetic surface counts the critical circles(which are  the closed trajectories of the system ) of the second independent Bott integral on this surface.

 A part of this monograph grew out of the joint papers of the author with V.B. Pilyugina and 
 R. Hill written in 1985-1995. Their collaboration is gratefully appreciated. 
 
 This book had its beginnings in talks which the author gave  at  Rochlin seminar in 1983 - 1990.
 We are happy to acknowledge the influence of V.A. Rochlin  on our approach to the subject of the book.
  
  The author would like to thank D. Anosov,  M.Gromov, B.Jiang, S. Patterson, V.Turaev, O. Viro,   P. Wong   for useful discussions and comments.
  
 The author is sincerely grateful to J. Eichhorn  for permanent support and help.  
  
Parts of this book were written while the author was visiting the University of G\"ottingen,
Institute des Hautes Etudes Scientifiques ( Bures-sur -Yvette), Max-Planck-Institut f\"ur Mathematik (Bonn).The author is indebted to these institutions for their invitations and 
hospitality.

\chapter{ Nielsen Fixed Point  Theory}
\markboth{\sc Nielsen theory}{\sc Nielsen theory}

  In this section we give a brief review of the  Nielsen theory.

\section{History} 
 
 Fixed point theory started in the early days of topology, because of its close relationship
 with other branches of mathematics.Existence theorems are often proved by converting 
 the problem into an appropriate fixed point problem.Examples are the existence of solutions
  for elliptic partial differential equations, and the existence of closed orbits in dynamical
  systems. In many problems, however, one is not satisfied with the mere existence of a
  solution. One wants to know the number, or at least a lower bound for the number of 
  solutions. But the actual number of fixed points of a map can hardly be the subject of an 
  interesting theory, since it can be altered by an arbitrarily small perturbation of the map.
  So, in topology, one proposes to determine  the minimal number of fixed points in a homotopy class.This is  what Nielsen fixed point theory about.
  Perhaps the best known fixed point theorem in topology is the Lefschetz fixed point theorem
  \begin{theorem}\cite{lef}
  Let $X$ be a compact polyhedron, and $ f: X\rightarrow X $ be a map.
  If the Lefschetz number $L(f)\not=0$, then every map homotopic to $f$ has a fixed 
  point.
  \label{lefschetz}
  \end {theorem}
  The Lefschetz number is the total algebraic count of fixed points.It is a homotopy invariant and is easily computable.

  So, the Lefschetz theorem, along with its special case, the Brouwer fixed point theorem,
  and its generalization, the widely used Leray-Schauder theorem in functional analysis,
  can tell existence only.
  In contrast, the chronologically first result of Nielsen theory has set a beautiful example of a 
  different type of theorem
   \begin{theorem}\cite{nie1}
   Let  $ f: T^2 \rightarrow T^2 $ be a map of the torus. Suppose that the endomorphism induced
   by $f$ on the fundamental group $\pi_1(T^2) \cong Z \oplus Z $ is represented by the
   $2\times2$ integral matrix $A$. Then the least number of fixed points in the homotopy class of $f$ equals the absolute value of the determinant of $E-A$, where $E$ is the identity matrix, i.e.
   $$
   Min\{\# \fix(g) \> | \> g \simeq f \}=|\det (E-A)|.
   $$
    \end {theorem}
It can be shown that $\det (E-A)$ is exactly $L(f)$ on tori and $|\det (E-A)|$ is the Nielsen number
$N(f)$ on tori.This latter theorem says much more than the Lefschetz theorem specialised to the torus, since it gives a lower bound for the number of fixed points, or it confirms the existence of a homotopic map which is fixed point free.The proof was via the universal covering space $R^2$
of the torus.From this instance evolved the central notions of Nielsen theory   -   the fixed
point classes and the Nielsen number.\\

Roughly speaking, Nielsen theory has two aspects. The geometric aspect concerns the comparison of the Nielsen number with the least number of fixed points in a homotopy class
of maps. The algebraic aspect deals with the problem of computation for the Nielsen number.
 Nielsen theory is based on the theory of covering spaces. An alternative way is to consider
 nonempty fixed point classes only, and use paths instead of covering spaces to define them.
 This is certainly more convenient for some geometric questions. But the covering space approach is theoretically more satisfactory, especially for computational problems, since the
 nonemptiness of certain fixed point classes is often the conclusion of the analysis, not the
 assumption.\\
  
  Now let us introduce the basic idea of Nielsen theory by an elementary example( see \cite{j})
  \begin{example}
 Let $f: S^1 \rightarrow S^1$ be a map of the circle. Suppose the degree of $f$ is $d$. Then
 the least number of fixed points in the homotopy class of $f$ is $ |1-d| $.
  \end{example}
  
   {\sc Proof}
Let $S^1$ be the unit circle on the complex plane, i.e. $ S^1= \{ z\in C \> | \> |z|=1 \}. $ 
Let $ p: R\rightarrow S^1$ be the exponential map $ p(\theta)= z =e^{i\theta}.$
Then $\theta$ is the argument of $z$, which is multi-valued function of $z$.
For every $f: S^1 \rightarrow S^1$, one can always find "argument expressions" ( or liftings)
$\tilde f: R \rightarrow  R $ such that $ f(e^{i\theta})=e^{i\tilde f(\theta)} $, in fact a whole 
series of them, differing from each other by integral multiples of $2\pi $.
For definiteness let us write $ \tilde f_0 $ for the argument expression with $ \tilde f_0(0)$ lying in $ [0,2\pi) $,and write  $\tilde f_k = \tilde f_0 + 2k\pi $. Since the degree  of $f$ is $d$, the functions
$\tilde f_k$ are such that $\tilde f_k(\theta + 2\pi) = \tilde f_k(\theta) + 2d\pi $.
For example, if $ f(z)= -z^d$, then $\tilde f_k(\theta)=d\theta + (2k+1)\pi $.
It is evident that if $z=e^{i\theta} $ is a fixed point of $f$, i.e. $z=f(z)$ ,then $ \theta $ is a fixed
point of some argument expression of $f$, i.e. $\theta=\tilde f_k(\theta) $ for some $k$.
On the other hand, if $\theta $ is a fixed point of $ \tilde f_k , \> q$ is an integer, then  $\theta +2q\pi $ is a fixed point of $ \tilde f_l$  iff $l-k=q(1-d) $. This follows from the calculation
$$
\tilde f_l(\theta + 2q\pi)=\tilde f_k(\theta + 2q\pi)+2(l-k)\pi= \tilde f_k(\theta) +2qd\pi + 2(l - k)\pi=
$$
$$
=(\theta + 2q\pi) + 2\pi \{(l - k) - q(1 - d )\}.
$$

Thus , if $l \not\equiv k \>  \mod \> (1 - d) $, then a fixed point of $\tilde f_k $ and a fixed point of 
$ \tilde f_l $  can never correspond to the same fixed point  of $f$ , i.e.
$ p( \fix(\tilde f_k)) \cap p( \fix(\tilde f_l)) = \emptyset.$
 So, the argument expressions fall into equivalence classes ( called lifting classes) by the relation $\tilde f_l\sim \tilde f_k$ iff $ k \equiv l \> mod \> (1-d) $, and the fixed points of $f$
 split into $|1-d| $ classes (called fixed point classes) of the form $p(\fix(\tilde f_k))$.
 That is , two fixed points are in the same class iff they come from fixed points of the same argument expression. Note that each fixed point class is by definition associated with a 
 lifting class, so that the number of fixed point classes is $ |1 - d | $ if $d \not = 1 $ and 
 is $ \infty $ if $ d=1 $. Also note that a fixed point class need not be nonempty.
 Now, to prove that a map $f$ of degree $d$ has at least $ |1-d| $ fixed points,
 we only have to show that every fixed point class is nonempty, or equivalently,
 that every argument expression has a fixed point, if $d \not = 1 $. In fact, 
 for each $k$, by means of the equality  $\tilde f_k(\theta + 2\pi) - \tilde f_k(\theta)=2d\pi $,
 it is easily seen that the function $\theta - \tilde f_k(\theta) $ takes different signs when
 $\theta $ approaches $ \pm \infty $, hence $ \tilde f_k(\theta)$ has at least one fixed point.
 That $|1-d| $ is indeed the least number of fixed points in the homotopy class is seen 
 by checking the special map $f(z) = - z^d $.\\
 
  The following sections can be considered as generalization of this simplest example.

 \section{ Lifting classes and fixed point classes}

 Let $f:X\rightarrow X$ be a continuous map of a compact connected polyhedron.
 Let $p:\tilde{X}\rightarrow X$ be the universal covering of $X$.
 A lifting of $f$ is a map  $\tilde{f}:\tilde{X}\rightarrow \tilde{X}$  such that $p\circ\tilde{f}=f\circ p$.
A covering translation is a map $\gamma : \tilde{X}\rightarrow \tilde{X}$ such that $p\circ\gamma= p$, i.e. a lifting of the identity map.
Now we describe standard facts from covering space theory
\begin{proposition}
 (1) For any $x_0 \in X $ and any $\tilde x_0, \tilde x^\prime_0 \in p^{-1}(x_0) $, there is a unique
 covering translation $\gamma : \tilde{X}\rightarrow \tilde{X}$ such that $\gamma(\tilde x_0)=\tilde x^\prime_0 $. The covering translations of  $\tilde{X} $ form a group $\Gamma $ which is isomorphic to $\pi_1(X)$ .\\
 (2) Let $f:X\rightarrow X$ be a continuous map. For given $x_0 \in X$ and $ x_1 =f(x_0) $,
 pick $\tilde x_0 \in p^{-1}(x_0) $ and $\tilde x_1 \in p^{-1}(x_1) $ arbitrarily. Then, there
 is a unique lifting of $f$ such that $\tilde{f}(\tilde x_0)=\tilde x_1$.\\
 (3) Suppose $\tilde{f}$ is a lifting of $f$, and $\alpha, \beta \in \Gamma $. Then $ \beta \circ \tilde{f} \circ \alpha $ is a lifting of $f$.\\
 (4) For any two liftings  $\tilde{f}$ and $\tilde{f^\prime}$ of $f$, there is a unique $ \gamma \in \Gamma$, such that $\tilde{f^\prime}=\gamma \circ \tilde{f}$.
 \end{proposition}

 \begin{lemma}
 Suppose $\tilde x \in p^{-1}(x) $ is a fixed point of lifting $\tilde{f}$ of $f$, and 
  $ \gamma \in \Gamma$ is a covering translation on $\tilde X $. Then, a lifting $\tilde{f^\prime}$ of $f$
  has $\gamma(\tilde x) \in p^{-1}(x) $ as a fixed point iff $\tilde{f^\prime}= \gamma \circ \tilde{f} \circ  \gamma^{-1}.$
 \end{lemma}
 
 {\sc Proof }
 "If" is obvious : $\tilde{f^\prime}(\gamma(\tilde x))= \gamma \circ \tilde{f} \circ  \gamma^{-1}(\gamma(\tilde x))=\gamma \circ \tilde{f}(\tilde x)=\gamma(\tilde x).$\\
  " Only if" : Both $\tilde{f^\prime}$ and $\gamma \circ \tilde{f} \circ  \gamma^{-1}$ have 
 $\gamma(\tilde x)$ as a fixed point , so they agree at the point $\gamma(\tilde x)$.
 By  Proposition 1 they are the same lifting.
 
 \begin{definition}
 Two liftings $\tilde{f^\prime}$ and $\tilde{f}$  of $f$ are said to be conjugate if there
 exists $ \gamma \in \Gamma$, such that $\tilde{f^\prime}= \gamma \circ \tilde{f} \circ  \gamma^{-1}.$
Lifting classes are equivalence classes by conjugacy.Notation: 
$$
[\tilde{f}]= \{ \gamma \circ \tilde{f} \circ  \gamma^{-1} \> | \> \gamma \in \Gamma \}
$$ 
 \end{definition}
 
 \begin{lemma}
 (1) $ \fix (f) = \cup_{\tilde{f}} p( \fix (\tilde{f})). $\\
 (2) $ p( \fix (\tilde{f})) =p( \fix (\tilde{f^\prime}))$ if $[\tilde{f}]=[\tilde{f^\prime}]$.\\
 (3) $p( \fix (\tilde{f}))  \cap p( \fix (\tilde{f^\prime})) = \emptyset $ if $[\tilde{f}] \not=[\tilde{f^\prime}]$. 
 \end{lemma}
 
  {\sc Proof }
 (1) If $x_0 \in \fix (f)$, pick $ \tilde x_0 \in p^{-1}(x_0) $. By proposition 1 there exists 
 $\tilde{f}$ such that $\tilde{f}(\tilde x_0)=\tilde x_0$. Hence $x_0 \in p( \fix (\tilde{f}))$.\\
 (2) If $\tilde{f^\prime}= \gamma \circ \tilde{f} \circ  \gamma^{-1}$, then by lemma    $\fix (\tilde{f^\prime})=
 \gamma \fix (\tilde{f})$, so  $ p (\fix (\tilde{f})) =p (\fix (\tilde{f^\prime}))$.\\
 (3) If $x_0 \in p( \fix (\tilde{f}))  \cap p( \fix(\tilde{f^\prime}))$, there are $ \tilde x_0,\tilde x^\prime_0  \in p^{-1}(x_0) $ such that $ \tilde x_0 \in \fix (\tilde{f})$ and $ \tilde x^\prime_0 \in \fix (\tilde{f^\prime})$.
 Suppose $\tilde x^\prime_0= \gamma \tilde x_0$.By lemma   , $\tilde{f^\prime}= \gamma \circ \tilde{f} \circ  \gamma^{-1},$ hence $[\tilde{f}]=[\tilde{f^\prime}] $.
 
 \begin{definition}
The subset $ p( \fix (\tilde{f}))$ of $ \fix (f)$ is called the fixed point class of $f$ determined by the lifting class $[\tilde{f}]$.
 \end{definition}
 We see that the fixed point set $\fix (f)$ splits into disjoint union of fixed point classes.

 \begin{example}
Let us consider the identity map $id_X: X \to X$. Then a lifting class is a usual conjugasy class in
$\Gamma ;p( \fix (id_{\tilde X})) = X$ and $p( \fix (\gamma))= \emptyset $ otherwise.
\end{example}

 \begin{remark}
 A fixed point class is always considered to carry a label - the lifting class determining it .
 Thus two empty fixed point classes are considered different  if they are determined by different
 lifting classes.
 \end{remark}
 Our definition of a fixed point class is via the universal covering space. It essentially 
 says: Two fixed point of $f$ are in the same class iff there is a lifting $\tilde f $ of $f$ having
 fixed points above both of them. There is another way of saying this, which does not use covering space explicitly, hence is  very useful in identifying fixed point classes.
 \begin{lemma}[\cite{j}]
 Two fixed points $x_0$ and $x_1$ of $f$ belong to the same fixed point class iff
 there is a path $c$ from $x_0$ to $x_1$ such that $c \cong f\circ c $ ( homotopy relative
    endpoints).
 \end{lemma}
 
 Lemma 3  can be considered as an equivalent definition of a non-empty fixed point class.
 Every map $f$  has only finitely many non-empty fixed point classes, each a compact
 subset of $X$.

\subsection{The influence of a homotopy}
Given a homotopy $H=\{h_t\}: f_0 \cong f_1$, we want to see its influence on fixed point classes
of $f_0$ and $f_1$. A homotopy $\tilde H=\{\tilde h_t\}: \tilde X \to \tilde X$ is called a lifting of the homotopy $H=\{h_t\}$, if $\tilde h_t$ is a lifting of $h_t$ for every $t\in I$.
Given a homotopy $H$ and a lifting $\tilde f_0$ of $f_0$, there is a unique lifting  $\tilde H$
of $H$ such that $ \tilde h_0=\tilde f_0$, hence by unique lifting property of covering spaces 
they determine a lifting $ \tilde f_1$ of $f_1$. Thus $H$ gives rise to a one-one correspondence
from liftings of $f_0$ to liftings  of $f_1$.
This correspondence preserves the conjugacy relation.Thus there is a one-to-one 
correspondence between lifting classes and fixed point classes of $f_0$ and those of $f_1$.

\section{ Reidemeister numbers}
\subsection{Reidemeister numbers of a continuous map}

\begin{definition}
The number of lifting classes of $f$ (and hence the number
of fixed point classes, empty or not) is called the $Reidemeister$ $number$ of $f$,
denoted by $R(f)$. It is a positive integer or infinity.
\end{definition}
The Reidemeister number $R(f)$ is a homotopy invariant.

\begin{example}
If $X$ is simply-connected then $R(f)=1$.
\end{example}

Let $f:X\rightarrow X$ be given, and let a
specific lifting $\tilde{f}:\tilde{X}\rightarrow\tilde{X}$ be chosen
as reference.
Let $\Gamma$ be the group of
covering translations of $\tilde{X}$ over $X$.
Then every lifting of $f$ can be written uniquely
as $\alpha\circ \tilde{f}$, with $\alpha\in\Gamma$.
So elements of $\Gamma$ serve as coordinates of
liftings with respect to the reference $\tilde{f}$.
Now for every $\alpha\in\Gamma$ the composition $\tilde{f}\circ\alpha$
is a lifting of $f$ so there is a unique $\alpha^\prime\in\Gamma$
such that $\alpha^\prime\circ\tilde{f}=\tilde{f}\circ\alpha$.
This correspondence $\alpha\rightarrow\alpha^\prime$ is determined by
the reference $\tilde{f}$, and is obviously a homomorphism.

\begin{definition}
The endomorphism $\tilde{f}_*:\Gamma\rightarrow\Gamma$ determined
by the lifting $\tilde{f}$ of $f$ is defined by
$$
  \tilde{f}_*(\alpha)\circ\tilde{f} = \tilde{f}\circ\alpha.
$$
\end{definition}

It is well known that $\Gamma\cong\pi_1(X)$.
We shall identify $\pi=\pi_1(X,x_0)$ and $\Gamma$ in the following way.
Pick base points $x_0\in X$ and $\tilde{x}_0\in p^{-1}(x_0)\subset \tilde{X}$
once and for all.
Now points of $\tilde{X}$ are in 1-1 correspondence with homotopy classes of paths
in $X$ which start at $x_0$:
for $\tilde{x}\in\tilde{X}$ take any path in $\tilde{X}$ from $\tilde{x}_0$ to 
$\tilde{x}$ and project it onto $X$;
conversely for a path $c$ starting at $x_0$, lift it to a path in $\tilde{X}$
which starts at $\tilde{x}_0$, and then take its endpoint.
In this way, we identify a point of $\tilde{X}$ with
a path class $<c>$ in $X$ starting from $x_0$. Under this identification,
$\tilde{x}_0=<e>$ is the unit element in $\pi_1(X,x_0)$.
The action of the loop class $\alpha = <a>\in\pi_1(X,x_0)$ on $\tilde{X}$
is then given by
$$
\alpha = <a> : <c>\rightarrow \alpha . c = <a.c>.
$$
Now we have the following relationship between $\tilde{f}_*:\pi\rightarrow\pi$
and
$$
f_*  :  \pi_1(X,x_0) \longrightarrow \pi_1(X,f(x_0)).
$$

\begin{lemma}[\cite{j}]
Suppose $\tilde{f}(\tilde{x}_0) = <w>$.
Then the following diagram commutes:
$$
\begin{array}{ccc}
  \pi_1(X,x_0)  &  \stackrel{f_*}{\longrightarrow}  &  \pi_1(X,f(x_0))  \\
                &  \tilde{f}_* \searrow \;\; &  \downarrow w_*   \\
                &                           &  \pi_1(X,x_0)
\end{array}
$$
\end{lemma}

We have seen that $\alpha \in \pi$ can be considered as the coordinate of the
lifting $\alpha \circ \tilde f$. Can we tell the conjugacy of two liftings from their coordinates?

\begin{lemma}
$[\alpha \circ\tilde f]=[\alpha^\prime \circ\tilde f] $  iff there is $\gamma  \in \pi$ such that
$\alpha^\prime=\gamma \alpha \tilde{f}_* (\gamma^{-1})$. 
\end{lemma}
{\sc Proof}
$[\alpha \circ\tilde f]=[\alpha^\prime \circ\tilde f] $  iff there is $\gamma  \in \pi$ such that
$\alpha^\prime\circ\tilde f=\gamma\circ(\alpha \circ\tilde f)\circ\gamma^{-1}=\gamma \alpha \tilde{f}_* (\gamma^{-1})\circ\tilde f$.

 \begin{theorem}[\cite{j}]
 Lifting classes of $f$ are in 1-1 correspondence with $\tilde{f}_*$-conjugacy classes in $\pi$,
 the lifting class $[\alpha\circ\tilde f]$ corresponds to the $\tilde{f}_*$-cojugacy class of $\alpha$.
  \end{theorem}
  
   By an abuse of language, we will say that the fixed point class $p( \fix(\alpha\circ\tilde f))$,
which is labeled with the lifting class $[\alpha\circ\tilde f]$,corresponds to the $\tilde{f}_*$-conjugacy class of $\alpha$. Thus the $\tilde{f}_*$-conjugacy classes in $\pi$ serve 
as coordinates for the fixed point classes of $f$, once a reference lifting $\tilde f$ is chosen.
  
  A reasonable approach to finding a lower bounds
for the Reidemeister number, is to consider a homomorphisms from $\pi$ sending
an $\tilde{f}_*$-conjugacy class to one element:

\begin{lemma}[\cite{j}]
The composition $\eta\circ\theta$,
$$
 \pi = \pi_1(X,x_0) \stackrel{\theta}{\longrightarrow} H_1(X)
       \stackrel{\eta}{\longrightarrow}
       \coker\left[H_1(X) \stackrel{1-f_{1*}}{\longrightarrow} H_1(X)\right] ,
$$
where $\theta$ is abelianization and $\eta$ is the natural projection,
sends every $\tilde{f}_*$-conjugacy class to a single element.
Moreover, any group homomorphism $\zeta:\pi\rightarrow G$ which
sends every $\tilde{f}_*$-conjugacy class to a single element,
factors through $\eta\circ\theta$.
\end{lemma}

The first part of this lemma is trivial.If $\alpha^\prime=\gamma \alpha \tilde{f}_* (\gamma^{-1})$ , then
$$
\theta(\alpha^\prime )=\theta(\gamma) + \theta(\alpha) + \theta(\tilde{f}_* (\gamma^{-1}))=
$$
$$
=\theta(\gamma) + \theta(\alpha) - f_{1*}(\theta(\gamma))=\theta(\alpha) +(1- f_{1*})\theta(\gamma), 
$$  
hence  $\eta\circ \theta(\alpha)= \eta\circ \theta(\alpha^\prime)$

This lemma shows the importance of the group $ \coker(1 -f_{1*})$. For example
$$
R(f)  \geq \# \coker(1 -f_{1*}).
$$

\begin{definition}
A map $f:X\rightarrow X$ is said to be {\it eventually commutative}
if there exists an natural number $n$ such that
$f^n_*(\pi_1(X,x_0))\;\; (\subset \pi_1(X,f^n(x_0)))$ is commutative.
\end{definition}
By means of Lemma 4, it is easily seen that $f$ is eventually
commutative iff $\tilde{f}_*$ is eventually commutative (see \cite{j})

\begin{theorem}[\cite{j}]
If $f$ is eventually commutative , then
$$
R(f)=\# \coker(1 -f_{1*}).
$$

\end{theorem}

 \subsection{Reidemeister numbers of a group endomorp\-hism}

Let $G$ be a group and $\phi:G\rightarrow G$ an endomorphism.
Two elements $\alpha,\alpha^\prime\in G$ are said to be
$\phi-conjugate$ iff there exists $\gamma \in G$ such that
$\alpha^\prime=\gamma \cdot \alpha \cdot \phi(\gamma)^{-1}$.
The number of $\phi$-conjugacy classes is called the $Reidemeister$
$number$ of $\phi$, denoted by $R(\phi)$.
We shall write $\{g\}$ for the $\phi$-conjugacy class of an element $g\in G$.
We shall also write $\cR(\phi)$ for the set of
 $\phi$-conjugacy classes of elements of $G$.
If $\phi$ is the identity map then the $\phi$-conjugacy classes
 are the usual conjugacy classes in the group $G$. 

An endomorphism $\phi:G\rightarrow G$ is said to be eventually
commutative if there exists a natural number $n$ such that the
subgroup $\phi^n(G)$ is commutative.

We are now ready to compare the Reidemeister number of an
endomorphism $\phi$ with the Reidemeister number
of $H_1(\phi):H_1(G)\rightarrow H_1(G)$,
where $H_1 = H_1^{Gp}$ is the first integral homology functor
from groups to abelian groups.

\begin{theorem}[\cite{j}]
If $\phi:G\rightarrow G$ is eventually commutative,
then
$$
R(\phi)
 = R(H_1(\phi))
 = \#\coker (1-H_1(\phi)).
$$
\end{theorem}

This means that to find out about the Reidemeister numbers
of eventually commutative endomorphisms, it is sufficient to
study the Reidemeister numbers of endomorphisms of abelian groups.

From theorem 3  it follows
\begin{corollary}
$R(f)=R(\tilde {f}_*)$
\end{corollary}
So we see that the homotopy invariant - the Reidemeister number of a continuous map is the same as the Reidemeister number of the induced endomorphism $\tilde {f}_*$ on the fundamental group.
The following are simple but very useful facts about $\phi$-conjugacy classes.
\begin{lemma}[\cite{j}]
If $G$ is a group and $\phi$ is an endomorphism of $G$ then an element $x\in G$ is 
always $\phi$-conjugate to its image $\phi(x)$.
\end{lemma}
{\sc Proof} 
If $g=x^{-1}$ then one has immediately $gx=\phi(x)\phi(g)$. The existence of a $g$
satisfying this equation implies that $x$ and $\phi(x)$ are $\phi$-conjugate.

\section{ Nielsen  numbers of a continuous map}

\subsection{ The fixed point index}

The fixed point index provides an algebraic count of fixed points.
We will introduce step-by-step construction of the index, and list without proof
 the most useful properties.The reader may consult the book of  Dold \cite{d}.

\bigskip

(A) THE INDEX OF AN ISOLATED FIXED POINT IN $R^n$.

\medskip

Suppose $R^n \supset U  \stackrel{f}{\longrightarrow} R^n $, and $a \in U$ is an
isolated fixed point of $f$.Pick a sphere ${S_a}^{n-1}$ centered at $a$, small enough to exclude other fixed points.On ${S_a}^{n-1}$ , the vector $x - f(x) \not=0$, so a direction field
$$
\phi:{S_a}^{n-1}\to S^{n-1}, \> \phi(x)=\frac{x- f(x)}{| x - f(x) |},
$$
is defined.
\begin{definition}
\ind(f,a)= degree of $\phi$
\end{definition}
\begin{example}
If $f$ is a constant map to the point $a$, then $ \ind(f,a)=1$
\end{example}

\begin{example}
Suppose $f$ is differentiable at $a$ with Jacobian $ A=(\frac{\partial f}{\partial x})_a$, and 
$\det(I-A)\not=0$.Then $a$ is an isolated fixed point and
$$
\ind(f,a)={\rm sign}\, \det(I-A)=(-1)^k,
$$
where $k$ is the number (counted with multiplicity) of real eigenvalues of $A$
greater than 1.
\end{example}

(B)  FIXED POINT INDEX IN $R^n$.

\begin{definition}
Suppose  $R^n \supset U \stackrel{f}{\longrightarrow} R^n $, and $ Fix(f) $ is compact.
Take any open set $V\subset U$ such that $ \fix (f) \subset V \subset \bar V \subset U $
and $\bar V$ is a smooth  n-manifold, then $\phi(\partial V)=i\cdot S^{n-1}$ in the homological sense for some $i \in \bbbz$. This $i$ is independent of the choice of $V$, and is defined
to be the fixed point index of $f$  on $U$, denoted $\ind (f,U)$. 
\end{definition}
Another way of defining $\ind(f,U)$ is by approximation, namely, approximate $f$ by a 
smooth map with only generic fixed points( isolated fixed points satisfying the condition
in example 5 ), and add up their indices.

\bigskip

(C) FIXED POINT INDEX FOR POLYHEDRA.

\medskip

Every compact polyhedron can be embedded in some Euclidean space as a neighborhood
retract.
Suppose now we are given a compact polyhedron $X$, and a map $X \supset U  \stackrel{f}{\longrightarrow} X $. $X$ can be imbedded in $R^N$ with inclusion $X\stackrel{i}{\longrightarrow} R^N$. And there is a neighborhood $W$ of $i(X)$ in $R^N$
and a retraction $W\stackrel{i}{\longrightarrow} X$ such that $r\circ i =id_X$. We have
a diagram

$$
 \begin{array}{cclclcccc}
                      X & \supset & U & \stackrel{f}{\longrightarrow} & X\\
                      \uparrow r  &     & \uparrow r& & \downarrow i \\
R^N\supset  W & \supset & r^{-1}(U) & \stackrel{i\circ f\circ r}{\longrightarrow} & R^N 
 \end{array}.
$$

\begin{definition}
When $Fix(f)$ is compact, define the fixed point index to be 
$$
\ind(f,U):= \ind(i\circ f\circ r, r^{-1}(U)),
$$
the later being the index in $R^N$. It is independent of the choice of $N, W, i $ and $r$.
\end{definition}

All the facts we need about the fixed point index are listed below.

(I) Existence of fixed points. If $\ind(f,U) \not=0$, then $f$ has at least one fixed point in $U$.

(II) Homotopy invariance. If $H=\{h_t\}: f_0\simeq f_1 : U\to X$ is a homotopy such that $\cup_{t\in I} \fix (h_t)$ is compact, then
$$
\ind(f_0,U)=\ind(f_1,U).
$$

(III) Additivity. Suppose $U_1, ...., U_s$ are disjoint open subsets of $U$, and $f$ has no
fixed points on  $ U - \cup_{j=1}^s U_j $. If $\ind(f,U)$ is defined, then $\ind(f,U_j)$, 
$ j=1,...,s$ are
all defined and
$$
\ind(f,U)=\sum_{j=1}^s \ind(f,U_j).
$$

(IV) Normalization. If $f: X\rightarrow X$ , then
$$
\ind(f, X)=  L(f) := \sum_{k=0}^{\dim X} (-1)^k \tr\Big[f_{*k}:H_k(X;\bbbq)\to H_k(X;\bbbq)\Big]
$$
where  $L(f)$ is the Lefschetz number of $f$.
\begin{lemma}[\cite{j}](Homotopy invariance)
Let $X$ be a connected, compact polyhedron, $H=\{h_t\}: f_0\simeq f_1 : X\to X$ be a homotopy, and $F_i$ be a fixed point class of $f_i, i=0,1$. If $F_0$ corresponds to $F_1$
via $H$, then 
$$
\ind(f_0,F_0)=\ind(f_1,F_1)
$$
\end{lemma}

\subsection{Nielsen numbers}

\begin{definition}
A fixed point class is called $essential$ if its index is nonzero.The number of essential fixed point classes is called the $Nielsen$ $number$ of $f$, denoted by $N(f)$.
\end{definition}
 The next lemma follows directly from the definitions and the properties of the index.
 \begin{lemma}
 (1) $N(f) \leq R(f) $.
 
 (2) Each essential fixed point class is non-empty.
 
 (3) $N(f) $ is non-negative integer, $0\leq N(f) < \infty$.
 
 (4) $N(f) \leq \# \fix(f) $.
 
 (5) The sum of the indices of all essential fixed point classes of $f$ equals to the Lefschetz
 number $L(f)$. Hence $L(f) \not=0$ implies $N(f)\geq 1$.
  \end{lemma} 

\begin{example}
$$
N(id_X)= \left \{
\begin{array}{ll}
1 & {\rm if \> Euler \> characteristic}\ \chi(X)\not=0,  \\
0 & {\rm if}\ \chi(X)=0 \\
\end{array}
\right.
$$ 
\end{example}

\begin{lemma}
$N(f)$ is a homotopy invariant of $f$. Every map homotopic to $f$ has at least $N(f)$ fixed points. In other words, $N(f)\leq Min\{ \# Fix (g)  |  g \simeq f\}.$ 
\end{lemma}
It is this lemma that made the Nielsen number so important in fixed point theory.
Lemma  tells us that an essential fixed point class can never disappear (i.e. become empty) via a
homotopy.

Two maps $f: X\to X$ and $g: Y\to Y $ are said to be of the same homotopy type if there
is a homotopy equivalence $h: X\to Y $ such that $h\circ f \simeq g\circ h $. 

\begin{theorem}[\cite{j}]( Homotopy type invariance of the Nielsen number). 
If $X, Y $ are compact connected polyhedra, and $f: X\to X$ and $g: Y\to Y $ are  of the same homotopy type, then $N(f)= N(g)$.
\end{theorem}

\subsection{The least number of fixed points}
Let $X$ be a compact connected polyhedron , and let $f: X\to X$ be a map.
Consider the number 
$$
MF[f]:=Min\{ \# \fix (g)  |  g \simeq f\},
$$
i.e. the least number of fixed points in the homotopy class $[f]$ of $f$. We know that
$N(f)$ is a lower bound for $MF[f]$. The importance of the Nielsen number in the fixed point 
theory lies in the fact that, under a mild restriction on the space involved , it is indeed
the minimal number of fixed points in the homotopy class.The equality $MF[f]=N(f)$ means 
that we can homotope the map $f$ so that each essential fixed point class is combined into a single fixed point and each inessential fixed point class is removed.

\begin{definition}
A point $x$ of a connected space $X$ is a (global) separating point of $X$ if $X-x$ is not connected. A point $x$ of a space $X$  is a local separating point if $x$ is a separating point
of some connected open subspace $U$ of $X$. 
\end{definition}

\begin{theorem}[\cite{j}]
Let $X$ be a compact connected polyhedron without local separating points. Suppose $X$ is not a surface (closed or with boundary) of negative Euler characteristic. Then $MF[f]=N(f)$ for any map $f: X\to X$.
\end{theorem}

This theorem is generalization of the classical theorem of Wecken \cite{wec} for manifolds of dimension $\geq 3$.

\begin{lemma}[\cite{j}](Geometric characterization  of the Nielsen number).
In the category of compact connected polyhedra, the Nielsen number of a self- map
equals the least number of fixed points among all self-maps having the same homotopy type.
\end{lemma}
{\sc Proof}
Let $f: X\to X$  be a self -map in the category , and consider the number
$ m= Min \{ \# \fix (g)  |  g \> has \>the \> same \> homotopy\> type \> as \> f\}.$

Then $N(f) \leq m $ by lemma 9 and theorem 6 . On the other hand , there always exists a 
manifold $M$ (with boundary) of dimension $\geq 3$  which has the same homotopy type as $X$
( for example , $M$= the regular neighborhood of $X$ imbedded in a Euclidean space). So , by the theorem 7, the lower bound $N(f)$ is realizable on $M$ by a map having the same homotopy type as $f$.

\chapter{The Reidemeister zeta function}
\markboth{\sc Reidemeister zeta function}{\sc Reidemeister zeta function}

\begin{quotation}
 {\sc PROBLEM}.
For which groups and endomorphisms is the Reidemeister zeta function
a rational function? When does it have a functional equation?
Is $R_\phi(z)$ an algebraic function?
\end{quotation}

\section{A Convolution Product}

When $R_\phi(z)$ is a rational function the infinite sequence
 $\{R(\phi^n)\}_{n=1}^\infty$ of Reidemeister numbers
is determined by a finite set of complex numbers - the zeros and
poles of $R_\phi(z)$.
 
\begin{lemma}
$R_\phi(z)$ is a rational function if and only if
there exists a finite set of complex numbers $\alpha_i$ and $\beta_j$
such that
$ R(\phi^n) = \sum_j \beta_j^n - \sum_i \alpha_i^n $
for every $n>0$.
\end{lemma}
 
{\sc Proof}
Suppose $R_\phi(z)$ is a rational function.
Then
$$
R_\phi(z) = \frac{\prod_i (1-\alpha_i z)}{\prod_j (1-\beta_j z)},
$$
where $\alpha_i, \beta_j \in \bbbc$.
Taking the logarithmic derivative of both sides and then using the
geometric series expansion we see that
$R(\phi^n) = \sum_j\beta_j^n - \sum_i \alpha_i^n$.
The converse is proved by a direct calculation.
 \vspace{0.2cm}
 
For two sequences $(x_n)$ and $(y_n)$ we may define the
corresponding zeta functions:
 $$X(z) := \exp\left( \sum_{n=1}^\infty \frac{x_n}{n} z^n \right),$$
 $$Y(z) := \exp\left( \sum_{n=1}^\infty \frac{y_n}{n} z^n \right).$$
Alternately, given complex functions $X$ and $Y$ (defined in a 
neighborhood of 0) we may define sequences
 $$ x_n := \frac{d^n}{dz^n}\log\left(X(z)\right)\mid_{z=0} ,$$
 $$ y_n := \frac{d^n}{dz^n}\log\left(Y(z)\right)\mid_{z=0} .$$
Taking the componentwise product of the two sequences gives
another sequence, from which we obtain another complex function.
We call this new function the {\it additive convolution} of $X$ and $Y$, and we write it
 $$ (X*Y)(z) := \exp\left( \sum_{n=1}^\infty \frac{x_n . y_n}{n} z^n \right).$$
It follows immediately from lemma 1 that if $X$ and $Y$ are
rational functions then $X*Y$ is a rational function.
In fact we may show using the same method the following
 
\begin{lemma}[Convolution of rational functions]
Let 
$$
X(z)=\prod_i (1-\alpha_i z)^{m(i)},\;\;\;
 Y(z)=\prod_j (1-\beta_j z)^{l(j)}
$$
be rational functions in $z$.
Then $X*Y$ is the following rational function
\begin{equation}
  (X*Y)(z) = \prod_{i,j} (1-\alpha_i \beta_j z)^{-m(i).l(j)}.
\end{equation}
\end{lemma}
A consequence of this is the following
 
\begin{lemma}[Functional equation of a convolution]
Let $X(z)$ and $Y(z)$ be rational functions satisfying
the following functional equations
$$
 X\left( \frac{1}{d_1.z} \right)
 = K_1 z^{-e_1} X(z)^{f_1},\;\;\;
 Y\left( \frac{1}{d_2.z} \right)
 = K_2 z^{-e_2} Y(z)^{f_2},\;\;\;
$$
with $d_i\in\bbbc^\times$, $e_i\in\bbbz$ ,$K_i\in\bbbc^\times$ and $f_i\in\{1,-1\}$.
Suppose also that $X(0)=Y(0)=1$.
Then the rational function $X*Y$ has the following functional equation:
 \begin{equation}
 (X*Y)\left(\frac{1}{d_1 d_2 z}\right)
 =
 K_3 z^{-e_1 e_2} (X*Y)(z)^{f_1 f_2}
\end{equation}
for some $K_3\in\bbbc^\times$.
\end{lemma}
 
{\sc Proof}
The functions $X$ and $Y$ have representations of the following form:
$$
X(z)=\prod_{i=1}^a (1-\alpha_i z)^{m(i)},\;\;\;
Y(z)=\prod_{j=1}^b (1-\beta_j z)^{l(j)}.
$$
The functional equation for $X$ means that if $z_0\in\bbbc^\times$ is a zero or pole
of $X$ with multiplicity $m$, then $\frac{1}{d_1.z_0}$ must be a
pole or zero of $X$ with multiplicity $f_1 m$.
We therefore have a map
 $$ i \longmapsto i^\prime$$
defined by the equation
 $$ \alpha_i = \frac{1}{d_1\alpha_{i^\prime}} $$
and with the property that
  $$ m(i) = f_1 m(i^\prime) .$$
Similarly there is a map $j\mapsto j^\prime$ with corresponding properties:
 $$ \beta_j = \frac{1}{d_2\beta_{j^\prime}},$$
 $$ l(j) = f_2 l(j^\prime) .$$
Putting these maps together, we obtain a map of pairs
 $$
 (i,j) \longmapsto (i^\prime,j^\prime)
 $$
with the following properties
 $$ \alpha_i \beta_j= \frac{1}{d_1 d_2 \alpha_{i^\prime} \beta_{j^\prime}},$$
 $$ m(i) l(j) = f_1 f_2 m(i^\prime) l(j^\prime).$$
It follows from the previous lemma
that the functions $(X*Y)(z)$ and $(X*Y)(\frac{1}{d_1 d_2 z})^{f_1 f_2}$ have
the same zeros and poles in the domain $\bbbc^\times$.
Since both functions are rational, we deduce that
 $X*Y$ has a functional equation of the form
 $$
  (X*Y)\left(\frac{1}{d_1 d_2 z}\right)
 =
 K_3 z^{-e_3} (X*Y)(z)^{f_1 f_2},
 $$
 for some $e_3\in\bbbz$.
It remains only to show that $e_3=e_1 e_2$.
  
By comparing the degrees at zero of the functions $X(z)$
 and $X(\frac{1}{d_1 z})$ we obtain from the functional equation for
 $X$:
 $$
 e_1 = \sum_{i=1}^a  m(i).
 $$
Similarly we have
 $$
 e_2 = \sum_{j=1}^b  l(j).
 $$
For the same reasons we have
\begin{eqnarray*}
 e_3 & = & \sum_{i,j} m(i) l(j) \\
     & = & \Big( \sum_{i=1}^a  m(i) \Big)\Big(\sum_{j=1}^b  l(j)\Big)\\
     & = & e_1 e_2 .
\end{eqnarray*}

\section{ Pontryagin Duality}

Let $G$ be a locally compact Abelian topological group.
We write $\hat{G}$ for the set of continuous
homomorphisms from $G$ to the circle $U(1)=\{z\in\bbbc : |z|=1\}$.
This is a group with pointwise multiplication. We call $\hat{G}$
the {\it Pontryagin dual} of $G$. When we equip $\hat{G}$ with
the compact-open topology it becomes a locally compact Abelian
topological group. The dual of the dual of $G$ is canonically
isomorphic to $G$.
 
A continuous endomorphism $f:G\rightarrow G$ gives rise
to a continuous endomorphism $\hat{f}:\hat{G}\rightarrow \hat{G}$ defined by
$$
 \hat{f}(\chi) := \chi\circ f.
$$
There is a 1-1 correspondence between the closed subgroups $H$ of $G$
and the quotient groups $\hat{G}/H^*$ of $\hat{G}$ for which
 $H^*$ is closed in $\hat{G}$.
This correspondence is given by the following:
 $$  H \leftrightarrow \hat{G}/H^*, $$
 $$
 H^* := \{\chi \in \hat{G} \mid H\subset\ker\chi\}.
 $$
Under this correspondence, $\hat{G}/H^*$ is canonically
isomorphic to the Pontryagin dual of $H$.
If we identify $G$ canonically with the dual of $\hat{G}$
then we have $H^{**}=H$.
 
If $G$ is a finitely generated free Abelian group then
a homomorphism $\chi:G\rightarrow U(1)$ is completely
determined by its values on a basis of $G$, and
these values may be chosen arbitrarily. The dual of $G$
is thus a torus whose dimension is equal to the rank of $G$.
 
If $G=\bbbz/n \bbbz$ then the elements of $\hat{G}$
are of the form
 $$ x \rightarrow e^{\frac{2\pi i y x}{n}} $$
with $y\in\{1,2,\ldots,n\}$.
A cyclic group is therefore (uncanonically) isomorphic to
itself.
  
The dual of $G_1\oplus G_2$ is canonically isomorphic
to $\hat{G}_1 \oplus\hat{G}_2$. From this we see that
any finite abelian group is (non-canonically) isomorphic to
its own Pontryagin dual group, and that the dual of any finitely generated
discrete Abelian group is the direct sum of a Torus and a
finite group.
 
Proofs of all these statements may be found, for example in
\cite{ru}.
We shall require the following statement:
 
\begin{proposition}
Let $\phi:G\to G$ be an endomorphism of an Abelian group $G$.
Then the kernel $\ker\left[\hat{\phi}:\hat{G}\to\hat{G}\right]$
is canonically isomorphic to the
Pontryagin dual of $\coker\phi$.
\end{proposition}
 
{\sc Proof}
We construct the isomorphism explicitly.
Let $\chi$ be in the dual of $\coker(\phi:G\to G)$.
In that case $\chi$ is a homomorphism
  $$  \chi : G / \im(\phi) \longrightarrow U(1).$$
There is therefore an induced map
 $$  \overline{\chi} : G \longrightarrow U(1) $$
which is trivial on $\im(\phi)$.
This means that $\overline{\chi}\circ\phi$ is trivial, or
in other words $\hat{\phi}(\overline{\chi})$ is the identity element of $\hat{G}$.
We therefore have $\overline{\chi}\in\ker(\hat{\phi})$.
 
If on the other hand we begin with $\overline{\chi}\in\ker(\hat{\phi})$,
then it follows that $\chi$ is trivial on $\im\phi$, and so
$\overline{\chi}$ induces a homomorphism
 $$ \chi : G / \im(\phi) \longrightarrow U(1)$$
and $\chi$ is then in the dual of $\coker\phi$.
The correspondence $\chi\leftrightarrow\overline{\chi}$ is
clearly a bijection.

\section { Eventually commutative endomorp\-hisms}

An endomorphism $\phi:G\rightarrow G$ is said to be eventually
commutative if there exists a natural number $n$ such that the
subgroup $\phi^n(G)$ is commutative.

\subsection{Trace formula for the Reidemeister numbers of eventually
commutative endomorphisms}
 We know from theorem 5 in Chapter 1 that if $\phi:G\rightarrow G$ is eventually commutative,
then
$$
 R(\phi) = R({H_1(\phi))}= \#\coker(1- H_1(\phi)) 
$$
 
This means that to find out about the Reidemeister numbers
of eventually commutative endomorphisms, it is sufficient to
study the Reidemeister numbers of endomorphisms of abelian groups.
For the rest of this section G will be a finitely generated
abelian group.

\begin{lemma}
Let $\phi:\bbbz^k\rightarrow\bbbz^k$ be a group endomorphism.
Then we have
 \begin{equation}
R(\phi) = (-1)^{r+p} \sum_{i=0}^k (-1)^i \tr(\Lambda^i\phi).
\end{equation}
where $p$ the number of $\mu\in\spec\phi$ such that
$\mu <-1$, and $r$ the number of real eigenvalues of $\phi$ whose
absolute value is $>1$.
$\Lambda^i$ denotes the exterior power.
\end{lemma}
 
{\sc Proof}
Since $\bbbz^k$ is Abelian, we have as before,
 $$  R(\phi) = \#\coker(1-\phi) .$$
On the other hand we have
 $$ \#\coker(1-\phi) = \mid\det(1-\phi)\mid ,$$
and hence $R(\phi)=(-1)^{r+p}\det(1-\phi)$ ( complex eigenvalues contribute nothing to the sign$\det(1-\phi)$ since they come in conjugate pairs and $ (1-\lambda)(1-\bar \lambda)=\mid 1-\lambda \mid^2 > 0$).
It is well known from linear algebra that
$\det(1-\phi)=\sum_{i=0}^k (-1)^i \tr (\Lambda^i\phi)$.
From this we have the  trace formula for Reidemeister number.
 
 Now let  $\phi$ be an endomorphism of finite Abelian group $G$.Let $V$ be the complex vector space of complex valued functions on the group $G$.The map $\phi$ induces a linear map $A:V\rightarrow V$ defined by
$$
A(f):=f\circ\phi.
$$

\begin{lemma}
Let $\phi:G\rightarrow G$ be an endomorphism
of a finite Abelian group $G$. Then we have
\begin{equation}
R(\phi) = TrA
\end{equation}
\end{lemma}
We give two proofs of this lemma .
The first proof is given here and the second
proof is a special case of the proof of theorem 15 
 
{\sc Proof}
The characteristic functions of the elements of $G$ form a basis of $V$,and are mapped to one another by $A$(the  map need not be a bijection).Therefore the trace of $A$ is the number of
elements of this basis which are fixed by $A$.On the other hand, since $G$ is Abelian, we  have,
\begin{eqnarray*}
 R(\phi)
 & = & 
 \#\coker(1-\phi) \\
 & = & 
 \# G / \#\im(1-\phi) \\
 & = & 
 \# G / \#(G/\ker(1-\phi)) \\
 & = & 
 \# G / (\#G /\#\ker(1-\phi)) \\
 & = & 
 \#\ker(1-\phi) \\
 & = & 
 \#\fix(\phi)
\end{eqnarray*}
We therefore have  $ R(\phi)=\#\fix(\phi)=\tr A $ .

For a finitely generated Abelian group $G$ we define the finite subgroup $G^{finite}$ to be the subgroup of
torsion elements of $G$. We denote the quotient $G^\infty:=G/G^{finite}$.
The group $G^\infty$ is torsion free.
Since the image of any torsion element by a homomorphism must be a torsion
element, the endomorphism $\phi:G\to G$ induces endomorphisms
 $$
 \phi^{finite}:G^{finite}\longrightarrow G^{finite},\;\;\;\;
 \phi^\infty:G^\infty\longrightarrow G^\infty.
 $$
 
As above, the map $\phi^{finite}$ induces a linear map $A:V\rightarrow V$, where $V$ be the complex vector space of complex valued  functions on the group $G^{finite}$.

\begin{theorem}
If $G$ is a finitely generated Abelian group and $\phi$ an
endomorphism of $G$ .Then we have
\begin{equation}
R(\phi) = (-1)^{r+p} \sum_{i=0}^k (-1)^i \tr (\Lambda^i\phi^\infty\otimes A).
\end{equation}
where $k$ is $rgG^\infty$, $p$ the number of $\mu\in\spec\phi^\infty$ such that
$\mu <-1$, and $r$ the number of real eigenvalues of $\phi^\infty$ whose
absolute value is $>1$.
\end{theorem}
 
{\sc Proof}
By proposition 2, the cokernel of $(1-\phi):G\rightarrow G$ is the Pontrjagin dual of the 
kernel of the dual map $\widehat{(1-\phi)}:\hat{G}\rightarrow \hat{G}$.
Since $\coker (1-\phi)$ is finite, we have
  $$
 \#\coker (1-\phi) = \#\ker \widehat{(1-\phi)} .
 $$
The map $\widehat{1-\phi}$ is equal to $\hat{1}-\hat{\phi}$.
Its kernel is thus the set of fixed points of the
map $\hat{\phi}:\hat{G}\rightarrow \hat{G}$.
We therefore have
\begin{equation}
 R(\phi) = \#\fix\left(\hat{\phi}:\hat{G}\rightarrow \hat{G}\right)
\end{equation}
The dual group of $G^\infty$ is a torus whose dimension is the
rank of $G$. This is canonically a closed subgroup of $\hat{G}$.
We shall denote it $\hat{G}_0$.
The quotient $\hat{G}/\hat{G}_0$ is canonically isomorphic to the
dual of $G^{finite}$. It is therefore finite.
From this we know that $\hat{G}$ is a union of finitely many disjoint
tori. We shall call these tori $\hat{G}_0,\ldots,\hat{G}_t$.
 
We shall call a torus $\hat{G}_i$ periodic if there is an iteration
$\hat{\phi}^s$ such that $\hat{\phi}^s(\hat{G}_i)\subset\hat{G}_i$.
If this is the case, then the map $\hat{\phi}^s:\hat{G}_i\rightarrow\hat{G}_i$
is a translation of the map $\hat{\phi}^s:\hat{G}_0\rightarrow\hat{G}_0$ and has
 the same number of fixed points as this map.
If $\hat{\phi}^s(\hat{G}_i)\not\subset\hat{G}_i$ then $\hat{\phi}^s$ has
no fixed points in $\hat{G}_i$.
From this we see
 $$
 \#\fix\left(\hat{\phi}:\hat{G}\rightarrow \hat{G}\right)
 =
 \#\fix\left(\hat{\phi}:\hat{G}_0\rightarrow \hat{G}_0\right)
 \times
 \#\{\hat{G}_i \mid \hat{\phi}(\hat{G}_i)\subset\hat{G}_i\}.
 $$
We now rephrase this
\begin{eqnarray*}
\lefteqn{ \#\fix\left(\hat{\phi}:\hat{G}\rightarrow \hat{G}\right)}\\
& = &
 \#\fix\left(
 \widehat{\phi^\infty}:\hat{G}_0\rightarrow \hat{G}_0
 \right)
 \times
 \#\fix\left(
 \widehat{\phi^{finite}}:\hat{G}/(\hat{G}_0) \rightarrow \hat{G}/(\hat{G}_0)
 \right).
\end{eqnarray*}
 From this we have product formula for Reidemeister numbers
 $$ R(\phi) = R(\phi^\infty) \cdot R(\phi^{finite}). $$
The trace formula for $R(\phi)$ follow from the previous two lemmas and formula
$$
\tr(\Lambda^i\phi^\infty)\cdot \tr(A)=\tr(\Lambda^i\phi^\infty\otimes A).
$$

\subsection{Rationality of Reidemeister zeta
functions of eventually commutative endomorphisms -  first proof.}

If we   compare the Reidemeister zeta function of an
endomorphism $\phi$ with the Reidemeister zeta function
of $H_1(\phi):H_1(G)\rightarrow H_1(G)$,
where $H_1 = H_1^{Gp}$ is the first integral homology functor
from groups to abelian groups, then we have from theorem 5 in chapter 1 following

\begin{theorem}
If $\phi:G\rightarrow G$ is eventually commutative,
then
$$
R_\phi(z)
 = R_{H_1(\phi)}(z)
 = \exp \left( \sum_{n=1}^\infty \frac{\#\coker (1-H_1(\phi)^n)}{n} z^n\right).
$$
\end{theorem}

This means that to find out about the Reidemeister zeta functions
of eventually commutative endomorphisms, it is sufficient to
study the zeta functions of endomorphisms of Abelian groups.

\begin{lemma}
Let $\phi:\bbbz^k\rightarrow\bbbz^k$ be a group endomorphism.
Then we have
\begin{equation}
 R_\phi(z)
 =
 \left(
 \prod_{i=0}^k
 \det(1-\Lambda^i\phi \cdot \sigma\cdot z)^{(-1)^{i+1}}
 \right)^{(-1)^r}
\end{equation}
where $\sigma=(-1)^p$ with $p$ the number of $\mu\in\spec\phi$ such that
$\mu <-1$, and $r$ the number of real eigenvalues of $\phi$ whose
absolute value is $>1$.
$\Lambda^i$ denotes the exterior power. 
\end{lemma}

{\sc Proof}
Since $\bbbz^k$ is Abelian, we have as before,
 $$  R(\phi^n) = \#\coker(1-\phi^n) .$$
On the other hand we have
 $$ \#\coker(1-\phi^n) = \mid\det(1-\phi^n)\mid ,$$
and hence $R(\phi^n)=(-1)^{r+pn}\det(1-\phi^n)$.
It is well known from linear algebra that
$\det(1-\phi^n)=\sum_{i=0}^k (-1)^i \tr (\Lambda^i\phi^n)$.
From this we have the following trace formula for Reidemeister numbers:
\begin{equation}
 R(\phi^n) = (-1)^{r+pn} \sum_{i=0}^k (-1)^i \tr (\Lambda^i\phi^n).
\end{equation}
We now calculate directly
\begin{eqnarray*}
R_\phi(z) & = & \exp\left(\sum_{n=1}^\infty \frac{R(\phi^n)}{n} z^n\right) \\
          & = & \exp\left(\sum_{n=1}^\infty 
                \frac{(-1)^{r}
                \sum_{i=0}^k (-1)^i
                \tr (\Lambda^i\phi^n)}{n} (\sigma z)^n\right) \\
          & = & \left(\prod_{i=0}^k\left(
                \exp\left(\sum_{n=1}^\infty 
                \frac{1}{n}\tr (\Lambda^i\phi^n)\cdot (\sigma z)^n\right)
                \right)^{(-1)^i} \right)^{(-1)^r} \\
          & = & \left(\prod_{i=0}^k
                \det\left(1-\Lambda^i\phi \cdot \sigma z\right)^{(-1)^{i+1}} 
                \right)^{(-1)^r}.
\end{eqnarray*}

\begin{lemma}
Let $\phi:G\rightarrow G$ be an endomorphism
of a finite Abelian group $G$. Then we have
\begin{equation}
 R_\phi(z) = \prod_{[\gamma]} \frac{1}{1-z^{\#[\gamma]}}
\end{equation}
where the product is taken over the periodic orbits of $\phi$
in $G$.
\end{lemma}

We give two proofs of this lemma.
The first proof is given here and the second
proof is a special case of the proof of theorem 16 .

{\sc Proof}
Since $G$ is Abelian, we again have,
\begin{eqnarray*}
 R(\phi^n)
 & = & 
 \#\coker(1-\phi^n) \\
 & = & 
 \# G / \#\im(1-\phi^n) \\
 & = & 
 \# G / \#(G/\ker(1-\phi^n)) \\
 & = & 
 \# G / (\#G /\#\ker(1-\phi^n)) \\
 & = & 
 \#\ker(1-\phi^n) \\
 & = & 
 \#\fix(\phi^n)
\end{eqnarray*}
We shall call an element of $G$ periodic if it is fixed by some
iteration of $\phi$.
A periodic element $\gamma$ is fixed by $\phi^n$ iff $n$ is divisible by
the cardinality the orbit of $\gamma$.
We therefore have
\begin{eqnarray*}
 R(\phi^n)
 & = & 
 \sum_{\gamma \ periodic \atop \#[\gamma]\mid n} 1 \\
 & = & 
 \sum_{[\gamma]\ such \ that, \atop \#[\gamma]\mid n} \#[\gamma].
\end{eqnarray*}
From this follows
\begin{eqnarray*}
 R_\phi(z)
 & = &
 \exp\left(\sum_{n=1}^\infty \frac{R(\phi^n)}{n} z^n\right) \\
 & = &
 \exp\left(\sum_{[\gamma]}
           \sum_{n=1\atop \#[\gamma]\mid n}^\infty
           \frac{\#[\gamma]}{n} z^n\right) \\
 & = &
 \prod_{[\gamma]}
 \exp\left(\sum_{n=1}^\infty
 \frac{\#[\gamma]}{\#[\gamma]n} z^{\#[\gamma]n}\right) \\
 & = &
 \prod_{[\gamma]}
 \exp\left(\sum_{n=1}^\infty
 \frac{1}{n} z^{\#[\gamma]n}\right) \\
 & = &
 \prod_{[\gamma]}
 \exp \left( - \log \left( 1-z^{\#[\gamma]}\right)\right) \\
 & = &
 \prod_{[\gamma]} \frac{1}{1-z^{\#[\gamma]}}.
\end{eqnarray*}

For a finitely generated Abelian group $G$ we define the finite subgroup $G^{finite}$ to be the subgroup of
torsion elements of $G$. We denote the quotient $G^\infty:=G/G^{finite}$.
The group $G^\infty$ is torsion free.
Since the image of any torsion element by a homomorphism must be a torsion
element, the function $\phi:G\to G$ induces maps
 $$
 \phi^{finite}:G^{finite}\longrightarrow G^{finite},\;\;\;\;
 \phi^\infty:G^\infty\longrightarrow G^\infty.
 $$

\begin{theorem}
If $G$ is a finitely generated Abelian group and $\phi$ an
endomorphism of $G$ then
$R_\phi(z)$ is a rational function and is equal to the following additive convolution:
\begin{equation}
 R_\phi(z)
 =
 R^\infty_\phi(z)
 * 
 R^{finite}_\phi(z).
\end{equation}
where $R^\infty_\phi(z)$ is the Reidemeister
zeta function of the endomorphism 
$\phi^\infty :G^\infty\rightarrow G^\infty$,
and $R^{finite}_\phi(z)$ is the Reidemeister
zeta function of the endomorphism 
$\phi^{finite} :G^{finite}\rightarrow G^{finite}$.
The functions $R^\infty_\phi(z)$ and $R^{finite}_\phi(z)$
are given by the formulae
\begin{equation}
  R^\infty_\phi(z)
  =
  \left(
  \prod_{i=0}^k
  \det(1-\Lambda^i\phi^\infty \cdot \sigma \cdot z)^{(-1)^{i+1}}
  \right)^{(-1)^r},
\end{equation}
\begin{equation}
  R^{finite}_\phi(z)
  =
  \prod_{[\gamma]}\frac{1}{1-z^{\#[\gamma]}}.
\end{equation}
with the product in (2.12) being taken over all periodic $\phi$-orbits
of torsion elements $\gamma\in G$.
Also, $\sigma=(-1)^p$ where $p$ is the number of real eingevalues
$\lambda\in\spec\phi^\infty$ such that $\lambda<-1$
and $r$ is the number of  real eingevalues
$\lambda\in\spec\phi^\infty$
such that $\mid\lambda\mid > 1$.
\end{theorem}

{\sc Proof}
By proposition 2, the cokernel of $(1-\phi^n):G\rightarrow G$ is the Pontrjagin dual of the 
kernel of the dual map $\widehat{(1-\phi^n)}:\hat{G}\rightarrow \hat{G}$.
Since $\coker (1-\phi^n)$ is finite, we have
 $$
 \#\coker (1-\phi^n) = \#\ker \widehat{(1-\phi^n)} .
 $$
The map $\widehat{1-\phi^n}$ is equal to $\hat{1}-\hat{\phi}^n$.
Its kernel is thus the set of fixed points of the
map $\hat{\phi}^n:\hat{G}\rightarrow \hat{G}$.
We therefore have
\begin{equation}
 R(\phi^n) = \#\fix\left(\hat{\phi}^n:\hat{G}\rightarrow \hat{G}\right).
\footnote{We shall use this formula again later
to connect the Reidemeister number of $\phi$ with
the Lefschetz number of $\hat{\phi}$.}
\end{equation}
The dual group of $G^\infty$ is a torus whose dimension is the
rank of $G$. This is canonically a closed subgroup of $\hat{G}$.
We shall denote it $\hat{G}_0$.
The quotient $\hat{G}/\hat{G}_0$ is canonically isomorphic to the
dual of $G^{finite}$. It is therefore finite.
From this we know that $\hat{G}$ is a union of finitely many disjoint
tori. We shall call these tori $\hat{G}_0,\ldots,\hat{G}_r$.

We shall call a torus $\hat{G}_i$ periodic if there is an iteration
$\hat{\phi}^s$ such that $\hat{\phi}^s(\hat{G}_i)\subset\hat{G}_i$.
If this is the case, then the map $\hat{\phi}^s:\hat{G}_i\rightarrow\hat{G}_i$
is a translation of the map $\hat{\phi}^s:\hat{G}_0\rightarrow\hat{G}_0$ and has
the same number of fixed points as this map.
If $\hat{\phi}^s(\hat{G}_i)\not\subset\hat{G}_i$ then $\hat{\phi}^s$ has
no fixed points in $\hat{G}_i$.
From this we see
 $$
 \#\fix\left(\hat{\phi}^n:\hat{G}\rightarrow \hat{G}\right)
 =
 \#\fix\left(\hat{\phi}^n:\hat{G}_0\rightarrow \hat{G}_0\right)
 \times
 \#\{\hat{G}_i \mid \hat{\phi}^n(\hat{G}_i)\subset\hat{G}_i\}.
 $$
We now rephrase this
\begin{eqnarray*}
\lefteqn{ \#\fix\left(\hat{\phi}^n:\hat{G}\rightarrow \hat{G}\right)}\\
& = &
 \#\fix\left(
 \widehat{\phi^\infty}^n:\hat{G}_0\rightarrow \hat{G}_0
 \right)
 \times
 \#\fix\left(
 \widehat{\phi^{finite}}^n:\hat{G}/(\hat{G}_0) \rightarrow \hat{G}/(\hat{G}_0)
 \right).
\end{eqnarray*}
From this we have that   $ R(\phi^n)=R(({\phi^{\infty}})^n)\cdot R(({\phi^{finite}})^n)$ for every $n$ and
 $$
  R_\phi(z) = R_{(\phi^\infty)}(z) * R_{(\phi^{finite})}(z). 
 $$
The rationality of $R_\phi(z)$ and the formulae
for $R_\phi^\infty(z)$ and $R_\phi^{finite}(z)$ follow from the previous two lemmas
and lemma 13.

\begin{corollary}
Let the assumptions of theorem 10 hold.
Then the poles and zeros of the Reidemeister zeta function
are complex numbers of the
form $\zeta^a b$ where $b$ is the reciprocal 
of an eigenvalue of one of the matrices
 $$  \Lambda^i (\phi^\infty)
     : \Lambda^i (G^\infty) 
       \longrightarrow
       \Lambda^i (G^\infty) \;\;\;\;\;\; 0\leq i\leq {\rm rank}\;G$$
and $\zeta^a$ is a $\psi^{th}$ root of unity
where $\psi$ is the number of periodic torsion elements in $G$.
The multiplicities of the roots or poles
$\zeta^a b$ and $\zeta^{a^\prime} b^\prime$ are the same
if $b=b^\prime$ and $hcf(a,\psi) = hcf(a^\prime,\psi)$.
\end{corollary}

\subsection{Functional equation for the Reidemeister zeta 
function of an eventually commutative endomorphism}

\begin{lemma}
[Functional equation for the torsion free part]
\hfill
\par
 Let $\phi: \bbbz^k \to \bbbz^k$ be an endomorphism.
The Reidemeister zeta function $R_\phi(z)$ has the
following functional equation:
\begin{equation}
 R_{\phi}\left(\frac{1}{dz}\right)
 =
 \epsilon_1 \cdot R_{\phi}(z)^{(-1)^k}.
\end{equation}
where $d=\det\phi$ and $\epsilon_1$
are a constants in $\bbbc^\times$.
\end{lemma}

{\sc Proof}
Via the natural nonsingular pairing
$(\Lambda^i \bbbz^k)\otimes(\Lambda^{k-i} \bbbz^k) \rightarrow\bbbc$
the operators $\Lambda^{k-i}\phi$ and $d\cdot (\Lambda^i \phi)^{-1}$ are adjoint
to each other. 

We consider an eigenvalue $\lambda$ of $\Lambda^i\phi$.
By lemma 17, this contributes a term $\left((1-\frac{\lambda\sigma}{dz})^{(-1)^{i+1}}\right)^{(-1)^r}$
to $R_\phi\left(\frac{1}{dz}\right)$.
We rewrite this term as
 $$
 \left(
 \left( 1 - \frac{d\sigma z}{\lambda} \right)^{(-1)^{i+1}}
 \left( \frac{-dz}{\lambda\sigma} \right)^{(-1)^i}
 \right)^{(-1)^r}
 $$
 and note that $\frac{d}{\lambda}$ is an eigenvalue of $\Lambda^{k-i}\phi$.
Multiplying these terms together we obtain,
 $$
 R_\phi\left(\frac{1}{dz}\right)
 =
 \left(
 \prod_{i=1}^k
 \prod_{\lambda^{(i)}\in\spec\Lambda^i\phi}
 \left(\frac{1}{\lambda^{(i)}\sigma}\right)^{(-1)^i}
 \right)^{(-1)^r}
 \times
 R_\phi(z)^{(-1)^k}.
 $$
The variable $z$ has disappeared because
 $$
 \sum_{i=0}^k (-1)^i \dim\Lambda^i\bbbz^k = \sum_{i=0}^k (-1)^i \cdot C^i_k = 0.
 $$

\begin{lemma}[Functional equation for the finite part]
 Let $\phi:G \to G$ be an endomorphism of a finite, Abelian group $G$.
The Reidemeister zeta function $R_\phi(z)$ has the
following functional equation:
\begin{equation}
 R_{\phi}\left(\frac{1}{z}\right)
 =
 (-1)^p z^q R_{\phi}(z),
\end{equation}
where $q$ is the number of periodic elements of $\phi$ in $G$
and $p$ is the number of periodic orbits of $\phi$ in $G$.
\end{lemma}

{\sc Proof}
This is a simple calculation.
We begin with formula (2.9).
\begin{eqnarray*}
	R_\phi\left(\frac{1}{z}\right)
	& = &
	\prod_{[\gamma]} \frac{1}{1-z^{-\#[\gamma]}} \\
	& = &
	\prod_{[\gamma]} \frac{z^{\#[\gamma]}}{z^{\#[\gamma]}-1} \\
	& = &
	\prod_{[\gamma]} \frac{-z^{\#[\gamma]}}{1-z^{\#[\gamma]}} \\
	& = &
	\prod_{[\gamma]} -z^{\#[\gamma]}\times
	\prod_{[\gamma]} \frac{1}{1-z^{\#[\gamma]}} \\
	& = &
	\prod_{[\gamma]} -z^{\#[\gamma]} \times R_\phi(z).
\end{eqnarray*}
The statement now follows because
$\sum_{[\gamma]}\#[\gamma] = q$.

\begin{theorem}[Functional equation]
Let $\phi:G\to G$ be an endomorphism
of a finitely generated Abelian group $G$.
If $G$ is finite the  functional equation
of $R_\phi$ is described in lemma 20.
If $G$ is infinite then $R_\phi$ has the
following functional equation:
\begin{equation}
 R_{\phi}\left(\frac{1}{dz}\right)
 =
 \epsilon_2 \cdot R_\phi(z)^{(-1)^{\rank G}}.
\end{equation}
where $d=\det\left(\phi^\infty:G^\infty\to G^\infty\right)$ and $\epsilon_2$
are a constants in $\bbbc^\times$.
\end{theorem}

{\sc Proof}
From theorem 10 we have
$R_\phi(z) = R_\phi^\infty(z) * R_\phi^{finite}(z)$.
In the previous two lemmas we have obtained functional
equations for the functions $R_\phi^\infty(z)$ and $R_\phi^{finite}(z)$.
Lemma 14 now gives the functional equation for $R_{\phi}(z)$.

\subsection{Rationality of Reidemeister zeta functions of eventually commutative endomorphisms - se\-cond proof.}

\begin{theorem}
Let $G$ is a finitely generated Abelian group and $\phi$ an
endomorphism of $G$ .Then $R_\phi(z)$ is a rational function and is equal to 
\begin{equation}
 R_\phi(z)
 =
 \left(
 \prod_{i=0}^k
 \det(1-\Lambda^i\phi^\infty\otimes A \cdot \sigma \cdot z)^{(-1)^{i+1}}
 \right)^{(-1)^r}
\end{equation}
where matrix $A$ is defined in lemma 16 , $\sigma=(-1)^p$, $p$ , $r$ and $k$ are constants described in theorem 8 .
\end{theorem} 
 
{\sc Proof}
If we repeat the proof of the theorem 8 for $\phi^n$ instead $\phi$ we obtain 
that $ R(\phi^n)= R((\phi^\infty)^n)\cdot R((\phi^{finite})^n)$.From this and lemmas 15 and 16   we have the trace formula for $ R(\phi^n)$:
\begin{eqnarray*}
 R(\phi^n) & = & (-1)^{r+pn} \sum_{i=0}^k (-1)^i \tr \Lambda^i(\phi^\infty)^n \cdot \tr A^n \\
           & = &  (-1)^{r+pn} \sum_{i=0}^k (-1)^i \tr (\Lambda^i(\phi^\infty)^n\otimes A^n) \\
         & = & (-1)^{r+pn} \sum_{i=0}^k (-1)^i \tr (\Lambda^i\phi^\infty\otimes A )^n.
\end{eqnarray*}
 
We now calculate directly
\begin{eqnarray*}
R_\phi(z) & = & \exp\left(\sum_{n=1}^\infty \frac{R(\phi^n)}{n} z^n\right) \\
          & = & \exp\left(\sum_{n=1}^\infty 
                \frac{(-1)^{r}
                \sum_{i=0}^k (-1)^i
                \tr (\Lambda^i\phi^\infty\otimes A )^n}{n} (\sigma \cdot z)^n\right) \\
          & = & \left(\prod_{i=0}^k\left(
 \exp\left(\sum_{n=1}^\infty 
                \frac{1}{n}\tr (\Lambda^i\phi^\infty\otimes A )^n\cdot(\sigma \cdot z)^n\right)
                \right)^{(-1)^i} \right)^{(-1)^r} \\
          & = & \left(\prod_{i=0}^k
                \det\left(1-\Lambda^i\phi^\infty\otimes A  \cdot\sigma \cdot z\right)^{(-1)^{i+1}} 
                \right)^{(-1)^r}.
\end{eqnarray*}

\begin{corollary}
Let the assumptions of theorem 12  hold.
Then the poles and zeros of the Reidemeister zeta function
are complex numbers which are the reciprocal 
of an eigenvalues of one of the matrices
 $$  \Lambda^i (\phi^\infty)\otimes A  \cdot\sigma 
     \;\;\;\;\;\; 0\leq i\leq {\rm rank}\;G$$
\end{corollary}

\subsection{Connection of the Reidemeister zeta function with the
Lefschetz zeta function of the dual map}

\begin{theorem}[Connection with Lefschetz numbers]
Let $\phi:G\to G$ be an endomorphism of a finitely
generated Abelian group.
Then we have the following
\begin{equation}
  R(\phi^n) = \mid L(\hat{\phi}^{n}) \mid,
\end{equation}
where $\hat{\phi}$ is the continuous endomorphism of $\hat{G}$
defined in section 2.2 and $L(\hat{\phi}^{n})$ is the Lefschetz number
of $\hat{\phi}$ thought of as a self-map of the topological space $\hat{G}$.
From this it follows:
\begin{equation}
  R_\phi(z) = L_{\hat{\phi}}(\sigma z)^{(-1)^r},
\end{equation}
where $r$ and $\sigma$ are the constants described in theorem 10.
If $G$ is finite then this reduces to
$$  R(\phi^n)=L(\hat{\phi}^n)  \; {\rm and} \; R_\phi(z)=L_{\hat{\phi}}(z).$$
\end{theorem}

The proof is similar to that of Anosov \cite{a} concerning
continuous maps of nil-manifolds.

{\sc Proof}
We already know from formula (2.13) in the proof of theorem 10
that $R(\phi^n)$ is the number of fixed points
of the map $\hat{\phi}^{n}$.
If $G$ is finite then $\hat{G}$ is a discrete
finite set, so the number of fixed points is equal to
the Lefschetz number. This finishes the proof in the
case that $G$ is finite.
In general it is only necessary to check that
the number of fixed points of $\hat{\phi}^n$ is equal to the
absolute value of its Lefschetz number.
We assume without loss of generality that $n=1$.
We are assuming that $R(\phi)$ is finite, so
the fixed points of $\hat{\phi}$ form a discrete set.
We therefore have
 $$
 L(\hat{\phi})
 =
 \sum_{x\in\fix\hat{\phi}} \ind(\hat{\phi},x).
 $$
Since $\phi$ is a group endomorphism, the zero element $0\in\hat{G}$ is always fixed.
Let $x$ be any fixed point of $\hat{\phi} $.
We then have a commutative diagram
 $$
 \begin{array}{ccccc}
 g & \hat{G} & \stackrel{\hat{\phi}}{\longrightarrow} & \hat{G} & g \\
 \updownarrow &  \updownarrow & & \updownarrow & \updownarrow \\
 g + x & \hat{G} & \stackrel{\hat{\phi}}{\longrightarrow} & \hat{G} & g + x
 \end{array}
 $$
in which the vertical functions are translations on $\hat{G}$ by $x$.
Since the vertical maps map $0$ to $x$, we deduce that
 $$ \ind(\hat{\phi},x) = \ind(\hat{\phi},0) $$
and so all fixed points have the same index.
It is now sufficient to show that $\ind(\hat{\phi},0)=\pm 1$.
This follows because the map on the torus
 $$ \hat{\phi}:\hat{G}_0\to\hat{G}_0 $$
lifts to a linear map of the universal cover, which is in this case the
Lie algebra of $\hat{G}$. The index is then the sign of the determinant of the identity map  minus this lifted map.
This determinant cannot be zero, because $1-\hat{\phi}$ must have finite
kernel by our assumption that the Reidemeister number of $\phi$ is
finite
(if $\det(1-\hat{\phi})=0$ then the kernel of $1-\hat{\phi}$
is a positive dimensional subgroup of $\hat{G}$, and therefore
infinite).

\section { Endomorphisms of finite groups}

In this section we consider finite non-Abelian groups.
We shall write the group law multiplicatively.
We generalize our results on endomorphisms
of finite Abelian groups to endomorphisms of finite non-Abelian groups.
We shall write $\{g\}$ for the
$\phi$-conjugacy class of an element $g\in G$. We shall write
$<g>$ for the ordinary conjugacy class of $g$ in $G$.
We continue to write $[g]$ for the $\phi$-orbit of $g\in G$,
and we also write now $[<g>]$ for the $\phi$-orbit
of the ordinary conjugacy class of $g\in G$.
We first note that if $\phi$ is an endomorphism of a group $G$
then $\phi$ maps conjugate elements to conjugate elements. It
therefore induces an endomorphism of the set of conjugacy classes of elements 
of $G$.
If $G$ is Abelian then a conjugacy class consists of a single element.
The following is thus an extension of lemma 16:
 
\begin{theorem}
Let $G$ be a finite group and let $\phi:G\to G$ be an endomorphism.
Then $R(\phi)$ is the number of ordinary conjugacy classes $<x>$ in
 $G$ such that $<\phi(x)>=<x>$.
\end{theorem}

{\sc Proof}
From the definition of the Reidemeister number we have,
\begin{eqnarray*}
 R(\phi)
 & = &
 \sum_{\{g\}} 1
\end{eqnarray*}
where $\{g\}$ runs through
the
set of $\phi$-conjugacy classes in $G$.
This gives us immediately
\begin{eqnarray*}
 R(\phi)
 & = &
 \sum_{\{g\}} \sum_{x\in\{g\}} \frac{1}{\#\{g\}}  \\
 & = &
 \sum_{\{g\}} \sum_{x\in\{g\}} \frac{1}{\#\{x\}}  \\
 & = &
 \sum_{x\in G} \frac{1}{\#\{x\}}.
\end{eqnarray*}
We now calculate for any $x\in G$ the order of $\{x\}$.
 The class $\{x\}$ is the orbit of $x$ under the $G$-action
 $$  (g , x)\longmapsto g x \phi(g)^{-1}  .$$
We verify that this is actually a $G$-action:
\begin{eqnarray*}
 (id,x) 
 & \longmapsto & id.x. \phi(id)^{-1} \\
 & = & x,  \\
 (g_1 g_2,x)
 & \longmapsto & g_1 g_2.x. \phi(g_1 g_2)^{-1} \\
 & = & g_1 g_2.x. (\phi(g_1) \phi(g_2))^{-1} \\
 & = & g_1 g_2.x. \phi(g_2)^{-1} \phi(g_1)^{-1} \\
 & = & g_1 (g_2.x. \phi(g_2)^{-1}) \phi(g_1)^{-1}.
\end{eqnarray*}
We therefore have from the orbit-stabilizer theorem,
 $$ \#\{x\} = \frac{\#G}{\#\{g\in G\mid g x \phi(g)^{-1} = x \} }.$$
The condition $g x \phi(g)^{-1} = x$
is equivalent to
\begin{eqnarray*}
x^{-1} g x \phi(g)^{-1} = 1
& \Leftrightarrow &
x^{-1} g x = \phi(g)
\end{eqnarray*}
We therefore have
 $$
 R(\phi)= \frac{1}{\#G} \sum_{x\in G} \#\{ g\in G\mid x^{-1}gx=\phi(g)\}.
$$
Changing the summation over $x$ to summation over $g$, we have:
$$
 R(\phi)= \frac{1}{\#G} \sum_{g\in G} \#\{ x\in G\mid x^{-1}gx=\phi(g)\}.
$$
If $<\phi(g)>\not=<g>$ then there are no elements $x$ such that
$x^{-1}gx=\phi(g)$.
We therefore have:
$$
 R(\phi)
 =
 \frac{1}{\#G}
 \sum_{g\in G \ {\rm such \ that} \atop <\phi(g)>=<g>}
 \#\{ x\in G\mid x^{-1}gx=\phi(g)\}.
$$
The elements $x$ such that $x^{-1}gx=\phi(g)$
form a coset of the subgroup satisfying $x^{-1}gx=g$.
This subgroup is the centralizer of $g$ in $G$ which we write $C(g)$.
With this notation we have,
\begin{eqnarray*}
 R(\phi)
  & = &
 \frac{1}{\#G}
 \sum_{g\in G \ {\rm such \ that} \atop <\phi(g)>=<g>}
 \#C(g) \\
 & = &
 \frac{1}{\#G}
 \sum_{<g>\subset G \ {\rm such \ that} \atop <\phi(g)>=<g>}
 \#<g>. \#C(g).
\end{eqnarray*}
The last identity follows because $C(h^{-1}gh)=h^{-1}C(g)h$.
From the orbit stabilizer theorem, we know that
$\#<g>. \#C(g)=\#G$.
We therefore have
$$
 R(\phi) = \#\{<g>\subset G \mid <\phi(g)>=<g>\}.
$$
\vspace{0.2cm}
 
 Let $W$ be the complex vector space of complex valued class functions on the group $G$.A class function is a function which takes the same value on every element of a usual congruency class.The map $\phi$ induces a linear map $B:W\rightarrow W$ defined by
$$
B(f):=f\circ\phi.
$$

\begin{theorem}(Trace formula)
Let $\phi:G\rightarrow G$ be an endomorphism
of a finite  group $G$. Then we have
\begin{equation}
R(\phi) = \tr B
\end{equation}
\end{theorem}
 
{\sc Proof}
The characteristic functions of the congruency classes in $G$ form a basis of $W$, and are mapped to one another by $B$ (the map need not be a bijection).Therefore the trace of $B$ is the number of elements of this basis which are fixed by $B$. By theorem 14, this is equal to the Reidemeister number of $\phi$.

From the theorem 14 we have immediately,
  
\begin{theorem}
Let $\phi$ be an endomorphism of a finite group $G$.
Then $R_\phi(z)$ is a rational function with a functional
equation. In particular we have,
$$
 R_\phi(z) = \prod_{[<g>]} \frac{1}{1-z^{\#[<g>]}},
$$
$$
 R_\phi\left(\frac{1}{z}\right)
 =
 (-1)^a z^b R_\phi(z).
$$
The product here is over all periodic $\phi$-orbits of
ordinary conjugacy classes of elements of $G$.
The number $\#[<g>]$ is the number of conjugacy classes
in the $\phi$-orbit of the conjugacy class $<g>$.
In the functional equation the numbers $a$ and $b$ are respectively
the number of periodic $\phi$-orbits of conjugacy classes of elements
of $G$ and the number of periodic conjugacy classes of elements
of $G$. A conjugacy class $<g>$ is called periodic
if for some $n>0$, $<\phi^n (g)>=<g>$
\end{theorem}
  
{\sc Proof}
From the  theorem 14 we
know that $R(\phi^n)$ is the number
of conjugacy classes $<g>\subset G$ such that $\phi^n(<g>)\subset<g>$.
We can rewrite this
$$
 R(\phi^n)
 =
 \sum_{
 \begin{array}{c}
 [<g>] \ {\rm such \ that} \\
 \#[<g>] \mid n
 \end{array}}
 \#[<g>].
$$
From this we have,
$$
 R_\phi(z)
 =
 \prod_{[<g>]}
 \exp\left(
 \sum_{
  \begin{array}{c}
  n=1 \ {\rm such \ that}\\
  \#[<g>] \mid n
 \end{array}
 }^\infty
 \frac{\#[<g>]}{n} z^n
 \right).
$$
The first formula now follows by using the power series expansion
for $\log(1-z)$.
The functional equation follows now in exactly
the same way as lemma 20 follows from lemma 18.

\begin{corollary}
Suppose that $\phi_1$ and $\phi_2$ are two
endomorphisms of a finite group $G$
with
 $$\forall g\in G,\;\phi_1(g) = h \phi_2(g) h^{-1} $$
for some fixed element $h\in G$.
Then $R_{\phi_1}(z)=R_{\phi_2}(z)$.
\end{corollary}

\begin{corollary}
Let $\phi$ be an inner automorphism.
Then
$$ R_\phi(z) = \frac{1}{(1-z)^b} $$
where $b$ is the number of conjugacy classes of elements in the group.
In particular, all but finitely many of the symmetric and
alternating groups have the property that any automorphism
is an inner automorphism, and so this corollary applies.
\end{corollary}
 
\begin{remark}
If we think of the set of conjugacy classes of elements of $G$ as a discrete set
then the Reidemeister number of $\phi$ is equal to
the Lefschetz number of the induced map on the set of the conjugacy classes of elements of $G$.
\end{remark}

Another proof of rationality of the Reidemeister zeta function for finite groups follows
 from  the trace formula for the Reidemeister numbers  in the theorem 15.
 \begin{theorem}
Let $\phi$ be an endomorphism of a finite group $G$.
Then $R_\phi(z)$ is a rational function and given by formula
 
 \begin{equation}
 R_\phi(z) = \frac{1}{\det(1-Bz)},
\end{equation}
Where $B$ is defined in theorem 15
\end{theorem}

{\sc Proof}
From theorem 15 it follows that $ R(\phi^n)=\tr B^n $ for every $n>o$.We now calculate directly
\begin{eqnarray*}
R_\phi(z) & = &  \exp\left(\sum_{n=1}^\infty \frac{R(\phi^n)}{n} z^n\right)=                          \exp\left(\sum_{n=1}^\infty \frac{\tr B^n}{n} z^n\right)=  \exp\left(\tr \sum_{n=1}^\infty  \frac{ B^n}{n} z^n\right)\\
          &  = &  \exp\left(\tr (-\log (1-Bz))\right)=  \frac{1}{\det(1-Bz)}.
\end{eqnarray*}

\section{ Endomorphisms of the direct sum of a free Abelian and a finite group}

In this section let $F$ be a finite group and $r$ a natural number. The group $G$ will be
$$
G = \bbbz^r \oplus F
$$
We shall describe the Reidemeister numbers of endomorphism $ \phi : G \rightarrow G$.
The torsion elements of $G$ are exactly the elements of the finite, normal subgroup $F$.
For this reason we have $\phi(F) \subset F $.Let $\phi^{finite}: F\rightarrow F $ be the restriction
of $\phi $ to $F$, and let $\phi ^{\infty}: G/F \rightarrow G/F $ be the induced map on the quotient group. 

 Let $ pr_{\bbbz^r}: G\rightarrow \bbbz^r $ and $pr_F :G\rightarrow F$ be the projections onto 
 $\bbbz^r $ and $F$ . Then the composition 
 $$
  pr_{\bbbz^r}\circ\phi: \bbbz^r \rightarrow G\rightarrow \bbbz^r
 $$
  is an endomorphism of $\bbbz^r$, which is given by some matrix $M \in M_r(\bbbz)$.
  We denote by $ \psi : \bbbz^r \rightarrow F$ the other component of the 
  restriction of $\phi$ to $\bbbz^r$, i.e.
  $$
  \psi(v)=pr_F(\phi(v)).
  $$
  We therefore have for any element $(v,f) \in G$
  $$
  \phi(v,f)=(M\cdot v,\psi(v)\phi(f)).
  $$
  \begin{lemma}
  Let $g_1=(v_1,f_1)$ and $g_2=(v_2,f_2)$ be two elements of $G$. Then $g_1$ and $g_2$
  are $\phi$-conjugate iff
  $$
   v_1\equiv v_2  \mod  (1-M)\bbbz^r
   $$
   and there is a $h\in F$ with
   $$
   hf_1=f_2 \phi((1-M)^{-1}(v_2-v_1))\phi(h).
   $$
   \end{lemma}

{\sc Proof }

Suppose $g_1$ and $g_2$ are $\phi$-conjugate. Then there is a  $g_3=(w,h) \in G$
with $g_3g_1= g_2\phi(g_3).$ Therefore 
$$
(w+v_1, hf_1)=(v_2+ M\cdot w, f_2\psi(w)\phi(h)).
$$
Comparing the first components we obtain $(1-M)\cdot w=v_2 -v_1 $ from which it follows that $v_1$ is congruent to $v_2$ modulo  $(1-M)\bbbz^r$.
Substituting $(1-M)^{-1}(v_2 -v_1)$ for $w$ in the second component we obtain the second relation in the lemma. The argument can easily be reversed to give the converse.

\begin{proposition}
In the notation described above 
$$
R(\phi)= R(\phi^{finite}) \times R(\phi ^{\infty}).
$$
\end{proposition}
{\sc Proof  }
We partition the set $\cR(\phi)$ of $\phi$-conjugacy classes of elements of $G$ into smaller sets:
$$
\cR(\phi)= \cup_{v\in \bbbz^r/(1-M)\bbbz^r} \cR(v)
$$
where $\cR(v)$ is the set of $\phi$-conjugacy classes $\{(w,f)\}_{\phi}$ for which $w$ 
is congruent to $v$ modulo $(1-M)\bbbz^r$. It follows from the previous lemma that this is a
partition.
 Now suppose $\{(w,f)\}_{\phi} \in \cR(v) $. We will show that $\{(w,f)\}_{\phi}=\{(v,f^*)\}_{\phi}$ 
for some $f^* \in F$. This follows by setting $ f^*=f \psi((1-M)^{-1}(w-v))$ and applying the 
previous lemma with $h=id$. Therefore $\cR(v)$ is the set of $\phi$-conjugacy classes  
$\{(v,f)\}_{\phi}$ with $f \in F$. 
From the previous lemma it follows that $(v,f_1)$ and $(v,f_2)$ are $\phi$-conjugate iff there 
is a $h\in F$ with
$$
hf_1=f_2\psi(0)\phi(h)=f_2\phi(h)
$$
This just means that $f_1$ and $f_2$ are $\phi^{finite}$-conjugate as elements of $F$. From this it
follows that $\cR(v)$ has cardinality $R(\phi^{finite})$. Since this is independent of $v$, we have
$$
R(\phi)=\sum_v R(\phi^{finite})= \mid \det(1-M)\mid \times R(\phi^{finite}).
$$ 

Now consider the map $\phi^{\infty}: G/F \rightarrow G/F $. We have 
$$
\phi^{\infty}((v,F))=(M\cdot v, \psi(v)F)= (M\cdot v,F).
$$
From this it follows that $\phi^{\infty}$ is isomorphic to map $ M: \bbbz^r \rightarrow \bbbz^r$.
This implies
$$
R(\phi ^{\infty})=R(M:\bbbz^r \rightarrow \bbbz^r)
$$
but it is known \cite{fh} that $R(M:\bbbz^r \rightarrow \bbbz^r)=\mid \det(1-M)\mid $.
Therefore $R(\phi)= R(\phi^{finite}) \times R(\phi ^{\infty})$ , proving proposition 3 .

Let $W$ be the complex vector space of complex valued class functions on the group $F$.The map $\phi$ induces a linear map $B:W\rightarrow W$ defined as above in theorem 15.
 
\begin{theorem}
If $G$ is the direct sum of a free Abelian and a finite group and $\phi$ an
 endomorphism of $G$ .Then we have
\begin{equation}
R(\phi) = (-1)^{r+p} \sum_{i=0}^k (-1)^i \tr (\Lambda^i\phi^\infty\otimes B).
\end{equation}
where $k$ is $rg(G/F)$, $p$ the number of $\mu\in\spec\phi^\infty$ such that
$\mu <-1$, and $r$ the number of real eigenvalues of $\phi^\infty$ whose
absolute value is $>1$.
\end{theorem}
 
{\sc Proof}
Theorem follows from lemmas 15 and theorem 15 , proposition 3 and formula
$$
\tr (\Lambda^i\phi^\infty)\cdot \tr (B)=\tr (\Lambda^i\phi^\infty\otimes B).
$$
 
 \begin{theorem}
Let $G$ is the direct sum of free Abelian  and a finite group and $\phi$ an
endomorphism of $G$ . If the numbers $R(\phi^n)$ are all finite then $R_\phi(z)$ is a rational function and is equal to 
\begin{equation}
  R_\phi(z)
 =
 \left(
 \prod_{i=0}^k
 \det(1-\Lambda^i\phi^\infty\otimes B \cdot\sigma\cdot z)^{(-1)^{i+1}}
 \right)^{(-1)^r}
\end{equation}
where matrix $B$ is defined in theorem 15, $\sigma=(-1)^p$,$p$ , $r$ and $k$ are constants described in lemma 17 .
\end{theorem}
 
{\sc Proof}
From proposition 3 it follows that $ R(\phi^n)= R((\phi^\infty)^n\cdot R((\phi^{finite})^n)$. 
From now on the proof repeat the proof of the theorem 12.

\begin{corollary}
Let the assumptions of theorem 19  hold.
Then the poles and zeros of the Reidemeister zeta function
are complex numbers which are the reciprocal 
of an eigenvalues of one of the matrices
 $$  \Lambda^i (\phi^\infty)\otimes B \cdot\sigma 
     \;\;\;\;\;\; 0\leq i\leq {\rm rank}\;G$$
\end{corollary}

Another proof of rationality of the Reidemeister zeta function gives

\begin{theorem}
If $G$ is the direct sum of a finitely generated free Abelian and a finite group  and $\phi$ an
endomorphism of $G$ then
$R_\phi(z)$ is a rational function and is equal to the following additive convolution:
\begin{equation}
 R_\phi(z)
 =
 R^\infty_\phi(z)
 * 
 R^{finite}_\phi(z).
\end{equation}
where $R^\infty_\phi(z)$ is the Reidemeister
zeta function of the endomorphism 
$\phi^\infty :G^\infty\rightarrow G^\infty$,
and $R^{finite}_\phi(z)$ is the Reidemeister
zeta function of the endomorphism 
$\phi^{finite} :G^{finite}\rightarrow G^{finite}$.
The functions $R^\infty_\phi(z)$ and $R^{finite}_\phi(z)$
are given by the formulae
\begin{equation}
  R^\infty_\phi(z)
  =
  \left(
  \prod_{i=0}^k
  \det(1-\Lambda^i\phi^\infty \cdot\sigma z)^{(-1)^{i+1}}
  \right)^{(-1)^r},
\end{equation}
$$
 R^{finite}_\phi(z) = \prod_{[<g>]} \frac{1}{1-z^{\#[<g>]}},
$$
The product here is over all periodic $\phi$-orbits of
ordinary conjugacy classes of elements of $G$.
The number $\#[<g>]$ is the number of conjugacy classes
in the $\phi$-orbit of the conjugacy class $<g>$.
Also, $\sigma=(-1)^p$ where $p$ is the number of real eingevalues
$\lambda\in\spec\phi^\infty$ such that $\lambda<-1$
and $r$ is the number of  real eingevalues
$\lambda\in\spec\phi^\infty$
such that $\mid\lambda\mid > 1$.
\end{theorem}

{\sc Proof}
From proposition 3  it follows that $ R(\phi^n)= R((\phi^\infty)^n\cdot R((\phi^{finite})^n)$.
From this we have
 $$ R_\phi(z) = R_{(\phi^\infty)}(z) * R^{finite}_\phi(z). $$
The rationality of $R_\phi(z)$ and the formulae
for $R_\phi^\infty(z)$ and $R_\phi^{finite}(z)$ follow from the lemma 13 , lemma 17
and theorem 16 .

\begin{theorem}[Functional equation]
Let $\phi:G\to G$ be an endomorphism
of a  group $G$ which is the direct sum of a finitely generated free Abelian and a finite group.
If $G$ is finite the  functional equation
of $R_\phi$ is described in theorem 16.
If $G$ is infinite then $R_\phi$ has the
following functional equation:
\begin{equation}
 R_{\phi}\left(\frac{1}{dz}\right)
 =
 \epsilon_2 \cdot R_\phi(z)^{(-1)^{\rank G}}.
\end{equation}
where $d=\det\left(\phi^\infty:G^\infty\to G^\infty\right)$ and $\epsilon_2$
are a constants in $\bbbc^\times$.
\end{theorem}

{\sc Proof}
From proposition 3 we have
$R_\phi(z) = R_\phi^\infty(z) * R_\phi^{finite}(z)$.
In the  lemma 19 and theorem 16 we have obtained functional
equations for the functions $R_\phi^\infty(z)$ and $R_\phi^{finite}(z)$.
Now,  lemma 14  gives us  the functional equation for $R_{\phi}(z)$.

\section{Endomorphisms of nilpotent groups}

In this section we consider finitely generated torsion free nilpotent group $\Gamma$.It is well known \cite{mal} that such group $\Gamma$ is a uniform discrete subgroup of a simply connected nilpotent Lie group $G$ (uniform means that the coset space $G/ \Gamma$ is compact).The coset space $M=G/ \Gamma$ is called a nilmanifold.Since $\Gamma=\pi_1(M)$ and $M$  is a $K(\Gamma,
1)$, every endomorphism $\phi:\Gamma \to \Gamma $ can be realized by a selfmap
$f:M\to M$ such that $f_*=\phi$ and thus $R(f)=R(\phi)$.Any endomorphism  $\phi:\Gamma \to \Gamma $ can be uniquely extended to an endomorphism  $F: G\to G$.Let $\tilde F:\tilde G\to \tilde G $ be the corresponding Lie algebra endomorphism induced from $F$. 
 
\begin{theorem}
If $\Gamma$ is a finitely generated torsion free nilpotent group and $\phi$ an
endomorphism of $\Gamma$ .Then 
\begin{equation}
R(\phi) = (-1)^{r+p} \sum_{i=0}^m (-1)^i \tr \Lambda^i\tilde F,
\end{equation}
 where $m$ is $rg\Gamma = \dim M$, $p$ the number of $\mu\in\spec\tilde F$ such that
$\mu <-1$, and $r$ the number of real eigenvalues of $\tilde F$ whose
absolute value is $>1$.
\end{theorem}
 
{\sc Proof:}
Let $f:M\to M$ be a map realizing $\phi$ on a compact nilmanifold $M$ of
dimension $m$.We suppose  that the Reidemeister number  $R(f)=R(\phi)$ is finite.The finiteness of $R(f)$ implies the nonvanishing of the Lefschetz number $L(f)$ \cite{fhw}.A strengthened version of Anosov's theorem \cite{a} is proven in
\cite{no} which states, in particular, that if
$L(f)\ne 0$ than $N(f)=|L(f)|=R(f)$.It is well known that $L(f)=\det (\tilde F -1)$ \cite{a}.From this we have
$$
R(\phi)=R(f)=|L(f)|=|\det (1- \tilde F)|=(-1)^{r+p}\det (1- \tilde F)=
$$
$$
= (-1)^{r+p} \sum_{i=0}^m (-1)^i \tr \Lambda^i\tilde F.  
$$
 
 \begin{theorem}
 If $\Gamma$ is a finitely generated torsion free nilpotent group and $\phi$ an
endomorphism of $\Gamma$ .Then $R_\phi(z)$ is a rational function and is equal to 
\begin{equation}
 R_\phi(z)
 =
 \left(
 \prod_{i=0}^m
 \det(1-\Lambda^i\tilde F \cdot\sigma\cdot z)^{(-1)^{i+1}}
 \right)^{(-1)^r}
\end{equation}
where $\sigma=(-1)^p$,$p$ , $r$, $m$ and  $\tilde F$ is defined in theorem 22.
\end{theorem}
 
{\sc Proof}
If we repeat the proof of the previous theorem  for $\phi^n$ instead $\phi$ we obtain 
that $ R(\phi^n)=(-1)^{r+pn}\det(1-\tilde F)$( we suppose that Reidemeister numbers  $ R(\phi^n)$ are finite for all $n$).Last formula implies the trace formula for $ R(\phi^n)$ :
 
$$
R(\phi^n) = (-1)^{r+pn} \sum_{i=0}^m (-1)^i \tr (\Lambda^i\tilde F)^n
$$
Now theorem follows  immediately by direct calculation as in lemma 17.

\begin{corollary}
Let the assumptions of theorem 23 hold.
Then the poles and zeros of the Reidemeister zeta function
are complex numbers which are reciprocal 
of an eigenvalue of one of the matrices
 $$  \Lambda^i (\tilde F)
     : \Lambda^i (\tilde G) 
       \longrightarrow
       \Lambda^i (\tilde G) \;\;\;\;\;\; 0\leq i\leq {\rm rank}\;G$$
\end{corollary}

\subsection{Functional equation}

\begin{theorem}
Let $\phi:\Gamma\to \Gamma$ be an endomorphism
of a finitely generated torsion free nilpotent group $\Gamma$.Then the Reidemeister zeta function $R_\phi(z)$  has the
 following functional equation:
\begin{equation}
 R_{\phi}\left(\frac{1}{dz}\right)
 =
 \epsilon_1 \cdot R_\phi(z)^{(-1)^{\rank G}}.
\end{equation}
where $d=\det\tilde F$ and $\epsilon_1$
are a constants in $\bbbc^\times$.
\end{theorem}
 
{\sc Proof}
Via the natural nonsingular pairing
$(\Lambda^i \tilde F)\otimes(\Lambda^{m-i} \tilde F) \rightarrow\bbbc$
the operators $\Lambda^{m-i}\tilde F$ and $d.(\Lambda^i \tilde F)^{-1}$ are adjoint
to each other. 
 
We consider an eigenvalue $\lambda$ of $\Lambda^i\tilde F$.
By theorem 23, This contributes a term
$$
\left((1-\frac{\lambda\sigma}{dz})^{(-1)^{i+1}}\right)^{(-1)^r}
$$
to $R_\phi\left(\frac{1}{dz}\right)$.

We rewrite this term as
 $$
 \left(
 \left( 1 - \frac{d\sigma z}{\lambda} \right)^{(-1)^{i+1}}
 \left( \frac{-dz}{\lambda\sigma} \right)^{(-1)^i}
 \right)^{(-1)^r}
 $$
 and note that $\frac{d}{\lambda}$ is an eigenvalue of $\Lambda^{m-i}\tilde F$.
Multiplying these terms together we obtain,
 $$
 R_\phi\left(\frac{1}{dz}\right)
 =
 \left(
 \prod_{i=1}^m
 \prod_{\lambda^{(i)}\in\spec\Lambda^i\tilde F}
 \left(\frac{1}{\lambda^{(i)}\sigma}\right)^{(-1)^i}
 \right)^{(-1)^r}
 \times
 R_\phi(z)^{(-1)^m}.
 $$
The variable $z$ has disappeared because
 $$
 \sum_{i=0}^m (-1)^i \dim\Lambda^i\tilde G = \sum_{i=0}^m (-1)^i  {C_k }^i = 0.
  $$

\section{ The Reidemeister zeta function and gro\-up exten\-sions.} 
Suppose we are given a commutative diagram
\begin{equation}
\begin{array}{lcl}
  G             &  \stackrel{\phi}{\longrightarrow}      &  G  \\
  \downarrow p  &                                        &  \downarrow p\\
  \overline{G}  &  \stackrel{\overline{\phi}}{\longrightarrow}      &  \overline{G}
\end{array}
\end{equation}
of groups and homomorphisms.
In addition let the sequence
\begin{equation}
  0 \rightarrow H \rightarrow G \stackrel{p}{\rightarrow} \overline{G} \rightarrow 0
\end{equation}
be exact.
Then $\phi$ restricts to an endomorphism
$\phi\mid_H : H \rightarrow H$.

\begin{definition}
The short exact sequence (2.31) of groups
is said to have a {\it normal splitting} if there is a section
$\sigma:\overline{G}\rightarrow G$ of $p$
such that $\im \sigma = \sigma(\overline{G})$ is a normal subgroup of $G$.
An endomorphism $\phi:G\rightarrow G$ is said to
{\it preserve} this normal splitting if $\phi$ induces a
morphism of (2.31) with $\phi(\sigma(\overline{G}))\subset\sigma(\overline{G})$.
\end{definition}

In this section we study the relation between the
Reidemeister zeta functions
$R_\phi(z)$, $R_{\overline{\phi}}(z)$ and $R_{\phi\mid_H}(z)$.

\begin{theorem}
Let the sequence (2.31)
have a normal splitting which is preserved by $\phi:G\rightarrow G$.
Then we have
$$
 R_\phi(z) = R_{\overline{\phi}}(z) * R_{\phi\mid_H}(z).
$$
In particular, if $R_{\overline{\phi}}(z)$ and $R_{\phi\mid_H}(z)$
are rational functions then so is $R_\phi(z)$.
If $R_{\overline{\phi}}(z)$ and $R_{\phi\mid_H}(z)$
are rational functions with functional equations as
described in theorems 21 and 24  then so is $R_\phi(z)$.
\end{theorem}

{\sc Proof}
From the assumptions of the theorem it follows that
for every $n>0$
$$  R(\phi^n)
  = R(\overline{\phi}^n) \cdot  R(\phi^n\mid_H)
  \;\;\;\;\; {\rm (see\ \cite{h}).}  $$

\section{The Reidemeister zeta function of a continuous map}

Using Corrolary 1    from Chapter 1  we may apply  all theorems about the Reidemeister zeta function of  group endomorphisms
to the Reidemeister zeta functions of continuous maps.
Theorem 16  yield

\begin{theorem}
Let $X$ be a polyhedron with finite fundamental group $\pi_1(X)$
and let $f:X\to X$ be a continuous map.
Then $R_f(z)$ is a rational function with a functional equation:
$$
 R_f(z)
 =
 R_{\tilde{f}_*}(z)
 =
 \prod_{[<g>]} \frac{1}{1-z^{\#[<g>]}},
$$
$$
 R_{f}\left(\frac{1}{z}\right)
 =
 (-1)^a z^b R_{
f}(z).
$$
The product in the first formula is over all periodic $\tilde{f}_*$-orbits
of ordinary conjugacy classes of elements of $\pi_1(X)$.
The number $\#[<g>]$ is the number of conjugacy classes in the
$\tilde{f}_*$-orbit of $<g>$.
In the functional equation the numbers $a$ and $b$ are respectively
the number of periodic $\tilde{f}_*$- orbits of cojugacy classes of
elements of $\pi_1(X)$, and the number of periodic conjugacy classes
of elements of $\pi_1(X)$.
\end{theorem}

Theorem 10  yield

\begin{theorem}
Let $f:X\to X$ be eventually commutative.
Then $R_f(z)$ is a rational function
and is given by:
$$
 R_f(z)=R_{\tilde{f}_*}(z)
       =R_{f_{1*}}(z)
       =R_{f_{1*}}^\infty(z) * R_{f_{1*}}^{finite}(z),
$$
where $R_{f_{1*}}^\infty(z)$ is the
Reidemeister zeta function of the endomorphism
$f_{1*}^\infty:H_1(X,\bbbz)^\infty\to H_1(X,\bbbz)^\infty$
and $R_{f_{1*}}^{finite}(z)$ is the
Reidemeister zeta function of the endomorphism
$f_{1*}^{finite}:H_1(X,\bbbz)^{finite}\to H_1(X,\bbbz)^{finite}$.
The functions $R_{f_{1*}}^\infty(z)$ and $R_{f_{1*}}^{finite}(z)$
are given by the formulae:
$$
R_{f_{1*}}^\infty(z)
=
\left(
\prod_{i=0}^k
\det\left( 1 - \Lambda^i f^\infty_{1*} \sigma z \right)^{(-1)^{i+1}}
\right)^{(-1)^r}
$$
$$
R_{f_{1*}}^{finite}(z)
=
\prod_{[h]} \frac{1}{1-z^{\#[h]}}
$$
With the product over $[h]$ being taken over all
periodic $f_{1*}$- orbits of torsion elements $h\in H_1(X,\bbbz)$,
and with $\sigma=(-1)^p$ where $p$ is the number of 
$\mu\in\spec f_{1*}^\infty$ such that $\mu<-1$. The number $r$ is the
number of real eigenvalues of $f_{1*}^\infty$ whose absolute
value is $>1$.
\end{theorem}
Theorem 13 yield
\begin{theorem}[Connection with Lefschetz zeta function]
Let $f:X\to X $ be eventually commutative.Then
$$
R_f(z)=R_{\tilde{f}_*}(z)=R_{f_{1*}}(z)= L_{\widehat{(f_{1*})}}(\sigma z)^{(-1)^r},
$$
where $r$ and $\sigma$ are constants as described in theorem 27 . If $X$ is a polyhedron with finite fundamental group then this reduces to
$$
R_f(z)=R_{\tilde{f}_*}(z)=R_{f_{1*}}(z)=L_{\widehat{(f_{1*})}}(z),
$$
\end{theorem}

\begin{theorem}[Functional equation]
Let $f:X\to X$ be eventually commutative.
If $H_1(X;\bbbz)$ is finite, then $R_f(z)$ has the following functional equation:
 $$
 R_f\left(\frac{1}{z}\right)
 =
 (-1)^p z^q R_f(z),
 $$
where $p$ is the number of periodic orbits of $f_{1*}$
in $H_1(X;\bbbz)$
and $q$ is the number of periodic elements of $f_{1*}$
in $H_1(X;\bbbz)$.

If $H_1(X;\bbbz)$ is infinite then $R_f(z)$ has the
following functional equation:
 $$
 R_f\left(\frac{1}{dz}\right)
 =
 \epsilon_2.R_f(z)^{(-1)^{\rank H_1(X;\bbbz)}},
 $$
where $d=\det (f_{1*}^\infty)\in\bbbc^\times$ and $\epsilon_2\in\bbbc^\times$ is a constant.
\end{theorem}

Theorem 20 yield
 \begin{theorem}
Let $X$ be a polyhedron whose fundamental group $\pi$ is the direct sum of a finitely generated free Abelian and a finite group.  Let $f : X \to X $ be a continuous map.Then
$R_f(z)$ is a rational function and is equal to the following additive convolution:
\begin{equation}
 R_f(z)
 =
 R^\infty_{\tilde{f}_*}(z)
 * 
 R^{finite}_{\tilde{f}_*}(z).
\end{equation}
where $R^\infty_{\tilde{f}_*}(z)$ is the Reidemeister
zeta function of the endomorphism 
${\tilde{f}_*}^\infty :\pi^\infty\rightarrow \pi^\infty$,
and $R^{finite}_{\tilde{f}_*}(z)$ is the Reidemeister
zeta function of the endomorphism 
${\tilde{f}_*}^{finite} :\pi^{finite}\rightarrow \pi^{finite}$.
The functions $R^\infty_{\tilde{f}_*}(z)$ and $R^{finite}_{\tilde{f}_*}(z)$
are given by the formulae
\begin{equation}
  R^\infty_{\tilde{f}_*}(z)
  =
  \left(
  \prod_{i=0}^k
  \det(1-\Lambda^i{\tilde{f}_*}^\infty \cdot\sigma z)^{(-1)^{i+1}}
  \right)^{(-1)^r},
\end{equation}
$$
 R^{finite}_{\tilde{f}_*}(z) = \prod_{[<g>]} \frac{1}{1-z^{\#[<g>]}},
$$
The product here is over all periodic ${\tilde{f}_*}$-orbits of
ordinary conjugacy classes of elements of $\pi$.
The number $\#[<g>]$ is the number of conjugacy classes
in the ${\tilde{f}_*}$-orbit of the conjugacy class $<g>$.
Also, $\sigma=(-1)^p$ where $p$ is the number of real eingevalues
$\lambda\in\spec{\tilde{f}_*}^\infty$ such that $\lambda<-1$
and $r$ is the number of  real eingevalues
$\lambda\in\spec{\tilde{f}_*}^\infty$
such that $\mid\lambda\mid > 1$.
\end{theorem}

\begin{theorem}
Let $f:X\rightarrow X$ be a self map of 
of a nilmanifold.
Then 
$$
R_{f}(z)
=
\left(
\prod_{i=0}^m
\det\left( 1 - \Lambda^i \tilde{F}\cdot \sigma \cdot z \right)^{(-1)^{i+1}}
\right)^{(-1)^r}
$$
$$
 R_f\left(\frac{1}{dz}\right)
 =
 \epsilon_1\cdot R_f(z)^{(-1)^{\rank \pi_1(X)}},
$$
where  $\sigma=(-1)^p$,$p$ , $r$, $m$ and  $\tilde F$ is defined in theorem 22,  $d=\det\tilde F$ and $\epsilon_1$
are a constants in $\bbbc^\times$ .
\end{theorem}

\subsection{The Reidemeister zeta function of a continuous map and Serre bundles.}

Let $p:E\rightarrow B$ be a Serre bundle in which $E$, $B$ and
every fibre are connected, compact polyhedra and $F_b=p^{-1}(b)$
is a fibre over $b\in B$.
A Serre bundle $p:E\rightarrow B$ is said to be 
{\it (homotopically) orientable} if for any two paths $w$, $w^\prime$ in $B$
with the same endpoints $w(0)= w^\prime(0)$ and $w(1)= w^\prime(1)$,
the fibre translations
$\tau_w , \tau_{w^\prime} : F_{w(0)}\rightarrow F_{w(1)}$ are homotopic.
A map $f:E\rightarrow E$ is called a {\it fibre map} if there
is an induced map $\bar{f}:B\rightarrow B$ such that
$p\circ f = \bar{f}\circ p$.
Let $p:E\rightarrow B$ be an orientable Serre bundle and
let $f:E\rightarrow E$ be a fibre map.
Then for any two fixed points $b,b^\prime$ of $\bar{f}:B\rightarrow B$
the maps $f_b=f\mid_{F_b}$ and $f_{b^\prime} = f\mid_{F_{b^\prime}}$
have the same homotopy type;
hence they have the same Reidemeister numbers
$R(f_b) = R(f_{b^\prime})$ \cite{j}.

The following theorem describes the relation between the Reidemeister
zeta functions $R_f(z)$, $R_{\bar{f}}(z)$ and $R_{f_b}(z)$ for a fibre
map $f:E \rightarrow E$ of an orientable Serre bundle $p:E\rightarrow B$.

\begin{theorem}
Suppose that $f:E \rightarrow E$ admits a Fadell splitting
in the sense that for some $e$ in $\fix f$ and $b=p(e)$ the
following conditions are satisfied:
\begin{enumerate}
\item
the sequence
$$  0 \longrightarrow \pi_1(F_b,e)
      \stackrel{i_*}{\longrightarrow} \pi_1(E,e)
      \stackrel{p_*}{\longrightarrow} \pi_1(B,e)
      \longrightarrow 0 $$
is exact,

\item
$p_*$ admits a right inverse (section) $\sigma$ such that $\im\sigma$
is a normal subgroup of $\pi_1(E,e)$ and $f_*(\im\sigma)\subset\im\sigma$.

\end{enumerate}
We then have 
$$R_f(z)=R_{\bar{f}}(z)*R_{f_b}(z).$$
If $R_{\bar{f}}(z)$ and $R_{f_b}(z)$ are rational functions
then so is $R_f(z)$.
If $R_{\bar{f}}(z)$ and $R_{f_b}(z)$ are rational functions
with functional equations as described in theorems 26 and 29  then so is $R_f(z)$.

\end{theorem}

{\sc Proof}
The proof follows from theorem 25 .

\chapter{ The Nielsen zeta function}
\markboth{\sc Nielsen zeta function}{\sc Nielsen zeta function}

\section{ Radius of Convergence of the Nielsen zeta function}

In this section we find a sharp estimate for the radius of convergence of the Nielsen zeta function
in terms of the topological entropy of the map. It follows from this estimate that the Nielsen zeta function has positive radius of convergence. 
\subsection{Topological entropy}

The most widely used measure for the complexity of a dynamical system is the topological
entropy. For the convenience of the reader, we include its definition.
 Let $ f: X \rightarrow X $ be a self-map of a compact metric space. For given $\epsilon > 0 $
 and $ n \in \bbbn $, a subset $E \subset X$ is said to be $(n,\epsilon)$-separated under $f$ if for
 each pair $x \not= y$ in $E$ there is $0 \leq i <n $ such that $ d(f^i(x), f^i(y)) > \epsilon$.
 Let $s_n(\epsilon,f)$  denote the largest cardinality of any $(n,\epsilon)$-separated subset $E$
 under $f$. Thus  $s_n(\epsilon,f)$ is the greatest number of orbit segments ${x,f(x),...,f^{n-1}(x)}$
 of length $n$ that can be distinguished one from another provided we can only distinguish 
 between points of $X$ that are  at least $\epsilon$ apart. Now let
 $$
 h(f,\epsilon):= \limsup_{n} \frac{1}{n}\cdot\log \,s_n(\epsilon,f)
 $$
 $$
 h(f):=\limsup_{\epsilon \rightarrow 0} h(f,\epsilon).
 $$
 The number $0\leq h(f) \leq \infty $, which to be independent of the metric $d$ used, is called the topological entropy of $f$.
 If $ h(f,\epsilon)> 0$ then, up to resolution $ \epsilon >0$, the number $s_n(\epsilon,f)$ of 
 distinguishable orbit segments of length $n$ grows exponentially with $n$. So $h(f)$
 measures the growth rate in $n$ of the number of orbit segments of length $n$
 with arbitrarily fine resolution.  
   A basic relation between Nielsen numbers and topological entropy was found by N.Ivanov
   \cite{i} and independently by Aronson and Grines. We present here a very short proof of Jiang \cite {j1} of the Ivanov's  inequality.
\begin{lemma}\cite{i}
$$
h(f) \geq \limsup_{n} \frac{1}{n}\cdot\log N(f^n)
 $$ 
\end{lemma}
{\sc Proof}
 Let $\delta$ be such that every loop in $X$ of diameter $ < 2\delta $ is contractible.
 Let $\epsilon >0$ be a smaller number such that $d(f(x),f(y)) < \delta $ whenever $ d(x,y)<2\epsilon $. Let $E_n \subset X $ be a set consisting of one point from each essential fixed point class of $f^n$. Thus $ \mid E_n \mid =N(f^n) $. By the definition of $h(f)$, it suffices
 to show that $E_n$ is $(n,\epsilon)$-separated.
 Suppose it is not so. Then there would be two points $x\not=y \in E_n$ such that $ d(f^i(x), f^i(y)) \leq \epsilon$ for $o\leq i< n$ hence for all $i\geq 0$. Pick a path $c_i$ from $f^i(x)$ to
 $f^i(y)$ of diameter $< 2\epsilon$ for $ o\leq i< n$ and let $c_n=c_0$. By the choice of $\delta$
 and $\epsilon$ ,  $f\circ c_i \simeq c_{i+1} $ for all $i$, so $f^n\circ c_0\simeq c_n=c_0$.
 This means $x,y$ in the same fixed point class of $f^n$, contradicting the construction of $E_n$.

 This inequality is remarkable in that it does not require smoothness of the map and provides a common lower bound for the topological entropy of all maps in a homotopy class.

   We denote by $R$
the radius of convergence of the Nielsen zeta function  $N_f(z)$. Let $h=\inf h(g) $ over all maps $g$ of the same homotopy type as $f$.
\begin{theorem}
For a continuous map of a compact polyhedron $X$ into itself,
\begin{equation}
 R \geq \exp(-h)>0.
\end{equation}

\end{theorem}
{\sc Proof}
The inequality $ R \geq \exp(-h)$ follows from the previous lemma, the Cauchy-Hadamard formula, and the homotopy invariance of the radius $R$ of the Nielsen zeta function $N_f(z)$. We consider a smooth compact manifold 
$M$ which is a regular neighborhood of $X$ and a smooth map $g:M\rightarrow M$ of the 
same homotopy type as $f$. As is known \cite{pz} , the entropy $h(g) $ is finite. Thus  
$\exp(-h) \geq \exp(-h(g)) > 0$.

\subsection{Algebraic lower estimation for the Radius of Convergence}

In this subsection we propose another prove of positivity of the radius $R$ and give an exact
algebraic lower estimation for the radius $R$ using the Reidemeister trace formula for 
generalized Lefschetz numbers.

The fundamental group  $\pi=\pi_1(X,x_0)$ splits into $\tilde{f}_*$-conjugacy classes.Let $\pi_f$ denote the set of $\tilde{f}_*$-conjugacy classes,and $\bbbz\pi_f$ denote the abelian group freely generated by $\pi_f$ . We will use the bracket notation $a\to [a]$ for both projections $\pi\to \pi_f$ and  $\bbbz\pi\to \bbbz\pi_f$.
Let $x$ be a fixed point of $f$.Take a path $c$ from $x_0$ to $x$.The  $\tilde{f}_*$-conjugacy class in $\pi$ of the loop $c\cdot (f\circ c)^{-1}$,which is evidently independent of the choice of $c$, is called the coordinate of $x$.Two fixed points are in the same fixed point class $F$ iff they have the same coordinates.This
$\tilde{f}_*$-conjugacy class is thus called the coordinate of the fixed point class $F$ and denoted $cd_{\pi}(F,f)$ (compare with  description in section 1).
 
The generalized Lefschetz number  or the Reidemeister trace \cite{re} is defined as
\begin{equation}
L_{\pi}(f):=\sum_{F} \ind (F,f).cd_{\pi}(F,f)  \in  \bbbz\pi_f ,
\end{equation}
 
the summation being over all essential fixed point classes $F$  of $f$.The Nielsen number $N(f)$ is the number of non-zero terms in $L_{\pi}(f)$,and the indices of the essential fixed point classes appear as the coefficients in $L_{\pi}(f)$.This invariant used to be called the Reidemeister trace because it can be computed as an alternating sum of traces on the chain level  as follows \cite{re},\cite {wec} .
Assume that $X$ is a finite cell complex and  $f:X\to X$ is a cellular map.A cellular decomposition ${e_j^d}$ of $X$ lifts to a $\pi$-invariant cellular structure on the universal covering $\tilde X$.Choose an arbitrary lift  ${\tilde{e}_j^d}$ for each ${e_j^d}$ . They constitute a free  $\bbbz\pi$-basis for the cellular chain complex of $\tilde{X}$.The lift $\tilde{f}$ of $f$ is also a cellular map.In every dimension $d$, the cellular chain map $\tilde{f}$ gives rise to a $\bbbz\pi$-matrix $ \tilde{F}_d $ with respect to the above basis,i.e $\tilde{F}_d=(a_{ij})$ if $ \tilde{f}(\tilde{e}_i^d)=\sum_{j}a_{ij}\tilde{e}_j^d $,where $ a_{ij}\in \bbbz\pi $.Then we have the Reidemeister trace formula
 
\begin{equation}
L_{\pi}(f)=\sum_{d}(-1)^d[ \tr \tilde{F}_d] \in \bbbz\pi_f .
\end{equation}

    Now we describe alternative approach to the Reidemeister trace formula proposed recently by Jiang \cite{j1}. This approach is useful when we study the periodic points of $f$, i.e. the fixed points of the iterates of $f$. 
 
 The mapping torus $T_f$ of $f:X\rightarrow X$ is the space obtained from $X\times [o,\infty )$ by identifying $(x,s+1)$ with $(f(x),s)$ for all $x\in X,s\in [0 ,\infty )$.On $T_f$ there is a natural semi-flow $\phi :T_f\times [0,\infty )\rightarrow T_f, \phi_t(x,s)=(x,s+t)$ for all $t\geq 0$.Then the map  $f:X\rightarrow X$ is the return map of the semi-flow $\phi $. A point $x\in X$ and a positive number $\tau >0$ determine the orbit curve $\phi _{(x,\tau )}:={\phi_t(x)}_{0\leq t \leq \tau}$ in $T_f$.
 
Take the base point $x_0$ of $X$ as the base point of $T_f$. It is known that the fundamental group $H:=\pi_ 1(T_f,x_0)$ is obtained from $\pi $ by adding a new generator $z$ and adding the relations $z^{-1}gz=\tilde f_*(g)$ for all $g\in \pi =\pi _1(X,x_0)$.Let  $H _c$ denote the set of conjugacy classes in $H $. Let $\bbbz H $ be the integral group ring of $H $, and let $\bbbz H_c $ be the free abelian group with basis $ H _c $.We again use the bracket notation $ a\rightarrow [a] $  for both projections $H \rightarrow H _c $ and $ \bbbz H \rightarrow \bbbz H _c $.  If  $F^n$ is a fixed point class  of $f^n$, then 
$f(F^n)$ is also fixed point class of $f^n$ and
 $\ind(f(F^n),f^n)=\ind(F^n,f^n)$.
Thus $f$ acts as an index-preserving permutation among fixed point classes of $f^n$. By definition, an $n$-orbit class $O^n$  of $f$ to be the union of elements of an orbit of this action. In other words, two points $x,x' \in \fix(f^n)$ are said to be in the same $n$-orbit class of $f$ if and only if some $f^i(x)$ and some $f^j(x')$ are in the same fixed point class of $f^n$.The set $\fix(f^n)$ splits into a disjoint union of $n$-orbits classes. Point $x$ is a fixed point of $f^n$  
or a periodic point of period $n$ if and only if orbit curve  $\phi _{(x,n)}$ is a closed curve. The free homotopy class of the closed curve $\phi _{(x,n)}$ will be called the $H$ -coordinate of point $x$,written $cd_{H }(x,n)=[\phi _{(x,n)}]\in H _c$. It follows that periodic points $x$ of period $n$ and $x'$ of period $n'$ have the same $H $-coordinate if and only if $n=n'$ and $x$, $x'$ belong to the same $n$-orbits class of $f$. Thus it is possible equivalently define $x,x'\in \fix f^n $ to be in the same $n$-orbit class if and only if they have the same $H-$coordinate.
 
Recently, Jiang \cite{j1} has considered generalized Lefschetz number with respect to $H $
\begin{equation}
L_{H }(f^n):= \sum_{O^n} \ind (O^n,f^n)\cdot cd_{H }(O^n) \in \bbbz H _c, \end{equation}
and proved following trace formula:
\begin{equation}
L_{H }(f^n)=\sum_{d}(-1)^d[\tr (z\tilde{F}_d)^n] \in \bbbz H_c,
\end{equation}
where $\tilde{F}_d$ be $\bbbz \pi$-matrices defined above and $z\tilde{F}_d$ is regarded as a $\bbbz H $-matrix.

 For any set $S$ let $\bbbz S$ denote the free abelian group with the specified basis $S$.The norm in $\bbbz S$ is defined by
\begin{equation}
\|\sum_i k_is_i\|:= \sum_i \mid k_i\mid \in \bbbz,
\end{equation}
when the $s_i$ in $S$ are all different.
 
For a $\bbbz H $-matrix $ A=(a_{ij}) $,define its norm by $ \| A \|:=\sum_{i,j}\| a_{ij} \| $.Then we have inequalities $\| AB \|\leq \| A\ |\cdot \| B \|$ when $A,B$ can be multiplied, and $\| \tr A \|\leq \| A \| $ when $A$ is a square matrix.For a matrix  $ A=(a_{ij}) $ in $\bbbz S$, its matrix of norms is defined to be the matrix $ A^{norm}:=(\| a_{ij} \|)$ which is a matrix of non-negative integers.In what follows, the set $S$ will be $\pi $, $H $ or
$H _c$.We denote by $s(A)$ the spectral radius of $A$, $s(A)=\lim_n (\| A^n \| |)^{\frac{1}{n}},$ which coincide with the largest  module of an eigenvalue of $A$.
 
\begin{theorem}
For any continuous map $f$ of any compact polyhedron   $X$ into itself the Nielsen zeta function has positive radius of convergence $R$,which admits following estimations 
\begin{equation}
 R\geq \frac{1}{\max_d \| z\tilde F_d \|} > 0
\end{equation}
and
\begin{equation}
R\geq \frac{1}{\max_d s(\tilde F_d^{norm})} > 0, 
\end{equation}
\end{theorem}
 
{\sc Proof}
By the homotopy type invariance of the invariants we can suppose that $f$ is a cell map of a finite cell complex.By the definition, the Nielsen number $N(f^n)$ is the number of non-zero terms in $L_{\pi}(f^n)$.The norm $\| L_{H}(f^n) \| $ is the sum of absolute values of the indices of all the $n$-orbits classes $O^n$ . It equals $\| L_{\pi}(f^n) \|$, the sum of absolute values of the indices of all the fixed point classes of $f^n$,  because any two fixed point classes of $f^n$ contained in the same $n$-orbit class $O^n$ must have the same index. From this we have 
$$
N(f^n)\leq\|L_{\pi}(f^n)\|=\|L_H(f^n)\|=\|\sum_d(-1)^d[\tr(z\tilde F_d)^n]\|\leq
$$
$$
\leq\sum_d\|[\tr(z\tilde F_d)^n]\|\leq\sum_d\|\tr(z\tilde F_d)^n\|\leq\sum_d\|(z\tilde F_d)^n\|\leq\sum_d\|(z\tilde F_d)\|^n 
$$
(see \cite{j1}).The radius of convergence $ R$ is given by Caushy-Adamar formula:
$$
 \frac{1}{R}=\limsup_n (\frac{N(f^n)}{n})^{\frac{1}{n}}=\limsup_n (N(f^n))^{\frac{1}{n}}.
$$
Therefore we have:
$$
R= \frac{1}{\limsup_n (N(f^n))^{\frac{1}{n}}}\geq \frac{1}{\max_d \| z\tilde F_d \|} > 0.
$$
Inequalities:
$$
N(f^n)\leq\|L_{\pi}(f^n)\|\!=\!\|L_H(f^n)\|\!=\!\|\sum_d(-1)^d[\tr(z\tilde F_d)^n]\|\!\leq\sum_d\| [\tr(z\tilde F_d)^n]\|\!\leq
$$
$$
\leq\sum_d{||}\tr(z\tilde F_d)^n{||}\leq \sum_d\tr((z\tilde
F_d)^n)^{norm}\leq\sum_d\tr((z\tilde F_d)^{norm})^n\leq
$$
$$
\leq \sum_d\tr((\tilde F_d)^{norm})^n
$$
and the definition of spectral radius give estimation:
$$
R= \frac{1}{\limsup_n (N(f^n))^{\frac{1}{n}}} \geq \frac{1}{\max_d s(\tilde F_d^{norm})} > 0.
$$
 
\begin{example}
 Let $X$ be surface with boundary, and  $f:X\rightarrow X$ be a map.Fadell and Husseini \cite{fahu} devised a method of computing the matrices of the lifted chain map for surface maps.Suppose $ \{a_1, .... ,a_r\} $ is a free basis for $\pi_1(X)$. Then $X$ has the homotopy type of a bouquet $B$  of $r$ circles which can be decomposed into one 0-cell and $r$  1-cells corresponding to the $a_i$,and $f$ has the homotopy type of a cellular map  $g:B\rightarrow B.$
By the homotopy type invariance of the invariants,we can replace $f$ with $g$ in computations.The homomorphism $\tilde f_*:\pi_1(X)\rightarrow \pi_1(X)$ induced by $f$ and $g$ is determined by the images $b_i=\tilde f_*(a_i), i=1,.. ,r $.The fundamental group $\pi_1(T_f)$ has a presentation $\pi_1(T_f)=<a_1,...,a_r,z| a_iz=zb_i, i=1,..,r>$.Let
$$
D=(\frac{\partial b_i}{\partial a_j})
$$
be the Jacobian in Fox calculus(see \cite{bi}).Then,as pointed out in \cite{fahu}, the matrices of the lifted chain map $\tilde g$ are
$$
\tilde F_0=(1), \tilde F_1=D=(\frac{\partial b_i}{\partial a_j}).
$$
Now, we can find estimations for the radius $R$ from (3.7) and (3.8).
 
\end{example}

\section{ Nielsen zeta function of a  periodic map}

 The following problem is of interest:  for which spaces and classes of maps is the Nielsen
 zeta function rational? When is it algebraic? Can  $N_f(z)$ be transcendental? 
 Sometimes one can answer these questions without directly calculating the Nielsen numbers
 $N(f^n)$, but using  the connection between Nielsen numbers of iterates. We denote
 $N(f^n)$ by $N_n$.We shall say  that $f:X\rightarrow X$ is a  periodic map of period $m$, if $f^m$ is  the identity map $id_X:X\rightarrow X$. Let $ \mu(d), d \in N$, be the M\"obius function of number theory. As is known, it is given by the following equations:  $\mu(d)=0 $ if $d$ is divisible by a square different from one ; $\mu(d)=(-1)^k $ if
 $d$ is not divisible  by a square different from one , where $k$ denotes the number of 
 prime divisors of $d$; $ \mu(1)=1$.

\begin{theorem}
Let $f$ be a  periodic map of least period $m$ of the connected
compact polyhedron $X$ . Then the Nielsen
 zeta function is equal to 
$$
N_f(z)=\prod_{d\mid m}\sqrt[d]{(1-z^d)^{-P(d)}}
$$
, where the product is taken over all divisors $d$ of the period $m$, and $P(d)$ is the integer
$$  P(d) = \sum_{d_1\mid d} \mu(d_1)N_{d\mid d_1} .  $$
\end{theorem}
{\sc Proof}
Since $f^m = id $, for each $j, N_j=N_{m+j}$. Since $(k,m)=1$, there exist positive integers $t$
and $q$ such that $kt=mq+1$. So $ (f^k)^t=f^{kt}= f^{mq+1}=f^{mq}f=(f^m)^{q}f =  f$.
Consequently, $  N((f^k)^t)=N(f) $. Let two fixed point $x_0$ and $x_1$ belong to the same  fixed point class. Then there exists a path $ \alpha $ from  $x_0$ to $x_1$ such that $ \alpha \ast (f\circ\alpha)^{-1} \simeq 0$. We have $f(\alpha \ast f\circ\alpha)^{-1})=(f\circ\alpha)\ast (f^2\circ\alpha)^{-1} \simeq 0$ and a product $ \alpha \ast (f\circ\alpha)^{-1}\ast (f\circ\alpha)\ast (f^2\circ\alpha)^{-1} =\alpha \ast (f^2\circ\alpha)^{-1}\simeq 0 $. It follows that $ \alpha \ast (f^k\circ\alpha)^{-1} \simeq 0$ is derived by the iteration of this process.So $x_0$ and $x_1$ belong to the same  fixed point class of $f^k$. If two point belong to the different  fixed point classes $f$, then they belong to the different  fixed point classes of $f^k$ . So, each essential class( class with nonzero index) for $f$  is an essential class for $f^k$;
in addition , different essential classes for $f$ are different essential classes for $f^k$.
So $ N(f^k)\geq N(f) $. Analogously, $ N(f)=N((f^k)^t) \geq  N(f^k) $. Consequently ,
$ N(f)=N(f^k) $. One can prove completely analogously that $ N_d= N_{di} $, if (i, m/d) =1,
where $d$ is a divisor of $m$. Using these series of equal Nielsen numbers, one can regroup
the terms of the series in the exponential of the Nielsen zeta function so as to get logarithmic
functions by adding and subtracting missing terms with necessary coefficient.
We show how to do this first for period $m=p^l$, where $p$ is a prime number . We have the
following series of equal  Nielsen numbers:
 $$
 N_1=N_k, (k,p^l)=1 (i.e., no\, N_{ip}, N_{ip^2},...
.., N_{ip^l},  i=1,2,3,....),
 $$
 $$ 
 N_p=N_{2p}=N_{3p}=.......=N_{(p-1)p}=N_{(p+1)p}= ... (no \,
N_{ip^2}, N_{ip^3}, ...  , N_{ip^l} ) 
$$
etc.; finally,
$$
N_{p^{l-1}}=N_{2p^{l-1}}=..... (no \, N_{ip^l})
$$
and separately the number $N_{p^l}$.\\
 Further,
\begin{eqnarray*}
\sum_{i=1}^\infty \frac{N_i}{i} z^i & = & \sum_{i=1}^\infty \frac{N_1}{i} z^i 
       +\sum_{i=1}^\infty \frac{(N_p -N_1)}{p}\frac{ {z^p}^i}{i} + \\
                                                    & +  &\sum_{i=1}^\infty \frac{(N_{p^2} -(N_p -N_1)- N_1)}{p^2} \frac{{z^{p^2}}^i}{i} + ...\\ 
 & + & \sum_{i=1}^\infty \frac{(N_{p^l} - ...- (N_p -N_1)- N_1)}{p^l}\frac{ {z^{p^l}}^i}{i}\\
 &=& -N_1\cdot \log (1-z) + \frac{N_1-N_p}{p}\cdot
						    \log (1-z^p) +\\
& + & \frac{N_p-N_{p^2}}{p^2}\cdot \log (1-z^{p^2}) + ...\\
& + & \frac{N_{p^{l-1}}-N_{p^l}}{p^l}\cdot \log (1-z^{p^l}).
\end{eqnarray*}
For an arbitrary period $m$ , we get completely analogously,

\begin{eqnarray*}
N_f(z) & = & \exp\left(\sum_{i=1}^\infty \frac{N(f^i)}{i} z^i \right)\\
           & = & \exp\left(\sum_{d\mid m} \sum _{i=1}^\infty \frac{P(d)}{d}\cdot\frac{{z^d}^i}{i}\right)\\
	   & = & \exp\left(\sum_{d\mid m}\frac{P(d)}{d}\cdot \log (1-z^d)\right)\\
	   & = & \prod_{d\mid m}\sqrt[d]{(1-z^d)^{-P(d)}}
\end{eqnarray*}
where the integers $P(d)$ are calculated recursively by the formula
$$
P(d)= N_d  - \sum_{d_1\mid d; d_1\not=d} P(d_1).
$$
Moreover, if the last formula is rewritten in the form
$$
N_d=\sum_{d_1\mid d}\mu(d_1)\cdot P(d_1)
$$ 
and one uses  the M\"obius Inversion law for real function in number theory, then
$$
P(d)=\sum_{d_1\mid d}\mu(d_1)\cdot N_{d/d_1},
$$
where $\mu(d_1)$ is the M\"obius function in number theory. The theorem is proved.

\begin{corollary}
If in Theorem 35 the period $m$ is a prime number, then
$$ 
N_f(z)  =   \frac{1}{(1-z)^{N_1}}\cdot \sqrt[m]{(1-z^m)^{N_1 - N_m}}.
$$
For an involution of a connected compact polyhedron, we get
$$
N_{in}(z)  =  \frac{1}{(1-z)^{N_1}}\cdot \sqrt[2]{(1-z^2)^{N_1 - N_2}}.
$$
\end{corollary}

\begin{remark}
Let $f: M^n \rightarrow M^n , \, n \geq  3$ be a homeomorphism of a compact hyperbolic manifold. Then by Mostow rigidity theorem $f$ is homotopic  to periodic homeomorphism $g$.
So theorem 35 applies and the Nielsen zeta function $N_f(z)$ is equal to
 $$
N_f(z)=N_g(z)= \prod_{d\mid m}\sqrt[d]{(1-z^d)^{-P(d)}}
$$
, where the product is taken over all divisors $d$ of the least period $m$ of $g$,  and $P(d)$ is the integer
$  P(d) = \sum_{d_1\mid d} \mu(d_1)N(g^{d\mid d_1}) .  $
 \end{remark}

\begin{remark}
Let $f:X\rightarrow X$ be a continuous map of a connected compact polyhedron $X$, homotopic to $id_X$ . Since the Lefschetz numbers $L(f^n)=L(id_X) = \chi(X)$, where $\chi(X)$
is the  Euler characteristic of $X$, then for $\chi(X)\not=0$ one has $ N(f^n)=N(id_X)=1$ for
all $ n>0$ , and $N_f(z)=\frac{1}{1-z}$; if $\chi(X)=0$, then $ N(f^n)=N(id_X)=0$for
all $ n>0$ , and $N_f(z)=1$
\end{remark}

\section{Orientation-preserving homeomorp\-hisms of a compact surface}

The proof of the following theorem is based on Thurston's theory of orienta\-tion-preserving
homeomorphisms of surfaces \cite{th}.

\begin{theorem}

The Nielsen zeta function of an orientation-preserving homeomorphism $f$ of a compact surface
$M^2$ is either a rational function or the radical of a rational function.
\end{theorem}

{\sc Proof} 
The case of an orientable surface with $\chi(M^2)\geq 0  ( S^2, T^2)$ is considered in subsection 3.8.
 In the case of an orientable surface with $\chi(M^2)< 0$ , according to Thurston's classification
 theorem, the homeomorphism $f$ is isotopic either to a periodic or a pseudo-Anosov, or  a reducible homeomorphism.In the first case the assertion of the theorem follows  from theorem 35.
 If $f$ is an orientation-preserving pseudo-Anosov homeomorphism of a compact surface(i.e. there is a number $\lambda >1$ and a pair of transverse measured foliations $(F^s,\mu^s)$ and $(F^u,\mu^u)$ such that $f(F^s,\mu^s)=(F^s,\frac{1}{\lambda}\mu^s)$ and $f(F^u,\mu^u)=(F^u,\lambda\mu^u)$),
 then for each $ n>0, N(f^n)=F(f^n)$ \cite{th}, \cite{i}, \cite{j}. Consequently, in this case the Nielsen zeta function coincides with the Artin-Mazur zeta function: $N_f(z)=F_f(z)$. Since in \cite{fash}  
 Markov partitions are constructed for a pseudo-Anosov homeomorphism, Manning's proof \cite{m} of the rationality  of the Artin-Mazur zeta function for diffeomorphisms satisfying Smale's axiom A carries over to the case of pseudo-Anosov homeomorphisms.Thus , the Nielsen zeta function $N_f(z)$  is also rational.Now if $f$ is isotopic to a reduced homeomorphism $\phi$,
 then there exists a reducing system $S$ of disjoint circles $S_1, S_2, ... , S_m$ on $ intM^2$
 such that \\
  
  1)  each circle $S_i$ does not bound a disk in $M^2$;\\
  
  2) $S_i$ is not isotopic to $S_j, i\not=j$;\\
    
  3)  the system of circles $S$ is invariant with respect to $\phi$; \\
  
  4)  the system $S$ has an open $\phi$-invariant tubular neighborhood $\eta(S)$ such that
       each $\phi$ -component $ \Gamma_j $  of the set $ M^2 - \eta(S)$ is mapped into itself by some                                                            
       iterate $\phi^{n_j}, n_j  >0$ of the map $\phi$; here  $\phi^{n_j}$ on $\Gamma_j$ is either a   
       pseudo-Anosov or a
       periodic homeomorphism; \\
  
  5)  each band $ \eta(S_i)$ is mapped into itself by some iterate $\phi^{m_i}, m_i > 0$; here      
       $\phi^{m_i}$
       on $ \eta(S_i)$ is a generalized twist (possibly trivial).\\ 
  
  Since the band $ \eta(S_i)$ is homotopically equivalent to the circle $ S^1$, as 
  will be shown in subsection 3.8 the Nielsen zeta function $N_{\phi^m_i}(z)$ is rational.
  The zeta functions $N_\phi (z)$ and $N_{\phi^m_i}(z)$ are connected on the $\phi$ - component
  $ \Gamma_j $ by the formula $N_\phi (z)=\sqrt[n_j]{ N_{\phi^n_j}(z^{n_j})};$ analogously,
  on the band $ \eta(S_i), N_\phi (z)=\sqrt[m_j]{ N_{\phi^m_j}(z^{m_j})}$.
  The fixed points of $\phi^n$, belonging to different components $ \Gamma_j $
  and bands $ \eta(S_i)$ are nonequivalent \cite{i1},so the Nielsen number $N(\phi^n)$
  is equal to the sum of the Nielsen numbers $ N(\phi^n / \Gamma_j)$ and $ N(\phi^n / \eta(S_i))$
  of $\phi$ -components and bands . Consequently, by the properties of the exponential,
  the Nielsen zeta function $N_{\phi } (z)=N_f(z)$ is equal to the product of the Nielsen zeta functions 
  of the $\phi$- components $ \Gamma_j $ and the bands $ \eta(S_i)$, i.e. is the radical of a
  rational function.
  \begin{remark}
  For an orientation -preserving pseudo-Anosov homeomorphism of a compact surface 
  the radius of the Nielsen zeta function $N_f(z)$ is equal to
   $$ R=\exp(-h(f))=\frac{1}{\lambda(A)}$$
 , where $\lambda(A)$ is the largest eigenvalue of the transition matrix of the topological Markov chain corresponding to $f$.
  \end{remark}

\subsection{Geometry of the Mapping Torus and Radius of Convergence }

Let $f:M^2\rightarrow M^2$ be an orientation-preserving homeomorphism of a compact orientable surface, $R$ the radius of convergence of Nielsen zeta function and $T_f$
be the mapping torus of $f$, i.e. $T_f$ is obtained from $M^2\times \lbrack 0,1\rbrack $
by identifying $(x,0)$ to $(f(x),1)$ , $x \in M^2$.
\begin{lemma}
Let the Euler characteristic $\chi(M^2) <0$. Then $R=1$ if and only if the Thurston normal form
for $f$ does not contain pseudo-Anosov components.

\end{lemma}  

{\sc Proof}
Let $\phi$  be a Thurston canonical form of $f$. Suppose that $\phi$ does not contain a 
pseudo-Anosov component. Then $\phi$ is periodic or reducible by $S$, where on each $\phi$-component $\phi$ is periodic. If $\phi$ is periodic then by theorem 35  $R=1$. If $\phi$
is reducible then in theorem 36 we have proved that the Nielsen zeta function $N_{\phi } (z)=N_f(z)$ is equal to the product of the Nielsen zeta functions 
  of the $\phi$- components $ \Gamma_j $ and the bands $ \eta(S_i)$, i.e. has radius of convergence $R=1$.

A three-dimensional manifold $M$ is called a graph manifold if there is a system of mutually disjoint two-dimensional tori $\{ T_i\}$ in $M$ such that the closure of each component of $M$
 cut along $\cup T_i$ is a (surface) $\times S^1$.

\begin{theorem}
 Let $\chi(M^2) <0 $. The mapping torus $T_f$ is a graph-manifold if and only if $R=1$.
 If $ Int T_f $ admits a hyperbolic structure of finite volume , then $R<1$. If $R<1$ then 
 $f$ has an infinite set of periodic points with pairwise different periods.
\end{theorem}
{\sc Proof}
T. Kobayashi \cite {koba} has proved that mapping torus $T_f$ is a graph-manifold if and only if the
Thurston normal form
for $f$ does not contain pseudo-Anosov components.So, lemma 23  implies the first statement of the theorem. Thurston has proved \cite {thu3}, \cite{su} that  $ Int T_f$  admits a hyperbolic structure of finite volume if and only if $f$ is isotopic to pseudo-Anosov homeomorp\-hism.
But for pseudo-Anosov homeomorphism  $1> R=\exp(-h(f))=\frac{1}{\lambda(A)}$, where $\lambda(A)$ is the largest eigenvalue of the transition matrix of the topological Markov chain corresponding to $f$.This proves the second statement of the theorem. If $R<1$ thenThurston normal form for $f$ does  contain pseudo-Anosov components. It is known \cite{koba} that
pseudo-Anosov homeomorphism has infinitely many periodic points those periods are
mutually distinct.

A link $L$ is a finite union of mutually disjoint circles in a three-dimensional manifold.The exterior of $L$ is the closure of the complement of a regular neighborhood of $L$. A link $L$ is a graph 
link if the exterior of $L$ is a graph manifold.Let $\Sigma $ a set consisting of a finite  number
of periodic orbits of $f$. The set $\Sigma \times \lbrack 0,1\rbrack $ projects to a link $L_{f,\Sigma}$ in the mapping torus $T_f$. 

\begin{corollary}
Let $\chi(M^2-\Sigma <0$.The link $L_{f,\Sigma}$ is a graph link if and only if $R=1$.
\end{corollary}
{\sc Proof} 
Homeomorphism $f$ is isotope $rel \Sigma$ to a diffeomorphism $g$. Let  $F$ be a surface
obtained from $M^2-\Sigma $ by adding a circle to each end . Since $g$ is differentiable 
at each point of $\Sigma$ , $g$ extends to $\tilde{g}: F \rightarrow F $ \cite{han}. 
 By theorem 37  $T_{\tilde{g}}$ is a graph manifold if and only if $R=1$. Hence 
$L_{f,\Sigma}$ is a graph link if and only if $R=1$.

\section{The Jiang subgroup and the Nielsen zeta function}

From the homotopy invariance theorem (see \cite{j})
it follows that if a homotopy $\{h_t\}:f\cong g:X\rightarrow X$
lifts to a homotopy
 $\{\tilde{h}_t\}:\tilde{f}\cong \tilde{g}:\tilde{X}\rightarrow \tilde{X}$,
 then we have
 $ \ind (f,p(\fix\ \tilde{f})) = \ind (g,p(\fix \tilde{g}))$.
Suppose $\{h_t\}$ is a cyclic homotopy $\{h_t\}:f\cong f$;
then this lifts to a homotopy from a given lifting $\tilde{f}$ to
another lifting $\tilde{f}^\prime = \alpha\circ\tilde{f}$, and we have
 $$
  \ind (f,p(\fix\ \tilde{f})) = \ind (f,p(\fix \alpha\circ\tilde{f})).
 $$
In other words, a cyclic homotopy induces a permutation of lifting classes
(and hence of fixed point classes);
those in the same orbit of this permutation have the same index.
This idea is applied to the computation of $N_f(z)$.

\begin{definition}
The {\it trace subgroup of cyclic homotopies} (the {\it Jiang subgroup})
$I(\tilde{f})\subset\pi$ is defined by
$$
 I(\tilde{f})
 =
 \left\{ \alpha\in\pi \ \Bigg|
 \begin{array}{l}
    {\rm there\ exists\ a\ cyclic\ homotopy\ } \\
    \{h_t\}:f\cong f
    {\rm which\ lifts\ to\ } \\
    \{\tilde{h}_t\}:
    \tilde{f}\cong\alpha\circ\tilde{f}
 \end{array}
 \right\}
$$
(see \cite{j}).
\end{definition}

Let $Z(G)$ denote the centre of a group $G$, and let $Z(H,G)$
denote the centralizer of the subgroup $H\subset G$.
The Jiang subgroup has the following properties:
\begin{enumerate}
\item
$$  I(\tilde{f})\subset Z(\tilde{f}_*(\pi),\pi); $$
\item
$$  I(id_{\tilde{X}}) \subset Z(\pi) ;$$
\item
$$  I(\tilde{g}) \subset I(\tilde{g}\circ\tilde{f});  $$
\item
$$  \tilde{g}_*(I(\tilde{f}))\subset I(\tilde{g}\circ\tilde{f});  $$
\item
$$  I(id_{\tilde{X}}) \subset I(\tilde{f}).  $$
\end{enumerate}
The class of path-connected spaces $X$ satisfying the condition
$I(id_{\tilde{X}})=\pi=\pi_1(X,x_0)$ is closed under homotopy equivalence
and the topological product operation, and contains the simply connected
spaces, generalized lens spaces, $H$-spaces and homogeneous spaces of the form
$G/G_0$ where $G$ is a topological group and $G_0$ a subgroup which is a connected, compact Lie group (for the proofs see \cite{j}).

From theorem 27 it follows:

\begin{theorem}
Suppose that $N(f^n)=R(f^n)$ for all $n>0$, and that $f$ is
eventually commutative. Then the Nielsen zeta function is rational, and is given by
\begin{eqnarray}
\lefteqn{
N_f(z)=R_f(z)=}\nonumber \\
& = &
\left(\left(
\prod_{i=0}^k
\det\left( 1 - \Lambda^i f^\infty_{1*} \sigma z \right)^{(-1)^{i+1}}
\right)^{(-1)^r}\right)
 *
\left(\prod_{[h]} \frac{1}{1-z^{\#[h]}}\right),
\end{eqnarray}
where $\sigma$, $r$, and $[h]$ are as in theorem 27.
The function written here has a functional equation
as described in theorem 29.
\end{theorem}

\begin{theorem}
Suppose that $\tilde{f}_*(\pi)\subset I(\tilde{f})$ and $L(f^n)\not=0$ for
every $n>0$. Then
$N_f(z)=R_f(z)$ is rational and is given by (3.9).
It has a functional
equation as described in theorem 29.
If  $L(f^n)=0$ only for finite number of $n$,then
$$ 
 N_f(z)=R_f(z)\cdot \exp\left(P(z)\right)
 $$
 where $R_f(z) $ is rational and is given by (3.9 ) and $P(z)$ is a polynomial.  
\end{theorem}

{\sc Proof}
We have $\tilde{f}^n_*(\pi)\subset I(\tilde{f}^n)$ for every $n>0$
(by property 4 and the condition $\tilde{f}_*(\pi)\subset I(\tilde{f})$).
For any $\alpha\in\pi$,
$p(\fix \alpha\circ\tilde{f}^n)=p(\fix \tilde{f}^n_*(\alpha)\circ\tilde{f}^n)$
by lemmas 2 and 5 and the fact  that $\alpha$ and $\tilde{f}^{n}_*\alpha$ are in the same $\tilde{f}_*^n$-conjugacy class( see lemma 7).
Since $\tilde{f}^n_*(\pi)\subset I(\tilde{f}^n)$, there is
a homotopy $\{h_t\}:f^n\cong f^n$ which lifts to
$\{\tilde{h}_t\}:\tilde{f}^n\cong \tilde{f}_*^n(\alpha)\circ\tilde{f}^n$.
Hence $\ind(f^n,p(\fix \tilde{f}^n))=\ind(f^n,p(\fix \alpha\circ\tilde{f}^n))$.
Since $\alpha\in\pi$ is arbitrary, any two fixed point classes of
$f^n$ have the same index.
It immediately follows that
$L(f^n)=0$ implies $N(f^n)=0$ and $L(f^n)\not=0$ implies
$N(f^n)=R(f^n)$.
By property 1, $\tilde{f}^n(\pi)\subset I(\tilde{f}^n)\subset Z(\tilde{f}^n_*(\pi),\pi)$, so $\tilde{f}^n_*(\pi)$ is abelian.
Hence $\tilde{f}_*$ is eventually commutative.
If  $L(f^n)\not=0$ for  every  $n>0$ then the first part of the theorem now follows from theorems 27 and 29 .
If  $L(f^n)=0$ only for finite number of $n$,then the fraction
$N_f(z)/R_f(z)=\exp(P(z))$, where $P(z)$ is a polynomial whose degree equal to maximal $n$, such that  $L(f^n)\not=0$. This gives the second part of the theorem.

\begin{corollary}
Let the assumptions of theorem 39 hold.
Then the poles and zeros of the Nielsen zeta function
are complex numbers of the
form $\zeta^a b$ where $b$ is the reciprocal 
of an eigenvalue of one of the matrices
 $$  \Lambda^i (f_{1*}^\infty)
     : \Lambda^i (H_1(X;\bbbz)^\infty) 
       \longrightarrow
       \Lambda^i (H_1(X;\bbbz)^\infty) \;\;\;\;\;\; 0\leq i\leq {\rm rank}\;G$$
and $\zeta^a$ is a $\psi^{th}$ root of unity
where $\psi$ is the number of periodic torsion elements in $H_1(X;\bbbz)$.
The multiplicities of the roots or poles
$\zeta^a b$ and $\zeta^{a^\prime} b^\prime$ are the same
if $b=b^\prime$ and $hcf(a,\psi) = hcf(a^\prime,\psi)$.
\end{corollary}

\begin{remark}
The conclusion of theorem 39  remains valid under the weaker precondition
``there is an integer $m$ such that $\tilde{f}_*^m(\pi)\subset I(\tilde{f}^m)$''
instead of $\tilde{f}_*(\pi)\subset I(\tilde{f})$,
but the proof is more complicated.
\end{remark}

From theorem 8 and results of Jiang \cite{j} it follows:

 \begin{theorem}(Trace formula for the Nielsen numbers)
Suppose that there is an integer $m$ such that $\tilde{f}_*^m(\pi)\subset I(\tilde{f}^m)$ and $L(f)\not=0$.Then
\begin{equation}
N(f)= R(f)= (-1)^{r+p} \sum_{i=0}^k (-1)^i \tr (\Lambda^i {f_{1*}}^\infty \otimes A).
\end{equation}
where $k$ is $rgH_1(X,\bbbz)^\infty$,$A$ is linear map on the complex vector space of complex valued functions on the group $TorsH_1(X,\bbbz)$, $p$ the number of $\mu\in\spec f_{1*}^\infty$ such that
$\mu <-1$, and $r$ the number of real eigenvalues of $f_{1*}^\infty$ whose
absolute value is $>1$.
\end{theorem}  

\begin{theorem}
Suppose that there is an integer $m$ such that $\tilde{f}_*^m(\pi)\subset I(\tilde{f}^m)$.If $L(f^n)\not=0$ for every $n>o$ ,then
\begin{equation}
  N_f(z)=R_f(z)= \left(
 \prod_{i=0}^k
 \det(1-\Lambda^if_{1*}^\infty\otimes A \cdot \sigma\cdot z)^{(-1)^{i+1}}
 \right)^{(-1)^r}
\end{equation}
If  $L(f^n)=0$ only for finite number of $n$,then
\begin{eqnarray}  
 N_f(z)&=& R_f(z)\cdot \exp P(z)
 \nonumber\\
 &=& \left(
 \prod_{i=0}^k
 \det(1-\Lambda^if_{1*}^\infty\otimes A \cdot\sigma\cdot z)^{(-1)^{i+1}}
 \right)^{(-1)^r}\cdot \exp P(z)
\end{eqnarray}
Where $P(z)$ is a polynomial ,$A, k, p,$ and $r$ are as in theorem 40.
 
 \end{theorem}
 
{\sc Proof}
 
If $L(f^n)\not=0$ for every $n>o$ ,then formula (3.11) follows from theorem 12.
If  $L(f^n)=0$, then $N(f^n)=0$ . If $L(f^n)\not=0$, then
 $N(f^n)=R(f^n)$(see proof of theorem 39 ).So the fraction
$N_f(z)/R_f(z)=\exp(P(z))$, where $P(z)$ is a polynomial whose degree equal to maximal $n$, such that  $L(f^n)\not=0$.

\begin{corollary}
Let the assumptions of theorem 41 hold.
Then the poles and zeros of the Nielsen zeta function
are complex numbers which are the reciprocal 
of an eigenvalue of one of the matrices
 $$  \Lambda^i (f_{1*}^\infty\otimes A \cdot\sigma) $$
\end{corollary}

\begin{corollary}
Let $I(id_{\tilde{X}})=\pi$ . If  $L(f^n)\not=0$ for all $n>0$
then formula (3.11)   is valid. If $L(f^n)=0$ for finite number of $n$ , then formula (3.12) is valid.
\end{corollary}

\begin{corollary}
Suppose that $X$ is aspherical, $f$ is eventually commutative. If
$L(f^n)\not=0$ for all $n>0$
then formula (3.11) is valid.If $L(f^n)=0$ for finite number of $n$ , then formula (3.12) is  valid
\end{corollary}

\section{Polyhedra with finite fundamental group.}

For a compact polyhedron $X$ with finite fundamental group,
$\pi_1(X)$, the universal cover $\tilde{X}$ is compact,
so we may explore the relation between $L(\tilde{f^n})$
and $\ind(p(\fix \tilde{f}^n))$.

\begin{definition}
The number $\mu([\tilde{f}^n])=\#\fix\ \tilde{f}^n_*$, defined to
be the order of the finite group $\fix\ \tilde{f}^n_*$, is called
the {\it multiplicity} of the lifting class $[\tilde{f}^n]$,
or of the fixed point class $p(\fix \tilde{f}^n)$.
\end{definition}

\begin{lemma}[\cite{j}]
$$
L(\tilde{f}^n) = \mu([\tilde{f}^n])\cdot \ind(f^n,p(\fix \tilde{f}^n)).
$$
\end{lemma}

\begin{lemma}[\cite{j}]
If $R(f^n) = \#\coker(1-f^n_{1*})$ (in particular if $f$ is eventually
commutative), then
 $$  \mu([\tilde{f}^n]) = \#\coker(1-f^n_{1*}).$$
\end{lemma}

\begin{theorem}
Let $X$ be a connected, compact polyhedron with finite fundamental group $\pi$.
Suppose that the action of $\pi$ on the rational homology of the
universal cover $\tilde{X}$ is trivial,
i.e. for every covering translation $\alpha\in\pi$,
$\alpha_*=id:H_*(\tilde{X},\bbbq)\rightarrow H_*(\tilde{X},\bbbq)$.
If
$L(f^n)\not=0$ for all
$n>0$ then $N_f(z)$ is a rational function given by
\begin{equation}
N_f(z) = R_f(z) = \prod_{[<h>]}\frac{1}{1-z^{\#[<h>]}},
\end{equation}
where the product is taken over all periodic $\tilde{f}_*$-orbits
of ordinary conjugacy classes in the finite group $\pi_1(X)$.
This function has a functional equation as described in theorem 26.
If  $L(f^n)=0$ only for finite number of $n$,then
$$  
 N_f(z)= R_f(z)\cdot \exp\left(P(z)\right),
$$
where $P(z)$ is a polynomial and $R_f(z)$ is given by formula (3.13)
\end{theorem}

{\sc Proof}
Under our assumption on $X$, any two liftings $\tilde{f^n}$
 and $\alpha\circ\tilde{f^n}$ induce the same homology homomorphism
 $H_*(\tilde{X},\bbbq)\rightarrow H_*(\tilde{X},\bbbq)$, and
 have thus the same value of $L(\tilde{f^n})$. From
 Lemma 24 it follows that any two fixed point classes $ f^n $ are either
 both essential or both inessential.
 If $L(f^n)\not=0$ for every $n>0$ then for every $n$ there is at least one
 essential fixed point class of $f^n$. Therefore for every $n$ all fixed point classes of $f^n$
 are essential and $N_f(z)=R_f(z)$.
 The formula ( 3.13) for $R_f(z)$ follows 
 from theorem 16 .
If  $L(f^n)=0$, then $N(f^n)=0$ . If $L(f^n)\not=0$, then
 $N(f^n)=R(f^n)$. So the fraction
$N_f(z)/R_f(z)=\exp(P(z))$, where $P(z)$ is a polynomial whose degree equal to maximal $n$, such that  $L(f^n)\not=0$. This gives the second statement of the theorem.

Let $W$ be the complex vector space of complex valued class functions on the fundamental group $\pi$.The map $\tilde{f}_*$ induces a linear map $B:W\rightarrow W$ defined by
$$
B(f):=f\circ\tilde{f}_*.
$$ 
 
\begin{theorem}(Trace formula for Nielsen numbers)
Let $X$ be a connected, compact polyhedron with finite fundamental group $\pi$.
Suppose that the action of $\pi$ on the rational homology of the
universal cover $\tilde{X}$ is trivial,
i.e. for every covering translation $\alpha\in\pi$,
$\alpha_*=id:H_*(\tilde{X},\bbbq)\rightarrow H_*(\tilde{X},\bbbq)$.
Let $L(f)\not=0$.Then 
\begin{equation}
N(f)=R(f)=\tr B,
\end{equation}

\end{theorem}
 
{\sc Proof}
Under our assumption on $X$ all fixed point classes of $f$
 are essential and  $N(f)=R(f)$  ( see proof of the previous theorem for n=1).The formula for $N(f)$ follows  now from
theorem 15.

\begin{theorem}
Let $X$ be a connected, compact polyhedron with finite fundamental group $\pi$.
Suppose that the action of $\pi$ on the rational homology of the
universal cover $\tilde{X}$ is trivial,
i.e. for every covering translation $\alpha\in\pi$,
$\alpha_*=id:H_*(\tilde{X},\bbbq)\rightarrow H_*(\tilde{X},\bbbq)$.
If $L(f^n)\not=0$ for every $n>o$ ,then
\begin{equation}
  N_f(z)=R_f(z) = \frac{1}{\det(1-Bz)},
\end{equation} 
If  $L(f^n)=0$ only for finite number of $n$,then 
\begin{equation}
  N_f(z)=R_f(z)\cdot \exp\left(P(z)\right)=\frac{\exp\left(P(z)\right)}{\det(1-Bz)},
\end{equation}
Where $P(z)$ is a polynomial,   $B$ is defined in theorem 15
\end{theorem}

{\sc Proof}
 
If $L(f^n)\not=0$ for every $n>o$ ,then 
 $N(f^n)=R(f^n)$ (see proof of theorem 42) and formula (3.16) follows from theorem 17 .
If  $L(f^n)=0$, then $N(f^n)=0$ . If $L(f^n)\not=0$, then
 $N(f^n)=R(f^n)$. So the fraction $N_f(z)/R_f(z)=\exp(P(z))$, where $P(z)$ is a polynomial whose degree equal to maximal $n$, such that  $L(f^n)\not=0$.

\begin{lemma}
Let $X$ be a polyhedron with finite fundamental group $\pi$ and let
$p:\tilde{X}\rightarrow X$ be its universal covering. Then the action
of $\pi$ on the rational homology of $\tilde{X}$ is trivial iff
$H_*(\tilde{X};\bbbq) \cong H_*(X;\bbbq)$.
\end{lemma}

\begin{corollary}
Let $\tilde{X}$ be a compact $1$-connected polyhedron which is a 
 rational homology $n$-sphere, where $n$ is odd.
Let $\pi$ be a finite group acting freely on $\tilde{X}$ and let
$X=\tilde{X}/\pi$.
Then theorems 42 and 44 applie.
\end{corollary}

{\sc Proof}
The projection $p:\tilde{X}\rightarrow X= \tilde{X}/\pi$ is a
universal covering space of $X$. For every $\alpha\in\pi$, the degree
of $\alpha:\tilde{X}\rightarrow\tilde{X}$ must be 1, because
$L(\alpha)=0$ ($\alpha$ has no fixed points). Hence
$\alpha_* = id: H_*(\tilde{X};\bbbq)\rightarrow H_*(\tilde{X};\bbbq)$.

\begin{corollary}
If $X$ is a closed 3-manifold with finite $\pi$, then theorems 42 and 44
applie.
\end{corollary}

{\sc Proof}
$\tilde{X}$ is an orientable, simply connected manifold, hence a 
 homology 3-sphere. We apply corollary 14 .

\begin{corollary}
Let $X=L(m,q_1,\ldots,q_r)$ be a generalized lens space
and $f:X\to X$ a continuous map with
$f_{1*}(1)=d$ where $\mid d \mid\not= 1$.
The Nielsen and Reidemeister zeta functions are then rational
and are given by the formula:
$$
 N_f(z)=R_f(z)=\prod_{[h]}\frac{1}{1-z^{\#[h]}}
 =
 \prod_{t=1}^{\varphi_d(m)} (1-e^{2\pi i t/ \varphi_d(m)} z)^{-a(t)}.
$$
where $[h]$ runs over the periodic $f_{1*}$-orbits of
elements of $H_1(X;\bbbz)$.
The numbers $a(t)$ are natural numbers given by the formula
 $$
 a(t)
 =
 \sum_{
 \begin{array}{c}
  s\mid m {\rm \ such \ that}\\
  \varphi_d(m) \mid t\varphi_d(s)
 \end{array}}
 \frac{\varphi(s)}{\varphi_d(s)},
 $$
where $\varphi$ is the Euler totient function and
$\varphi_d(s)$ is the order of the multiplicative subgroup of $(\bbbz/s\bbbz)^\times$
generated by $d$.
\end{corollary}

{\sc Proof}
By corollary 14 we see that theorem 42
applies for lens spaces.
Since $\pi_1(X)=\bbbz/m\bbbz$,
the map $f$ is eventually commutative.
A lens space has a structure as a CW complex with one
cell $e_i$ in each dimension $0\leq i\leq 2n+1$.
The boundary map is given by $\partial e_{2k}=m.e_{2k-1}$
for even cells, and $\partial e_{2k+1}=0$ for odd cells.
From this we may calculate the Lefschetz numbers:
$$
 L(f^n) = 1-d^{(l+1)n} \not= 0.
$$
This is true for any $n$ as long as $\mid d\mid\not=1$.
Then by theorem 42 we have
$$
 N(f^n) = R(f^n) = \#\coker(1-f^n_{1*})
$$
where $f_{1*}$ is multiplication by $d$.
One then sees that $(1-f_{1*}^n)(\bbbz/m\bbbz)=(1-d^n)(\bbbz/m\bbbz)$
and therefore $\coker(1-f_{1*}^n)=(\bbbz/m\bbbz)/(1-d^n)((\bbbz/m\bbbz)$.
The cokernel is thus a cyclic group of order $hcf(1-d^n,m)$.

We briefly investigate the sequence $n\mapsto hcf(1-d^n,m)$.
It was originally this calculation which
lead us to the results of section 2.4.

Let $\varphi:\bbbn\to\bbbn$ be the Euler totient function,
ie. $\varphi(r) = \#(\bbbz/r\bbbz)^\times$.
In addition we define $\varphi_d(r)$ to be the order
of the multiplicative subgroup of $(\bbbz/r\bbbz)^\times$
generated by $d$.
One then has by Lagrange's theorem $\varphi_d(r)\mid\varphi(r)$.
The number $\varphi(r)$ is the smallest $n>0$ such that
$d^n\equiv 1\mod r$.

The sequence $n\mapsto hcf(1-d^n,m)$ is
periodic in $n$ with least period $\varphi_d(m)$.
It can therefore be expressed as a finite Fourier series:
 $$
 hcf(1-d^n,m)
 =
 \sum_{t=1}^{\varphi_d(m)}
 a(t) \exp\left(\frac{2\pi i n t}{\varphi_d(m)}\right).
 $$
The coefficients $a(t)$ are given by Fourier's inversion formula:
 $$
 a(t)
 =
 \frac{1}{\varphi_d(m)}
 \sum_{n=1}^{\varphi_d(m)}
 hcf(1-d^n,m) \exp\left(\frac{-2\pi i nt}{\varphi_d(m)}\right).
 $$
After a simple calculation, one obtains the formula:
 $$
 \exp\left(\sum_{n=1}^\infty \frac{hcf(1-d^n,m)}{n} z^n \right)
 =
 \prod_{t=1}^{\varphi_d(m)} (1-e^{2\pi i t/ \varphi_d(m)} z)^{-a(t)}.
 $$
We now calculate the coefficients $a(t)$ more explicitly.
We have $hcf(d^n-1,m)=r$ iff $d^n\equiv 1\mod r$ and for all
primes $p\mid\frac{m}{r}$, $d^n\not\equiv 1\mod pr$.
This is the case if and only if $n\equiv 0\mod\varphi_d(r)$ and for all
primes $p\mid\frac{m}{r}$, $n\not\equiv 0\mod\varphi_d(pr)$.
Using this we partition the sum in the expression for $a(t)$:
 $$
 a(t)
 =
 \frac{1}{\varphi_d(m)}
 \sum_{r\mid m}
 r
 \sum_{
 \begin{array}{c}
 n=1 \ {\rm such \ that} \\
 n\equiv 0\mod\varphi_d(r)\ {\rm and} \\
 \forall p\mid\frac{m}{r} ,\ n\not\equiv 0\mod\varphi_d(pr)
 \end{array}}^{\varphi_d(m)}
 \exp\left(\frac{-2\pi i nt}{\varphi_d(m)}\right).
 $$
We define
 $$
 g(r)
 :=
 \sum_{
 \begin{array}{c}
 n=1 \ {\rm such \ that} \\
 n\equiv 0\mod\varphi_d(r)
 \end{array}}^{\varphi_d(m)}
 \exp\left(\frac{-2\pi i nt}{\varphi_d(m)}\right).
 $$
Using this we rewrite the inner sum in the expression for $a(t)$.
\begin{eqnarray*}
\lefteqn{
 \sum_{
 \begin{array}{c}
 n=1 \ {\rm such \ that} \\
 n\equiv 0\mod\varphi_d(r)\ {\rm and} \\
 \forall p\mid\frac{m}{r} ,\ n\not\equiv 0\mod\varphi_d(pr)
 \end{array}}^{\varphi_d(m)}
 \exp\left(\frac{-2\pi i nt}{\varphi_d(m)}\right) = }  \\
 & = &
 g(r)
 - \sum_{p_1\mid\frac{m}{r}} g(p_1 r)
 + \sum_{p_1,p_2\mid\frac{m}{r}} g(p_1 p_2 r)
 - \ldots
\end{eqnarray*}
Here the second sum is over all 2-element sets of primes $\{p_1,p_2\}$ such that $p_1$ and $p_2$ both divide $\frac{m}{r}$.
Substituting this into our expression for $a(t)$ we obtain,
 $$
 a(t)
 =
 \frac{1}{\varphi_d(m)}
 \sum_{r\mid m}
 r
 \left(
 g(r)
 - \sum_{p_1\mid\frac{m}{r}} g(p_1 r)
 + \sum_{p_1,p_2\mid\frac{m}{r}} g(p_1 p_2 r)
 - \ldots
 \right).
 $$
We now change the variable of summation:
 $$
 a(t)
 =
 \frac{1}{\varphi_d(m)}
 \sum_{s\mid m}
 g(s)
 \left(
 s
 - \sum_{p_1\mid s} \frac{s}{p_1}
 + \sum_{p_1,p_2\mid s} \frac{s}{p_1 p_2}
 - \ldots
 \right).
 $$
The large bracket here can be factorized into a kind of
 Euler product:
 $$
 a(t)
 =
 \frac{1}{\varphi_d(m)}
 \sum_{s\mid m}
 g(s)
 \prod_{p\mid s} \left( 1 - \frac{1}{p} \right).
 $$
This is then seen to be exactly the Euler totient function:
 $$
 a(t)
 =
 \frac{1}{\varphi_d(m)}
 \sum_{s\mid m}
 g(s)
 \varphi(s).
 $$
On the other hand, a simple calculation shows
 $$
 g(s)
 =
 \left\{
 \begin{array}{ll}
 \frac{\varphi_d(m)}{\varphi_d(s)} &
 {\rm if \ } \varphi_d(m) \mid t\varphi_d(s) \\
 0 &
 {\rm otherwise.}
 \end{array}
 \right.
 $$
From this we have
 $$
 a(t)
 =
 \sum_{
 \begin{array}{c}
  s\mid m {\rm \ such \ that}\\
  \varphi_d(m) \mid t\varphi_d(s)
 \end{array}}
 \frac{\varphi(s)}{\varphi_d(s)}.
 $$
It is now clear that the coefficients $a(t)$ are
natural numbers, and we have the formula stated.

\begin{remark}
Let $X=L^{2l+1}(m,q_1,q_2,\ldots,q_r)$ and let $f:X\to X$ be
an continuous map.
If $f$ induces the trivial map on the cyclic group $\pi_1(X)$ then we have,
$$
 N_f(z) = 1, \;\;\;\; R_f(z) = \frac{1}{(1-z)^m}.
$$
If $f$ induces the map $g\mapsto -g$ on $\pi_1(X)$ then we have
$$
 R_f(z)
 =
 \left\{
 \begin{array}{ll}
   \frac{1}{(1-z)^{\frac{m}{2}+1} (1+z)^{\frac{m}{2}-1}} & {\rm if} \ m
    \ {\rm is \ even},  \vspace{0.2cm} \\
   \frac{1}{(1-z)^{\frac{m+1}{2}} (1+z)^{\frac{m-1}{2}}} & {\rm if} \ m
    \ {\rm is \ odd},
 \end{array}
 \right.
$$
$$
 N_f(z)
 =
 \left\{
 \begin{array}{ll}
    1 &  {\rm if} \ l \ {\rm is \ odd},  \vspace{0.2cm} \\
   \frac{1}{1-z^2} & {\rm if} \ l \ {\rm is\ even\ and}\ m\ {\rm is\ even},\vspace{0.2cm} \\
   \sqrt{\frac{1+z}{1-z}} & {\rm if} \ l \ {\rm is \ even \ and}\ m\ {\rm is\ odd}.
 \end{array}
 \right.
$$
We have now described explicitly all
the Reidemeister and  Nielsen zeta functions of all continuous
maps of lens spaces. Apart from one exception they are all rational.
\end{remark}

\section{Nielsen zeta function in other special cases}
 
Theorems 22 and 31 implie
\begin{theorem}
Let $f$ be any continuous map of a nilmanifold $M$ to itself.If $R(f^n)$ is finite for all $n$
then
\begin{equation}
N_f(z)= R_f(z)
 =
  \left(
 \prod_{i=0}^m
 \det(1-\Lambda^i\tilde F .\sigma.z)^{(-1)^{i+1}}
 \right)^{(-1)^r}
\end{equation}
where $\sigma=(-1)^p$,$p$ , $r$, $m$ and  $\tilde F$ is defined in theorem 31.
\end{theorem}

\begin{theorem}
 Suppose $M$ is a orientable compact connected 3-manifold such that int$M$ admits a complete hyperbolic structure with finite volume and $ f: M \rightarrow M $ is orientation preserving homeomorphism.Then Nielsen zeta function is rational and
 $$
N_f(z)=L_f(z)
$$
 
\end{theorem}
 
 {\sc Proof} 
 B.Jiang and S. Wang \cite{jw} have proved that $N(f)=L(f)$. This is also true for all iterations.

 \subsection{ Pseudo-Anosov homeomorphism of a compact surface}
Let $X$ be a compact surface of negative Euler characteristic and
 $f:X\rightarrow X$ is a pseudo-Anosov homeomorphism,i.e. there is a number $\lambda >1$ and a pair of transverse measured foliations $(F^s,\mu^s)$ and $(F^u,\mu^u)$ such that $f(F^s,\mu^s)=(F^s,\frac{1}{\lambda}\mu^s)$ and $f(F^u,\mu^u)=(F^u,\lambda\mu^u)$.
Fathi and Shub \cite{fash} has proved the existence of Markov partitions for a
pseudo-Anosov homeomorphism.The existence of Markov partitions implies that there is a symbolic dynamics for $(X,f)$.This means that there is a finite set
$N$, a matrix $A=(a_{ij})_{(i,j)\in N\times N}$ with entries $0$ or $1$ and a surjective map $p:\Omega\rightarrow X$,where
$$
\Omega=\{(x_n)_{n\in \bbbz}: a_{x_nx_{n+1}}=1  ,  n\in \bbbz \}
$$
such that $p\circ \sigma =f\circ p$ where $\sigma$ is the shift (to the left) of the sequence $(x_n)$ of symbols.We have first \cite{bl}:
$$
\# \fix\sigma ^n=\tr A^n.
$$
In general $p$ is not bijective.The non-injectivity of $p$ is due to the fact that the rectangles of the Markov partition can meet on their boundaries.To cancel the overcounting of periodic points on these boundaries,we use Manning's combinatorial arguments \cite{m} proposed in the case of Axiom A diffeomorphism (see also \cite {pp}) . Namely, we construct finitely many subshifts of finite type ${\sigma _i}, i=0,1,..,m$, such that $\sigma_0=\sigma$, the other shifts semi-conjugate with restrictions of $f$ \cite {pp} ,and signs $ \epsilon _i\in \{-1,1\}$ such that for each $n$
$$
\# \fix f^n=\sum_{i=0}^m \epsilon_i\cdot\#\fix \sigma_i^n =\sum_{i=0}^m \epsilon_i\cdot\tr A_i^n,
$$
where $A_i$ is transition matrix, corresponding to subshift of finite type $\sigma_i$.
For pseudo-Anosov homeomorphism of compact surface  $N(f^n)=\# \fix (f^n)$ for each $n>o$ \cite{th}.So we have following trace formula for Nielsen numbers
 
\begin{lemma}
Let $X$ be a compact surface of negative euler characteristic and
$f:X\rightarrow X$ is a pseudo-Anosov homeomorphism.Then
$$
N(f^n)=\sum_{i=0}^m \epsilon_i\cdot\tr A_i^n.
$$
\end{lemma}
 
This lemma implies 
 
\begin{theorem}
Let $X$ be a compact surface of negative Euler characteristic and
$f:X\rightarrow X$ is a pseudo-Anosov homeomorphism.Then
\begin{equation}
N_f(z)=\prod_{i=0}^m\det(1-A_iz)^{-\epsilon_i}
\end{equation}
where $A_i$ and $\epsilon_i$
 the same as in lemma 27.
\end{theorem}

\section{The Nielsen zeta function and Serre bundles.}

Let $p:E\rightarrow B$ be a orientable Serre bundle in which $E$, $B$ and
every fibre are connected, compact polyhedra and $F_b=p^{-1}(b)$
is a fibre over $b\in B$ ( see section  ).
 Let $f:E\rightarrow E$ be a fibre map.
Then for any two fixed points $b,b^\prime$ of $\bar{f}:B\rightarrow B$
the maps $f_b=f\mid_{F_b}$ and $f_{b^\prime} = f\mid_{F_{b^\prime}}$
have the same homotopy type;
hence they have the same Nielsen numbers
$N(f_b) = N(f_{b^\prime})$ .
 The following theorem describes the relation between the Nielsen
zeta functions $N_f(z)$, $N_{\bar{f}}(z)$ and $N_{f_b}(z)$ for a fibre
map $f:E \rightarrow E$ of an orientable Serre bundle $p:E\rightarrow B$.

\begin{theorem}
Suppose  that for every $n>0$ \\

1) $ KN(f_b^n)=N(f_b)$ , where $b \in \fix ({\bar f}^n),
K=K_b= Ker(i_*: \pi_1(F_b) \to \pi_1(E))$;\\

2) in every essential fixed point class of $f^n$, there is  a point $e$ such that
$$
p_*( \fix( \pi_1(E,e) \stackrel{(f^n)*}{\rightarrow} \pi_1(E,e)=
= \fix(\pi_1(B, b_0) \stackrel{(\bar f^n)*}{\rightarrow}  \pi_1(B, b_0),
$$
where $b_0=p(e)$.
We then have 
$$N_f(z)=N_{\bar{f}}(z)*N_{f_b}(z).$$
If $N_{\bar{f}}(z)$ and $N_{f_b}(z)$ are rational functions
then so is $N_f(z)$.
If $N_{\bar{f}}(z)$ and $N_{f_b}(z)$ are rational functions
with functional equations as described in theorem 38 and 42  then so is $N_f(z)$.
\end{theorem}
{\sc Proof}
From the conditions of the theorem it follows that 
$$ N(f^n)=N(\bar f^n)\cdot N(f_b^n)$$
for every $n$( see \cite{j}).
From this we have 
$$
N_f(z)=N_{\bar{f}}(z)*N_{f_b}(z).
$$
The rationality of $N_f(z)$  and functional equation for it follow from lemmas 13  and 14  .

\begin{corollary}
Suppose that for every $n>0$ homomorphism $1-(\bar f^n)_*: \pi_2(B,b) \to \pi_2(B,b)$
is an epimorphism. Then the condition 1) above is satisfied.

\end{corollary}

\begin{corollary}
Suppose that $f:E \rightarrow E$ admits a Fadell splitting
in the sense that for some $e$ in $\fix f$ and $b=p(e)$ the
following conditions are satisfied:
\begin{enumerate}
\item
the sequence
$$  0 \longrightarrow \pi_1(F_b,e)
      \stackrel{i_*}{\longrightarrow} \pi_1(E,e)
      \stackrel{p_*}{\longrightarrow} \pi_1(B,e)
      \longrightarrow 0 $$
is exact,

\item
$p_*$ admits a right inverse (section) $\sigma$ such that $\im\sigma$
is a normal subgroup of $\pi_1(E,e)$ and $f_*(\im\sigma)\subset\im\sigma$.

\end{enumerate}
 Then theorem 48 applies.  
\end{corollary}

\section{Examples}

Let $f:X\rightarrow X$ be a continuous map of a simply connected,
connected, compact polyhedron.
Then $R_f(z)=\frac{1}{1-z}$.

Let  $X=S^1$ and $f:  S^1 \rightarrow S^1$ be continuous map of degree $d$.Then 
$N(f^n)=\mid 1-d^n \mid $, and the Nielsen zeta function is rational  and is equal to
$$
 N_f(z)
 =
 \left\{
 \begin{array}{ll}
   \frac{1-z}{1 - dz}  &  {\rm if} \ d>0\   \vspace{0.2cm} \\
   \frac{1}{1-z} & {\rm if} \ d=0 \ \vspace{0.2cm} \\
   \frac{1+z}{1+dz} & {\rm if} \ d<0 \ .
 \end{array}
 \right.
$$
  If $X=S^{2n}$, and $f:  S^{2n} \rightarrow S^{2n}$ is a continuous map of degree $d$
  then
  $$ 
 N_f(z)
 =
 \left\{
 \begin{array}{ll}
   \frac{1}{\sqrt{1 - z^2}}  &  {\rm if} \ d=-1\   \vspace{0.2cm} \\
   \frac{1}{1-z} & {\rm if} \ d\not=-1. \ \vspace{0.2cm} \\
 \end{array}
 \right.
$$
Now if $X=S^{2n+1}$, and $f:  S^{2n+1} \rightarrow S^{2n+1}$ is a continuous map of degree $d$
  then
 $$ 
 N_f(z)
 =
 \left\{
 \begin{array}{ll}
   1 &  {\rm if} \ d=1\   \vspace{0.2cm} \\
    \sqrt{\frac{1+z}{1-z}} & {\rm if} \ d=-1 \  \vspace{0.2cm} \\
   \frac{1}{1-z} & {\rm if} \ \mid d \mid \not=1. \ \vspace{0.2cm} \\
 \end{array}
 \right.
$$
Thus, even on a simply connected space the Nielsen zeta function can be the radical of a 
rational function.
 In the next example $X=T^n$ is torus and $f: T^n  \rightarrow T^n $ is a hyperbolic 
 endomorphism of the torus . Hyperbolic means that the covering linear map
 $\tilde f : R^n \rightarrow R^n $ has no eigenvalues of modulus one. Then
 $R(f^n)=N(f^n)=\mid \det(E-\tilde {f^n}) \mid =\mid L(f^n) \mid $ \cite {bbpt}. Thus 
 $ R[f^n)=N(f^n)=(-1)^{r+pn}\cdot \det(E-\tilde{f^n})$, where $r$ is equal to the number of 
 $\lambda_i \in Spec(\tilde f) $ such that $ \mid \lambda_i \mid > 1$, and $p$
 is equal to the number of $\mu_i \in Spec(\tilde f)$ such that $\mu_i <-1$.Consequently,
$R[f^n)=N(f^n)=(-1)^{r+pn}\cdot L(f^n)$ and the Reidemeister and Nielsen zeta function are rational and equal to 
$R_f(z)=N_f(z)=(L_f(\sigma\cdot z))^{(-1)^r}$ , where $\sigma=(-1)^p.$
It follows from the results of Franks, Newhouse and Manning \cite{sds} that the following
diffeomorphisms $g$ are topologically conjugate to hyperbolic automorphisms of the torus
$\Gamma$: a Anosov diffeomorphism  of the torus, a Anosov diffeomorphism of codimension
one \cite{sds} of manifold , which is metrically decomposable \cite{sds} , a Anosov diffeomorphism of a manifold , whose fundamental group is commutative.Consequently,
by the topological conjugacy of $g$ and $\Gamma$, the Nielsen zeta function $N_g(z)$
is rational and equal to $N_g(z)= N_{\Gamma}(z)=(L_f(\sigma\cdot z))^{(-1)^r}$.
In this example the Reidemeister and Nielsen zeta functions coincide with the Artin-Mazur zeta function.
In fact, the covering map $\tilde{\Gamma} $ has a unique fixed point, which is the origin;
hence, by the covering homotopy theorem \cite{rf} , the fixed points of $\Gamma$ are
pairwise nonequivalent.The index of each equivalence class, consisting of one fixed
point, coincides with its Lefschetz index, and by the hyperbolicity of $\Gamma$, the later 
is not equal to zero.Thus $R(\Gamma)= N(\Gamma)= F(\Gamma)$. Analogously,$R(\Gamma^n)= N(\Gamma^n)= F(\Gamma^n)$ for each $n>o$ and $ R_{\Gamma}(z)= N_{\Gamma}(z)= F_{\Gamma}(z)$. Since $\Gamma$
satisfies axiom A of  Smale, by Manning theorem \cite{m}  we get another proof of the rationality of $R_{\Gamma}(z)= N_{\Gamma}(z)= F_{\Gamma}(z)$.\\

let $X=RP^{2k+1}, k\not=0$ be projective space of odd dimension.
 Then for each $n>0, N(f^n)=0$, if $d=1$, and $N(f^n)=(2,1-d^n)$, if $\mid d \mid \not=1$.
 Consequently, $N(f^n)=2$, for all $n>0$, if $ \mid d \mid \not=1$ is odd , and $N(f^n)=1$
 for all $n>0$, if $d$ is even, and the Nielsen zeta function is rational and equal to:

  $$ 
 N_f(z)
 =
 \left\{
 \begin{array}{ll}
   1 &  {\rm if} \ d=1\   \vspace{0.2cm} \\
    {\frac{1}{(1-z)^2}} & {\rm if} \ \mid d \mid \not=1 \ {\rm is \ odd } \vspace{0.2cm} \\
   \frac{1}{1-z} & {\rm if }\ d \ {\rm is\ even}. \ \vspace{0.2cm} \\
 \end{array}
 \right.
$$

Now if $d=-1$, then for even $n, N(f^n)=0$, for odd $n, N(f^n)=2$, and $N_f(z)=(1+z)/(1-z)$.
For $RP^{2k}$, the projective spaces of even dimension, one gets exactly the same result for 
$N_f(z)$.\\

 Now let $f: M\rightarrow M$ be an expanding map\cite{sh} of the orientable smooth compact
 manifold $M$. Then $M$ is aspherical and is $K(\pi_1(M),1)$ and the fundamental group $\pi_1(M)$ is torsion free \cite{sh} . If $pr: \tilde M \rightarrow M $ is the universal covering,
 $\tilde f: \tilde M \rightarrow \tilde M$ is an arbitrary map , covering $f$, then according 
 to Shub \cite{sh} $ \tilde f $ has exactly one fixed point. From this and the covering homotopy theorem \cite{rf} it follows that the fixed points of $f$ are pairwise nonequivalent.
The index of each equivalence class, consisting of one fixed
point, coincides with its Lefschetz index,.If $f$ preserves the orientation of $M$, then the 
Lefschetz index $L(p,f)$ of the fixed point $p$ is equal to  $L(p,f)= (-1)^r$, where $r=\dim M$.
Then by Lefschetz trace formula $R(f)=N(f)=F(f)=(-1)^r\cdot L(f)$.Since the iterates $f^n$ are also
 orientation preserving expanding maps, we get analogously that
  $R(f^n)=N(f^n)=F(f^n)=(-1)^r\cdot L(f^n)$ for every $n$, and the Nielsen zeta function is rational and equal to $R_f(z)=N_f(z)=F_f(z)=L_f(z)^{(-1)^r}$. Now if $f$ reverses the orientation of the manifold 
   $M$, then $L(p,f^n)=(-1)^{r+n}$. Hence , $R(f^n)=N(f^n)=F(f^n)=(-1)^{r+n}\cdot L(f^n)$ and
  $R_f(z)=N_f(z)=F_f(z)=L_f(-z)^{(-1)^r}$.\\

\begin{example}[\cite{bb}]
 
Let $f:S^2\vee S^4\rightarrow S^2\vee S^4$ to be a continuous map of the bouquet of spheres such that the restriction $f/_{S^4}=id_{S^4}$ and the degree of the restriction $f/_{S^2}:S^2\rightarrow S^2$ equal to $-2$.Then $L(f)=0$, hence
$N(f)=0$ since $ S^2\vee S^4$ is simply connected.For $k>1$ we have $L(f^k)=2+(-2)^k\not=0$,therefore $N(f^k)=1$.From this we have by direct calculation that 
\begin{equation}
N_f(z)=\exp(-z)\cdot \frac{1}{1-z} . 
\end{equation}
\end{example}
 
\begin{remark}
 
We would like to mention that in all known cases the Nielsen zeta function is a
nice function. By this we mean that it is a product of an exponential of a polynomial with a function some power of which is rational. May be this is a
general pattern; it could however be argued that this just reflects our inability to calculate the Nielsen numbers in general case.
\end{remark}

\chapter{ Reidemeister and Nielsen zeta functions modulo normal subgroup, minimal
dynamical zeta functions}

\section{Reidemeister and Nielsen zeta functions  modulo
a normal subgroup}

In the theory of (ordinary) fixed point classes,
we work on the universal covering space.
The group of covering transformations plays a key role.
It is not surprising that this theory can be generalized
to work on all regular covering spaces.
Let $K$ be a normal subgroup of the fundamental group $\pi_1(X)$.
Consider the regular covering $p_K:\tilde{X}/K \to X$
corresponding to $K$.
A map $\tilde{f}_K:\tilde{X}/K \to\tilde{X}/K$
is called a lifting of $f:X\to X$ if
$p_K\circ\tilde{f}_K = f\circ p_K$.
We know from the theory of covering spaces that such liftings exist
if and only if $f_*(K)\subset K$.
If $K$ is a fully invariant subgroup of $\pi_1(X)$ ( in the
sense that every endomorphism sends $K$ into $K$)
such as, for example the commutator subgroup of $\pi_1(X)$,
then there is a lifting $\tilde{f}_K$ of any continuous map $f$.

 We can  develop a theory  which is  similar to the
theory in Chapters I - II  by simply replacing $\tilde{X}$ and
$\pi_1(X)$ by $\tilde{X}/K$ and $\pi_1(X)/K$ in every definition,
every theorem and every proof, since everything was done in terms
of liftings and covering translations. 
 What follows is a  list of definitions and some basic facts.

 Two liftings $\tilde{f}_K$ and $\tilde{f}_K^\prime$ are called
{\it conjugate} if there is a $\gamma_K\in\Gamma_K\cong\pi_1(X)/K$
such that $\tilde{f}_K^\prime = \gamma_K\circ\tilde{f}_K\circ\gamma_K^{-1}$.
The subset $p_K(\fix(\tilde{f}_K))\subset \fix(f)$ is called
the mod K  fixed point class of $f$ determined by the lifting class $[\tilde{f}_K]$ on $\tilde{X}/K$ .
The fixed point set $\fix(f)$ splits into a disjoint union of mod K fixed point classes. Two fixed points $x_0$ and $x_1$ belong to the same mod K class iff there is a path $c$ from $x_0$ to $x_1$ such that $c\ast (f\circ c)^{-1} \in K$. Each mod K fixed point class is a disjoint union
of ordinary fixed point classes. So the index of a mod K fixed point class can be defined in obvious way.
A mod K fixed point class is called $essential$ if its index is nonzero.
 The number of lifting classes of $f$ on $\tilde{X}/K$  (and hence the number
of mod K fixed point classes, empty or not) is called the  mod K Reidemeister Number of $f$,
denoted $KR(f)$. This is a positive integer or infinity.
The number of essential mod K fixed point classes is called the mod K Nielsen number
of $f$, denoted by $KN(f)$.
The mod K Nielsen number is always finite. $KR(f)$ and $KN(f)$ are homotopy
type invariants.
 The mod K Nielsen number was introduced by G.Hirsch in 1940 , primarily for purpose of estimating Nielsen number from below.
 The mod K Reidemeister zeta functions of $f$ 
and the mod K  Nielsen zeta function of $f$ was  defined in \cite{f2}, \cite{f4}
as power series:
\begin{eqnarray*}
KR_f(z) & := & \exp\left(\sum_{n=1}^\infty \frac{KR(f^n)}{n} z^n \right), \\
KN_f(z) & := & \exp\left(\sum_{n=1}^\infty \frac{KN(f^n)}{n} z^n \right).
\end{eqnarray*}
$KR_f(z)$ and $KN_f(z)$ are homotopy invariants. If $K$ is the trivial subgroup of $\pi_1(X)$
then $KN_f(z) $ and $KR_f(z)$ coincide with the Nielsen and Reidemeister zeta functions respectively.

 \subsection{ Radius of Convergence of the mod K Nielsen zeta function}

We show that the mod K Nielsen zeta function has positive radius of convergence.We denote by $R$
the radius of convergence of the  mod K Nielsen zeta function  $N_f(z)$,\begin{theorem}
Suppose that $f: X \rightarrow X $ be a continuous map of a compact polyhedron and
$f_*(K)\subset K$. Then
\begin{equation}
 R \geq \exp(-h)>0
\end{equation}
and
\begin{equation}
R\geq \frac{1}{\max_d \| z\tilde F_d \|} > 0,
\end{equation}
and
\begin{equation}
R\geq \frac{1}{\max_d s(\tilde F_d^{norm})} > 0, 
\end{equation}
where $\tilde F_d$ and $h$ is the same as in section 3.1.1 and 3.1.2.

\end{theorem}
{\sc Proof}
The theorem follows from  inequality $N(f^n) \geq KN(f^n)$ , Caushy-Adamar formula and   theorems 33    and  34     .

\begin{remark}
Let $f$ be a $C^1$-mapping of a compact, smooth, Riemannian manifold $M$.
Then $ h(f) \leq  \log\,sup \parallel Df(x)*\parallel $ \cite{ruelle2}, where $ Df(x)*$ is a mapping 
between exterior algebras of the tangent spaces $T(x)$ and $ T(f(x))$, induced by $Df(x)$,
$\parallel  \cdot \parallel $ is the norm on operators, induced from the Riemann metric.
Now from the inequality $$h(f) \geq \limsup_{n} \frac{1}{n}\cdot\log\, KN(f^n) $$ and  the Cauchy-Adamar formula we have 
\begin{equation}
R\geq \frac{1}{\sup_{x\in M}  \| D(f)*(x) \|} , 
\end{equation}
\end{remark}

\subsection{mod K  Nielsen zeta function of a  periodic map}

 We denote
 $KN(f^n)$ by $KN_n$.Let $ \mu(d), d \in N$, be the M\"obius function

\begin{theorem}
Let $f$ be a  periodic map of least period $m$ of the connected
compact polyhedron $X$ and $f_*(K)\subset K$ . Then the mod K Nielsen
 zeta function is equal to 
$$
KN_f(z)=\prod_{d\mid m}\sqrt[d]{(1-z^d)^{-P(d)}},
$$
 where the product is taken over all divisors $d$ of the period $m$, and $P(d)$ is the integer
$$  P(d) = \sum_{d_1\mid d} \mu(d_1)KN_{d\mid d_1} .  $$
\end{theorem}
{\sc Proof}
Since $f^m = id $, for each $j, KN_j=KN_{m+j}$. Since $(k,m)=1$, there exist positive integers $t$
and $q$ such that $kt=mq+1$. So $ (f^k)^t=f^{kt}= f^{mq+1}=f^{mq}f=(f^m)^{q}f =  f$.
Consequently, $  KN((f^k)^t)=KN(f) $. 
Let two fixed point $x_0$ and $x_1$ belong to the same mod K fixed point class. Then there exists a path $ \alpha $ from  $x_0$ to $x_1$ such that $ \alpha \ast (f\circ\alpha)^{-1} \in K$.Since  $f_*(K)\subset K$, we have $f(\alpha \ast f\circ\alpha)^{-1})=(f\circ\alpha)\ast (f^2\circ\alpha)^{-1} \in K$ and a product $ \alpha \ast (f\circ\alpha)^{-1}\ast (f\circ\alpha)\ast (f^2\circ\alpha)^{-1} =\alpha \ast (f^2\circ\alpha)^{-1}\in K $. It follows that $ \alpha \ast (f^k\circ\alpha)^{-1} \in K$ is derived by the iteration of this process. So $x_0$ and $x_1$ belong to the same mod K fixed point class of $f^k$. If two point belong to the different mod K fixed point classes $f$, then they belong to the different mod K fixed point classes of $f^k$.So, each essential class( class with nonzero index) for $f$  is an essential class for $f^k$;
in addition , different essential classes for $f$ are different essential classes for $f^k$.
So $ KN(f^k)\geq KN(f) $. Analogously, $ KN(f)=KN((f^k)^t) \geq  KN(f^k) $.Consequently ,
$ KN(f)=KN(f^k) $. One can prove completely analogously that $ KN_d= KN_{di} $, if (i, m/d) =1,
where $d$ is a divisor of $m$. Using these series of equal mod K Nielsen numbers, one can regroup
the terms of the series in the exponential of the mod K Nielsen zeta function so as to get logarithmic
functions by adding and subtracting missing terms with necessary coefficient:
\begin{eqnarray*}
KN_f(z) & = & \exp\left(\sum_{i=1}^\infty \frac{KN(f^i)}{i} z^i \right)\\
           & = & \exp\left(\sum_{d\mid m} \sum _{i=1}^\infty \frac{P(d)}{d}\cdot\frac{{z^d}^i}{i}\right)\\
	   & = & \exp\left(\sum_{d\mid m}\frac{P(d)}{d}\cdot \log (1-z^d)\right)\\
	   & = & \prod_{d\mid m}\sqrt[d]{(1-z^d)^{-P(d)}}
\end{eqnarray*}
where the integers $P(d)$ are calculated recursively by the formula
$$
P(d)= KN_d  - \sum_{d_1\mid d; d_1\not=d} P(d_1).
$$
Moreover, if the last formula is rewritten in the form
$$
KN_d=\sum_{d_1\mid d}\mu(d_1)\cdot P(d_1)
$$ 
and one uses  the M\"obius Inversion law for real function in number theory, then
$$
P(d)=\sum_{d_1\mid d}\mu(d_1)\cdot KN_{d/d_1},
$$
where $\mu(d_1)$ is the M\"obius function. The theorem is proved.

\section{Minimal dynamical zeta function}
 
\subsection{Radius of Convergence of the minimal zeta function}

In the Nielsen theory for periodic points, it is well known that $N(f^n)$ is sometime poor as a lower bound for the number of fixed points of $f^n$. A good homotopy invariant lower bound $ NF_n(f)$,called the Nielsen type number for $f^n$,is defined in \cite{j}.Consider any finite set of periodic orbit classes
$\{O^{k_j}\}$ of varied period $k_j$ such that every essential periodic $m$-orbit class, $m|n$, contains at least one class in the set.Then $NF_n(f)$ is the minimal sum $ \sum_j k_j$ for all such finite sets.
Halpern (see \cite{j}) has proved that for all $n$ $ NF_n(f) =\min\{\#\fix(g^n)| g$ has the same homotopy type as $f$ \}.Recently, Jiang \cite{j1} found that as far as asymptotic growth rate is concerned,these Nielsen type numbers are no better than the Nielsen numbers.
\begin{lemma}[\cite{j1}]
\begin{equation}
\limsup_n (N(f^n))^{\frac{1}{n}}=\limsup_n (NF_n(f))^{\frac{1}{n}}
\end{equation} 
\end{lemma}
We define minimal dynamical zeta function as power series
$$
 M_f(z) := \exp\left(\sum_{n=1}^\infty \frac{NF_n(f)}{n} z^n \right),
$$

\begin{theorem}
For any continuous map $f$ of any compact polyhedron   $X$ into itself the minimal zeta function has positive radius of convergence $R$,which admits following estimations 
\begin{equation}
R \geq \exp(-h) > 0,
\end{equation}
 \begin{equation}
R\geq \frac{1}{\max_d \| z\tilde F_d \|} > 0,
\end{equation}
and
\begin{equation}
R\geq \frac{1}{\max_d s(\tilde F_d^{norm})} > 0, 
\end{equation}
where $\tilde F_d$ and $h$ is the same as in section 3.1.1 and 3.1.2.
 
\end{theorem}
 
{\sc Proof}
The theorem follows from Caushy-Adamar formula,lemma 28 and theorems 33  and 34.
 
\begin{remark}
Let us consider a smooth compact manifold $M$, which is a regular neighborhood of $X$
and a smooth map $g: M \rightarrow M$ of the same homotopy type as $f$. There is a smooth
map $\phi: M \rightarrow M$ homotopic to $g$ such that for every $n$ iteration $\phi^n$
has only a finite number of fixed points $F(\phi^n)$(see  \cite{j} , p.62).According to Artin and 
Mazur \cite{am} there exists constant $c=c(\phi) < \infty $, such that $F(\phi^n)< c^n $for every $n>0$. Then due to Halperin result $c^n>F(\phi^n)\geq  NF_n(f)$ for every $n>0$. Now  the Cauchy-Adamar  formula gives us the second proof that the radius of the convergence $R$ is positive.   
\end{remark}

\chapter {Congruences for  Reidemeister and Nielsen numbers}
\markboth{\sc Congruences}{\sc Congruences}

\section{Irreducible Representation and the Unitary Dual of $G$}
 Let $V$ be a Hilbert space.
A unitary representation of $G$ on $V$
 is a homomorphism $\rho: G \to \U(V)$
 where $\U(V)$ is the group of unitary transformations of $V$.
Two of these $\rho_1: G \to \U(V_1)$ and $\rho_2: G \to \U(V_2)$
 are said to be equivalent if there is a Hilbert space
 isomorphism $V_1 \cong V_2$ which commutes with
 the $G$-actions.
A representation $\rho:G \to \U(V)$ is
 said to be irreducible if there is
 no decomposition
$$
 V \cong V_1 \oplus V_2
$$
 in which $V_1$ and $V_2$ are non-zero, closed
 $G$-submodules of $V$.

One defines the unitary dual $\hat{G}$ of $G$
 to be the set of all equivalence classes of
 irreducible, unitary representations of $G$.

If $\rho:G\to \U(V)$ is a representation
 then $\rho\circ\phi:G\to \U(V)$ is also
 a representation, which we shall denote $\hat{\phi}(\rho)$.
If $\rho_1$ and $\rho_2$ are equivalent then
 $\hat{\phi}(\rho_1)$ and $\hat{\phi}(\rho_2)$ are equivalent.
Therefore the endomorphism $\phi$
 induces a map $\hat{\phi}:\hat{G}\to\hat{G}$
 from the unitary dual to itself.

\begin{definition}
Define the number $\#\Fix(\hat\phi)$ to be the number of fixed points
 of the induced map $\hat{\phi}:\hat{G}\to\hat{G}$.
We shall write $\cS(\phi)$ for the set
 of fixed points of $\hat{\phi}$.
Thus $\cS(\phi)$ is the set of
 equivalence classes of irreducible representations $\rho:G\to \U(V)$
 such that there is a transformation $M\in \U(V)$
 satisfying
\begin{equation}
 \forall x\in G,\;\;\;\;
 \rho(\phi(x)) = M \cdot \rho(x) \cdot M^{-1}.
\end{equation}
\end{definition}

Note that if $\phi$ is an inner automorphism $x\mapsto g x g^{-1}$
 then we have for any representation $\rho$,
$$
 \rho(\phi(x)) = \rho(g) \cdot \rho(x) \cdot \rho(g)^{-1},
$$
 implying that the class of $\rho$ is fixed by the induced map.
Thus for an inner automorphism the induced map is
 trivial and $\#\Fix(\hat\phi)$ is the cardinality of $\hat{G}$.
When $G$ is Abelian
 the group $\hat G$ is  the Pontryagin dual of $G$.

\section{ Endomorphism of the Direct Sum of a Free Abelian and a Finite Group}
In this section let $F$ be a finite group and $r$ a natural number. The group $G$ will be the direct sum
$$
G = \bbbz^r \oplus F
$$
We shall describe the Reidemeister numbers of endomorphism $ \phi : G \rightarrow G$.
The torsion elements of $G$ are exactly the elements of the finite, normal subgroup $F$.
For this reason we have $\phi(F) \subset F $.Let $\phi^{finite}: F\rightarrow F $ be the restriction
of $\phi $ to $F$, and let $\phi ^\infty: G/F \rightarrow G/F $ be the induced map on the quotient group. 
We have proved  in proposition 3 that 
$$
R(\phi)= R(\phi^{finite}) \times R(\phi^\infty).
$$
We shall prove the following result:
\begin{proposition}
In the notation described above 
$$
\#\fix(\hat\phi)=\#\fix(\hat\phi^{finite})\times \#\fix(\hat\phi^\infty)
$$
\end{proposition}
 {\sc Proof }

  Consider  the dual $\hat G$. This is cartesian product of the duals of $ \bbbz^r$ and
 $F$:
 $$
  \hat G=\hat \bbbz^r\times\hat F,   \>  \rho=\rho_1\otimes \rho_2
 $$
where $\rho_1$ is an irreducible representation of $\bbbz^r$ and $\rho_2$ is an irreducible
representation of $F$. Since $\bbbz^r$ is abelian, all of its irreducible  representations are
1-dimensional , so $\rho(v)$ for $v\in \bbbz^r$ is always a scalar matrix, and  $\rho_2$ is the 
restriction of $rho$ to $F$.
If $\rho=\rho_1\otimes \rho_2 \in \cS(\phi)$ then there is a matrix $T$ such that 
$$
\rho\circ\phi=T\cdot \rho \cdot T^{-1}.
$$
This implies 
$$
\rho^{finite}\circ\phi^{finite}=T\cdot \rho^{finite} \cdot T^{-1},
$$
so $\rho_2=\rho^{finite}$ is in $\cS(\phi^{finite})$.
For any fixed $\rho_2\in \cS(\phi^{finite})$, the set of $\rho_1$ with $\rho_1\otimes \rho_2 \in \cS(\phi^{finite})$ is the set of $\rho_1$ satisfying 
$$
\rho_1(M\cdot v)\rho_2(\psi(v))= T\cdot \rho_1(v) \cdot T^{-1}
$$
for some matrix $T$ independent of $v\in \bbbz^r$. Since $\rho_1(v)$ is a scalar matrix,
the equation is equivalent to 
$$
\rho_1(M\cdot v)\rho_2(\psi(v))=\rho_1(v),
$$
i.e.
$$
\rho_1((1-M)v)=\rho_2(\psi(v)).
$$
Note that $\hat \bbbz^r$ is isomorphic to the torus $T^n$, and the transformation 
$ \rho_1\rightarrow \rho_1\circ(1-M)$ is given by the action of the matrix $1-M$ on the torus $T^n$. Therefore  the number of $\rho_1$ satisfying the last equation is the degree
of the map $(1-M)$ on the torus , i.e. $\mid \det(1-M)\mid $. From this it follows that 
$$
\#\fix(\hat\phi)=\#\fix(\hat\phi^{finite})\times \mid \det(1-M)\mid.
$$
As in the proof of proposition  3 we have $R(\phi ^{\infty})= \mid \det(1-M)\mid.$Since $\phi ^\infty$
is an endomorphism of an abelian group we have $\#\fix(\hat\phi^\infty)=R(\phi ^\infty)$.Therefore
$$
\#\fix(\hat\phi)=\#\fix(\hat\phi^{finite})\times \#\fix(\hat\phi^\infty). 
$$

As a consequence we have the following 
\begin{theorem}
If $\phi$ be any endomorphism of $G$ where $G$ is the direct sum of a finite group $F$
with a finitely generated free Abelian group, then
$$
R(\phi)=\#\fix(\hat\phi)
$$
\end{theorem}
{\sc Proof }
Let  $\phi^{finite}$ is an endomorphism of a finite group $F$
and  $V$ be the complex vector space of class functions
 on the group $F$. A class function is a function which
 takes the same value on every element of a (usual)
 congruence class.
The map $\phi^{finite}$ induces a map
\begin{eqnarray*}
 \varphi:V & \to & V \\
 f & \mapsto & f\circ \phi^{finite}
\end{eqnarray*}
We shall calculate the trace of $\varphi$ in two ways.
The characteristic functions of the congruence
 classes in $F$ form a basis of $V$, and are
 mapped to one another by $\varphi$ (the map need not
 be a bijection). Therefore the trace
 of $\varphi$ is the number of elements of this
 basis which are fixed by $\varphi$.
By Theorem 14 , this is equal to the Reidemeister number.

Another basis of $V$, which is also mapped to
 itself by $\varphi$ is the set of traces
 of irreducible representations of $F$ (see \cite{la} chapter XVIII).
From this it follows that the trace of $\varphi$
 is the number of irreducible representations $\rho$
 of $F$ such that $\rho$ has the same trace as
 $\hat\phi^{finite}(\rho)$.
However, representations of finite groups
 are characterized up to equivalence by their
 traces. Therefore the trace of $\varphi$
 is equal to the number of fixed points of $\hat\phi^{finite}$.

So,   we have $ R(\phi^{finite})=\#\fix(\hat\phi^{finite})$.
Since $\phi^\infty$ is an endomorphism of the finitely generated free Abelian group we have 
$R(\phi ^\infty)=\#\fix(\hat\phi^\infty)$ ( see formula (2.13) ). It now follows from propositions 3   and 4  that
$R(\phi)=\#\fix(\hat\phi)$.\\

\begin{remark}
By specialising to the case when $G$ is finite and $\phi $ is the identity map, we obtain the 
classical result equating the number of irreducible representation of a finite group with the number of conjugacy classes of the group.
\end{remark}

\section{ Endomorphism of almost Abelian \break \hbox{groups}}

In this section let $G$ be an almost Abelian  and  finitely generated group.A group will be called
almost Abelian if it  has an Abelian subgroup of finite index.
We shall prove in this section an  analog of theorem 52    for almost Abelian group founded by Richard Hill \cite{hi}. It seems plausible that one could prove the
 same theorem for the so - called ``tame'' topological groups (see \cite{ki}).
However we shall be interested mainly in discrete groups, and it is known that
 the discrete tame groups are almost Abelian.

We shall introduce the profinite completion $\Gb$ of $G$ and the
corresponding endomorphism $\phib:\Gb\to\Gb$. This is a compact
totally disconnected group in which $G$ is densely embedded.
The proof will then follow in three steps:
$$
 R(\phi) = R(\phib), \quad
 \#\fix(\hat\phi) = \#\fix(\hat \phib),\quad
 R(\phib) = \#\fix(\hat \phib).
$$
If one omits the requirement that $G$ is almost Abelian then
 one can still show that $R(\phi)\ge R(\phib)$ and $\#\fix(\hat \phi)\ge \#\fix(\hat \phib)$.
The third identity is a general fact for compact groups
 (Theorem \ref{compact}).


\subsection{Compact Groups}

Here we shall prove the third of the above identities.

Let $K$ be a compact topological group
 and $\phi$ a continuous endomorphism of $K$.
We define the number $\#\fix^{\rm top}(\hat\phi)$
 to be the number of fixed points of $\hat\phi$ in the
 unitary dual of $K$, where we only consider continuous representations
 of $K$. The number $R(\phi)$ is defined as usual.

\begin{theorem}[\cite{hi}]
For a continuous endomorphism $\phi$ of a compact group $K$
 one has
 $R(\phi) = \#\fix^{\rm top}(\hat\phi)$.
 \label{compact}
\end{theorem}

The proof uses the Peter-Weyl Theorem:

\begin{theorem}[Peter - Weyl]
If $K$ is compact then there is the following
 decomposition of the space $L^2(K)$
 as a $K\oplus K$-module.
$$
 L^2(K)
 \cong
 \bigoplus_{\lambda\in\hat{K}}
 {\rm Hom}_{C}( V_\lambda , V_\lambda ) .
$$
\end{theorem}

and Schur's Lemma:

\begin{lemma}[Schur]
If $V$ and $W$ are two
 irreducible unitary representations
 then
 $$
 {\rm Hom}_{C K} ( V , W )
 \cong
 \left\{
 \begin{array}{ll}
 0 & V \not\cong W \\
 {C} & V \cong W .
 \end{array}
 \right.
 $$
\end{lemma}

{\sc Proof of Theorem} \ref{compact}.
The $\phi$-conjugacy classes,
 being orbits of a compact group, are compact.
Since there are only finitely many of them,
 they are also open subsets of $K$ and thus have positive Haar measure.

We embed $K$ in $K\oplus K$ by the map
 $g\mapsto (g , \phi(g) ) $.
This makes $L^2(K)$ a $K$-module with
 a twisted action.
By the Peter-Weyl Theorem we have (as $K$-modules)
$$
 L^2(K)
 \cong
 \bigoplus_{\lambda\in\hat{K}}
 {\rm Hom}_{C}( V_\lambda , V_{\hat\phi (\lambda)} ) .
$$
We therefore have a corresponding decomposition of the
 space of $K$-invariant elements:
$$
 L^2(K)^K
 \cong
 \bigoplus_{\lambda\in\hat{K}}
 {\rm Hom}_{C K}( V_\lambda , V_{\hat\phi (\lambda)}) .
$$
We have used the well known identity
 ${\rm Hom}_{C}(V,W)^K = {\rm Hom}_{C K}(V,W)$.

The left hand side consists of functions $f:K\to C$ satisfying
 $f(g x \phi(g)^{-1})=f(x)$ for all $x,g\in K$.
These are just functions on the $\phi$-conjugacy classes.
The dimension of the left hand side is thus $R(\phi)$.
On the other hand by Schur's Lemma the dimension of the right hand
 side is $\#\fix^{\rm top}(\hat\phi)$.

\subsection{Almost Abelian groups}

Let $G$ be an almost Abelian group with an
 Abelian subgroup $A$ of finite index $[G:A]$.
Let $A^0$ be the intersection of all subgroups of $G$
 of index $[G:A]$.
Then $A^0$ is an Abelian normal subgroup of finite index in $G$
 and one has $\phi(A^0)\subset A^0$ for every endomorphism $\phi$ of $G$.

\begin{lemma}
 If $R(\phi)$ is finite then so is $R(\phi|_{A^0})$.
 \label{c}
\end{lemma}

{\sc Proof.}
A $\phi$-conjugacy class is an orbit of the group $G$.
A $\phi|_{A^0}$-conjugacy class is an orbit of the group $A^0$.
Since $A^0$ has finite index in $G$ it follows that
 every $\phi$-conjugacy class in $A^0$ can be the union
 of at most finitely many $\phi|_{A^0}$-conjugacy classes.
This proves the lemma.
\medskip

Let $\Gb$ be the profinite completion of $G$
 with respect to its normal subgroups of finite index.
There is a canonical injection
 $G\to \Gb$ and the
 map $\phi$ can be extended to
 a continuous endomorphism
 $\bar\phi$ of $\Gb$.

There is therefore a canonical map
 $$
 \cR(\phi) \to \cR(\bar\phi).
 $$
Since $G$ is dense in $\Gb$,
 the image of a $\phi$-conjugacy class $\{x\}_\phi$
 is its closure in $\Gb$.
From this it follows that the above map is surjective.
We shall actually see that the map is bijective.
This will then give us
 $$
 R(\phi)= R(\bar\phi).
 $$
However $\bar\phi$ is an endomorphism of the
 compact group $\Gb$ so by Theorem 53
 $$
 R(\bar\phi) = \#\fix^{\rm top}(\hat{\bar \phi}).
 $$
It thus suffices to prove the following two lemmas:

\begin{lemma}
 If $R(\phi)$ is finite then
 $\#\fix^{\rm top}(\hat{\bar\phi}) = \#\fix(\hat\phi)$.
 \label{a}
\end{lemma}

\begin{lemma}
 If $R(\phi)$ is finite then
 the map
 $\cR(\phi) \to \cR(\bar\phi)$
 is injective.
 \label{b}
\end{lemma}

{\sc Proof of Lemma} \ref{a}.
By Mackey's Theorem (see \cite{ki}),
 every representation $\rho$ of $G$ is contained in a representation
 which is induced by a 1-dimensional representation $\chi$ of $A$.
If $\rho$ is fixed by $\hat\phi$
 then for all $a\in A^0$
 we have
 $\chi(a)=\chi(\phi(a))$.
Let $A^1=\{a\cdot\phi(a)^{-1} : a\in A^0\}$.
By Lemma \ref{c} $R(\phi|_{A^0})$ is finite
 and by Theorem 5    $R(\phi|_{A^0}) = [A^0 : A^1]$.
Therefore $A^1$ has finite index in $G$.
However we have shown that $\chi$ and therefore also $\rho$ is constant
 on cosets of $A^1$.
Therefore $\rho$ has finite image,
 which implies that
 $\rho$ is the restriction to $G$ of
 a unique continuous irreducible representation $\bar\rho$
 of $\Gb$. One verifies by continuity
 that $\hat{\bar{\phi}}(\bar\rho)= \bar\rho$.

Conversely if $\bar\rho\in\cS(\bar\phi)$ then
 the restriction of $\bar\rho$ to $G$ is in $\cS(\phi)$.
\medskip

{\sc Proof of Lemma} \ref{b}.
We must show that the intersection with $G$ of the closure
 of $\{x\}_\phi$ in $\Gb$ is equal to $\{x\}_\phi$.
We do this by constructing a coset of a normal subgroup of finite
 index in $G$ which is contained in $\{x\}_\phi$.
For every $a\in A^0$ we have $x\sim_\phi xa$ if
 there is a $b\in A^0$ with
 $x^{-1}bx\phi(b)^{-1}=a$.
It follows that $\{x\}_\phi$ contains
 a coset of the group $A^2_x := \{x^{-1}bx\phi(b)^{-1} : b\in A^0\}$.
It remains to show that $A^2_x$ has finite index in $G$.

Let $\psi(g) = x \phi(g) x^{-1}$.
Then by Corollary 4 we have $R(\psi)=R(\phi)$.
This implies $R(\psi)<\infty$ and therefore by Lemma \ref{c}
 that $R(\psi|_{A_0})<\infty$.
However by Theorem 5 we have
 $R(\psi|_{A_0}) = [A^0:A^2_x]$.
This finishes the proof.

\begin{theorem}[\cite{hi}]
 If $\phi$ be any endomorphism of $G$ where $G$ is an almost abelian group, then
$$
R(\phi)=\#\fix(\hat\phi)
$$
\end{theorem}
{\sc Proof }
The proof follows from lemmas 31 , 32 and theorem 53

\section{Endomorphisms of nilpotent groups}

In this section, we shall extend the computation of the Reidemeister number
 to endomorphisms of finitely generated torsion free nilpotent groups via
topological techniques. 
Let $\Gamma$ be a finitely generated torsion free nilpotent group. It is well
known \cite{mal} that $\Gamma=\pi_1(M)$ for some compact nilmanifold $M$.
In fact, the {\it rank} (or {\it Hirsch number}) of $\Gamma$  is equal to $dimM$, the dimension of $M$. Since $M$ is a $K(\Gamma,
1)$, every endomorphism $\phi:\Gamma \to \Gamma$ can be realized by a selfmap
$f:M\to M$ such that $f_{\#}=\phi$ and thus $R(f)=R(\phi)$.

\begin{theorem}
Let $\Gamma$ be a finitely generated torsion free nilpotent group of rank $n$.
For any endomorphism $\phi:\Gamma \to \Gamma$ such that $R(\phi)$ is finite, there
exists an endomorphism $\psi:{\bbbz }^n\to {\bbbz}^n$ such that $R(\phi)=\#
\fix \hat{\psi}$.
\end{theorem}
 
\noindent
{\sc Proof:}
Let $f:M\to M$ be a map realizing $\phi$ on a compact nilmanifold $M$ of
dimension $n$.
Following \cite{fahu1}, $M$ admits a principal torus bundle $T\to M
\stackrel{p}{\rightarrow} N$ such that $T$ is a torus and $N$ is a nilmanifold
of lower dimension. Since every selfmap of $M$ is homotopic to a fibre
preserving map of $p$, we may assume without loss of generality that $f$ is
fibre preserving such that the following diagram commutes.
 
\begin{eqnarray*}
T &\stackrel{f_b}{\rightarrow} &T \\
\downarrow &  &\downarrow \\
M &\stackrel{f}{\rightarrow} &M \\
p \downarrow &\hspace {.2in} &\downarrow p \\
N &\stackrel{\bar f}{\rightarrow} &N
\end{eqnarray*}
  
A strengthened version of Anosov's theorem \cite{an} is proven in
\cite{no} which states, in particular, that $|L(f)|=R(f)$ if
$L(f)\ne 0$. Since the bundle $p$ is orientable, the product formula
$L(f)=L(f_b)\cdot L(\bar f)$ holds and
thus yields a product formula for the Reidemeister numbers, i.e.,
$R(f)=R(f_b)\cdot R(\bar f)$. To prove the assertion, we proceed by induction
on the rank of $G$ or $dimM$.
 
The case where $n=1$ follows from the theorem 52 since $M$ is
the unit circle . To prove the inductive step, we assume
that $R(\bar f)=\#\fix (\widehat{\bar {\psi}} :\widehat{\pi_1(T^m)}
(=\widehat{{\bbbz}^m})
\to \widehat {\pi_1(T^m)})$ where $m<n$ and $T^m$ is an $m$-torus. Since
$R(f_b)=\#\fix (\hat {\phi_b}:\widehat {\pi_1(T)}\to \widehat {\pi_1(T)})$, if
$L(f)\ne 0$ then the product formula for Reidemeister numbers gives
 
\begin{eqnarray*}
R(f)&=&\#\fix(\hat{\phi_b})\cdot \#\fix(\widehat{\bar {\psi}}) \\
    &=&\#\fix(\hat{\psi})
\end{eqnarray*}
where $\psi =\phi_b \times \bar {\psi}:\pi_1(T^n)=\pi_1(T) \times
\pi_1(T^m) \to \pi_1(T) \times \pi_1(T^m)$ with $T^n=T\times T^m$.

 \section{Main Theorem} 
 
The following lemma is useful for calculating Reidemeister numbers.
It will also be used in the proof of the Main Theorem

\begin{lemma}
Let $\phi:G\to G$ be any endomorphism
 of any group $G$, and let $H$ be a subgroup
 of $G$ with the properties
$$
  \phi(H) \subset H
$$
$$
  \forall x\in G \; \exists n\in \N \hbox{ such that } \phi^n(x)\in H.
$$
Then
$$
 R(\phi) = R(\phi_H),
$$
 where $\phi_H:H\to H$ is the restriction of $\phi$ to $H$.

\end{lemma}

{\it Proof }
Let $x\in G$. Then there is an $n$ such that $\phi^n(x)\in H$.
From Lemma 7 it is known that $x$ is $\phi$-conjugate
 to $\phi^n(x)$.
This means that the $\phi$-conjugacy class $\{x\}_\phi$
 of $x$ has non-empty intersection with $H$.

Now suppose that $x,y\in H$ are $\phi$-conjugate,
 ie. there is a $g\in G$ such that
 $$gx=y\phi(g).$$
We shall show that $x$ and $y$ are $\phi_H$-conjugate,
 ie. we can find a $g\in H$ with the above property.
First let $n$ be large enough that $\phi^n(g)\in H$.
Then applying $\phi^n$ to the above equation we obtain
 $$ \phi^n(g) \phi^n(x) = \phi^n(y) \phi^{n+1}(g). $$
This shows that $\phi^n(x)$ and $\phi^n(y)$ are $\phi_H$-conjugate.
On the other hand, one knows by Lemma 7 that $x$ and $\phi^n(x)$ are
 $\phi_H$-conjugate, and $y$ and $\phi^n(y)$ are $\phi_H$ conjugate,
 so $x$ and $y$ must be $\phi_H$-conjugate.

We have shown that the intersection with $H$ of a 
 $\phi$-conjugacy class in $G$ is a $\phi_H$-conjugacy class
 in $H$.
We therefore have a map
$$
\begin{array}{cccc}
 Rest : & \cR(\phi) & \to & \cR(\phi_H)\\
        & \{x\}_\phi & \mapsto & \{x\}_\phi \cap H
\end{array}
$$
This clearly has the two-sided inverse
$$
 \{x\}_{\phi_H} \mapsto \{x\}_\phi.
$$
Therefore $Rest$ is a bijection and $R(\phi)=R(\phi_H)$.
\medskip

\begin{corollary}
 Let $H=\phi^n(G)$. Then $R(\phi) = R(\phi_H)$.
\end{corollary}
Let $\mu(d)$, $d\in\bbbn$ be the Moebius function,
i.e.
$$
\mu(d) =
\left\{
\begin{array}{ll}
1 & {\rm if}\ d=1,  \\
(-1)^k & {\rm if}\ d\ {\rm is\ a\ product\ of}\ k\ {\rm distinct\ primes,}\\
0 & {\rm if}\ d\ {\rm is\ not\ square-free.} 
\end{array}
\right.
$$

\begin{theorem}[Congruences for the Reidemeister numbers]

Let $\phi:G$ $\to G$ be an endomorphism of the group $G$ such that all numbers $R(\phi^n)$ are finite and let $H$ be a subgroup
 of $G$ with the properties
$$
  \phi(H) \subset H
$$
$$
  \forall x\in G \; \exists n\in \N \hbox{ such that } \phi^n(x)\in H.
$$
If one of the following conditions is satisfied: \\
(I) $H$ is finitely generated Abelian, \\
(II) $H$ is finite,  \\
(III) $H$ is a direct sum of a finite group and a finitely generated free Abelian group, \\
or more generally  \\
(IV) $H$ is finitely generated almost Abelian group,\\
 or \\
(V) $H$  is finitely generated, nilpotent and torsion free ,
then one has for all natural numbers $n$,
 $$
 \sum_{d\mid n} \mu(d)\cdot R(\phi^{n/d}) \equiv 0 \mod n.
 $$
\end{theorem}

{\sc Proof}
From theorems 52, 55, 56        and  lemma 33 it follows immediately that , in cases I - IV, 
 for every $n$ 
$$
 R(\phi^n)=\#\fix\left[ \hat{\phi_H}^n:\hat{H}\to\hat{H} \right].
$$

Let $P_n$ denote the number of periodic points of $\hat{\phi_H}$
of least period $n$. One sees immediately that
$$
 R(\phi^n)=\#\fix\left[ \hat{\phi_H}^n \right] = \sum_{d\mid n} P_d.
$$
Applying M\"obius' inversion formula, we have,
$$
 P_n = \sum_{d\mid n} \mu(d) R(\phi^{n/d}).
$$
On the other hand, we know that $P_n$ is always divisible
be $n$, because $P_n$ is exactly $n$ times the number of
$\hat{\phi_H}$-orbits in $\hat{H}$ of length $n$.
In the case V when  $H$ is finitely generated, nilpotent and torsion free ,we know from theorem 56 that  there
exists an endomorphism $\psi:{\bf Z}^n\to {\bf Z}^n$ such that $R(\phi^n)=\#
\fix \hat{\psi^n}$.
 The proof then follows as in previous cases.

\begin{remark}
For finite groups,  congruences for Reidemeister numbers  follow from those of Dold for Lefschetz numbers since
we have identified in  remark 2 the Reidemeister
numbers with the Lefschetz numbers of induced dual maps.
\end{remark}

\section{Congruences for  Reidemeister  numbers of a continuous map}

Using corollary 1 we may apply the theorem 57  to the Reidemeister
numbers of continuous maps.

\begin{theorem}
  Let $f:X\to X$ be a self-map such that all numbers $R(f^n)$ are finite.Let $f_*:\pi_1(X)\to \pi_1(X)$ be an induced endomorphism of the group $\pi_1(X)$  and let $H$ be a subgroup
 of $\pi_1(X)$ with the properties
$$
  f_*(H) \subset H
$$
$$
  \forall x\in \pi_1(X) \; \exists n\in \N \hbox{ such that } f_*^n(x)\in H.
$$
If one of the following conditions is satisfied : \\
(I) $H$ is finitely generated Abelian, \\
(II) $H$ is finite,  \\
(III) $H$ is a direct sum of a finite group and a finitely generated free Abelian group \\
or more generally,  \\
(IV) $H$ is finitely generated almost Abelian group,\\
 or \\
(V) $H$  is finitely generated, nilpotent and torsion free ,\\
then one has for all natural numbers $n$,
$$
 \sum_{d\mid n} \mu(d)\cdot  R(f^{n/d}) \equiv 0 \mod n.
$$
\end{theorem}

\section { Congruences for Reidemeister numbers of equivariant group endomorphisms}

Let $G$ be a compact Abelian topological group acting on a topological group
$\pi$ as automorphisms of
$\pi$, i.e., a homomorphism $\nu :G\to Aut(\pi)$. For every $\sigma\in \pi$,
the isotropy subgroup of $\sigma$ is given by $G_{\sigma}=\{g\in G|g(\sigma)
 =\sigma\}$ where $g(\sigma)=\nu(g)(\sigma)$. For any closed subgroup $H\le G$,
the fixed point set of the $H$-action, denoted by
$$
\pi^H=\{\sigma\in \pi|h(\sigma)=\sigma,\forall h\in H\},
$$
is a subgroup of $\pi$. Since $G$ is Abelian, $G$ acts on $\pi^H$ as
automorphisms of $\pi^H$. Denote by ${\cal C}(\pi)$ (and ${\cal C}(\pi^H)$)
the set of
conjugacy classes of elements of $\pi$ (and $\pi^H$, respectively). Note
that the group $G$ acts on the conjugacy classes via
$$
<\sigma>\mapsto <g(\sigma)>
$$
for $<\sigma>\in {\cal C}(\pi)$ (or ${\cal C}(\pi^H)$).
 
\medskip
 
Let $End_G(\pi)$ be the set of $G$-equivariant endomorphisms of $\pi$. For
any $\phi \in End_G(\pi)$ and $H\le G$, $\phi^H\in End_{G}(\pi^H)$ where
$\phi^H=\phi|\pi^H:\pi^H\to \pi^H$. Furthermore, $\phi$ induces a $G$-map
on ${\cal C}(\pi)$ defined by
$$
\phi_{conj}:{\cal C}(\pi)\to {\cal C}(\pi)
 $$
via
$$
<\sigma>\mapsto <\phi(\sigma)>.
$$
Similarly, $\phi^H$ induces
$$
(\phi^H)_{conj}:{\cal C}(\pi^H)\to {\cal C}(\pi^H).
$$
 
\bigskip
 
In the case where $\pi$ is Abelian, $G$ acts on $\hat \pi$ via
$$
\chi(\sigma)\mapsto \chi(g(\sigma))
$$
for any $\chi \in \hat \pi, g\in G$. Thus the dual $\hat \phi$ of $\phi \in
End_G(\pi)$ is also $G$-equivariant, i.e., $\hat \phi \in End_G(\hat \pi)$.
Similarly, $\widehat {\phi^H}\in End_{G}(\widehat {\pi^H})$.
 
\bigskip
 
To establish the congruence relations in this section, we need the following
 basic counting principle.
 
\medskip
 
\begin{lemma}
Let $G$ be an Abelian topological group and $\Gamma$ be a $G$-set with finite
isotropy types. For any $G$-map $\psi:\Gamma \to \Gamma$,
$$
\fix \psi =\bigsqcup_{K\in Iso(\Gamma)} \fix \psi_{K}
$$
where $Iso(\Gamma)$ is the set of isotropy types of $\Gamma$,
$\Gamma_{K}=\{\gamma\in \Gamma|G_{\gamma}=K\},
\psi_{K}=\psi|\Gamma_{K}:\Gamma_{K}\to \Gamma^{K}$ and
$\fix\psi_K=(\fix\psi)\cap \Gamma_K$. In particular, if $\#\fix \psi <\infty$
then
\begin{equation}
\#\fix \psi =\sum_{K\in Iso(\Gamma)} \#\fix \psi_{K}.
\end{equation}
 
\end{lemma}
 
\noindent
{\sc Proof:}
 It follows from the decomposition
$$
\Gamma =\bigsqcup_{K\in Iso(\Gamma)} \Gamma_K. \Box
$$
 
\bigskip
 
\noindent
{\bf Remark~} Suppose that $G$ is a compact Abelian Lie group.
If $\fix \psi$ is finite then
it follows that $\#\fix \psi_K=I(\psi_K)$, the fixed point index of $\psi_K$
which is divisible by $\chi (G/K)$, the Euler characteristic of $G/K$ (see
\cite{d}, \cite{komia}, \cite{ulrich}).
 
\bigskip
 
\begin{theorem}
Let $G$ be an Abelian compact Lie group.
For any subgroup $H\in Iso(\pi)$ and $\phi\in End_G(\pi)$, if
 
{\em (I)} $\pi$ is finitely generated and $\phi$ is eventually commutative or
 
{\em (II)} $\pi$ is finite or
{\em (III)} $\pi$ is finitely generated torsion free nilpotent,
then
 
$$
\sum_{H\le K\in Iso(\pi)} \mu(H,K) R(\phi^K) \equiv 0
\quad mod~\chi (G/H)
$$
and
$$
\sum_{H\le K\in Iso(\pi)} \varphi (H,K) R(\phi^K)\equiv 0 \quad mod~\chi (G/H)
$$
where $\mu (,)$ denotes the M\"obius function on $Iso(\pi)$ and
$$\varphi (H,K)=\sum_{H\le L\le K} \chi (L/H)\mu (L,K).$$
 
\end{theorem}
 
\noindent
{\sc Proof:}
(I): Since $\phi$ is eventually commutative, so is $\phi^H$ for every
$H\le G$. With the canonical $G$-action on $H_1(\pi)$, $H_1(\phi)$ is a
$G$-equivariant endomorphism of $H_1(\pi)$. Similarly, we have $H_1(\phi^H)
\in End_G(H_1(\pi^H))$ and hence $\widehat {H_1(\phi^H)}\in End_G(\widehat
 {H_1(\pi^H)})$.
It follows from theorem 52  and formula (5.1) that
 
$$
R(\phi^H)=\# \fix (\widehat {H_1(\phi^H)})=\sum_{H\le K\in Iso(\pi)}\# \fix
(\widehat {H_1(\phi^H)})_K.
$$
Hence, by M\"obius inversion \cite{aigner}, we have
 
$$
\sum_{H\le K\in Iso(\pi)} \mu(H,K) R(\phi^K) =\# \fix(\widehat
{H_1(\phi^K)})_H\equiv 0 \quad mod~\chi (G/H).
$$
 
\medskip
 
\noindent
(II): Following theorem 14 , we have
 
$$
R(\phi^H)=\#\fix(\phi^H)_{conj}=\sum_{H\le K\in Iso(\pi)}\#\fix
((\phi^H)_{conj})_K.
$$

The assertion follows from formula (5.1) and the M\"obius inversion formula.
 
\medskip
 
\noindent
(III): The commutative diagram in the proof of Theorem 56 gives rise to the
following commutative diagram in which the rows are short exact sequences
of groups where $\pi'$ is free abelian and $\bar {\pi}$ is nilpotent.
 
\[
\begin{array}{llllll}
\mbox{1} \rightarrow
& \pi' \stackrel{i_*}{\rightarrow} &\pi
\stackrel{p_*}{\rightarrow} &\bar {\pi} \rightarrow \mbox{1} \\
& \uparrow \phi'
& \uparrow \phi &
\uparrow \bar {\phi} \\
\mbox{1} \rightarrow
& \pi' \stackrel{i_*}{\rightarrow} &\pi
\stackrel{p_*}{\rightarrow} &\bar {\pi} \rightarrow \mbox{1}
\end{array}
\]
 
 Note that $G$ acts on both $\pi'$ and $\bar {\pi}$ as automorphisms and so
$\phi'$ and $\bar {\phi}$ are both $G$-equivariant. Similarly, we have
$\phi'^H\in End_G(\pi'^H)$ and $\bar {\phi}^H\in End_G(\bar {\phi}^H)$.
It follows from the proof of Theorem 56 that the endomorphism $\psi$ can be
made $G$-equivariant by the diagonal action on the product of the fibre and
the base. Hence,
$$
R(\phi^H)=\#\fix \widehat {\psi^H}=\sum_{H\le K\in Iso(\pi)}\# \fix
(\widehat {\psi^H})_K.
$$
Again, the assertion follows from M\"obius inversion.
 
Finally, for the congruences with $\varphi (,)$, we proceed as in
Theorem 6 of \cite{komia}. $\Box$

\bigskip
 
\begin{remark} In \cite{komia}, Komiya considered the $G$-invariant set
$X^{(H)}=GX^H$ for arbitrary $G$ (not necessarily Abelian). In our case,
$\pi^{(H)}$ need not be a subgroup of $\pi$.
For example, take $\pi =G=S_3$ to be the symmetric group on three letters
 and let $G$ act on $\pi$ via conjugation. For any subgroup $H\le G$ of
order two, it is easy to see that $\pi^{(H)}$ is not a subgroup of $\pi$.
 \end{remark}
\bigskip

\section{Congruences for Reidemeister numbers of equivariant maps}

Unlike Komiya's generalization \cite{komia} of Dold's result, the
congruence relations among the Reidemeister numbers of $f^n$ established
in  section   cannot be generalized without further assumptions as we
illustrate in the following example.
 
\bigskip
 
\begin{example}
{\em
Let $X=S^2\subset {\bf R}^3,G={\bf Z}_2$ act on $S^2$ via

$$
\zeta (x_1,x_2,x_3)=(x_1,x_2,\zeta x_3)
$$
so that $X^G=S^1$.
Let $f:X\to X$ be defined by $(x_1,x_2,x_3)\mapsto (-x_1,x_2,x_3)$.
It follows that
$$
R(f)=1; \qquad R(f^G)=2.
$$
Hence,
$$
\sum_{(1)\le K} \mu ((1),K)R(f^K)=\mu (1)R(f)+\mu (2)R(f^G)=-1
$$
which is not congruent to $0~mod~2$.
}
\end{example}
 
\bigskip
 
In order to apply the previous result , we need $R(f^H)=
R((f_{\ast})^H)$, for all $H\le G$.
In the above example, $R(f^G)=2$ but $R((f_{\ast})^G)=1$.
 
 \bigskip
 
Recall that a selfmap $f:X\to X$ is {\it eventually commutative}
\cite{j} if the
induced homomorphism $f_{\ast}:\pi_1(X)\to \pi_1(X)$ is eventually
commutative. The following is immediate from Theorem 59.
 
\medskip
 
\begin{theorem}
Let $f:X\to X$ be a $G$-map on a finite $G$-complex $X$ where $G$ is an
abelian compact Lie group, such that $X^H$ is connected for all
$H\in Iso(X)$.
If $f$ is eventually commutative, or $\pi_1(X^H)$ is finite or nilpotent, and
$R(f^H)=R((f_{\ast})^H)$ for all $H\in Iso(X)$, then
$$
\sum_{H\le K\in Iso(\pi)} \mu (H,K) R(f^K)\equiv 0 \quad mod~\chi (G/H)
$$
and
$$
\sum_{H\le K\in Iso(\pi)} \varphi (H,K) R(f^K)\equiv 0 \quad mod~\chi (G/H).
$$

\end{theorem}
 
\bigskip

\section{Congruences for Nielsen  numbers of a continuous map}

\begin{theorem}
Suppose that there is a natural number $m$ such that $\tilde{f}^m_*(\pi)\subset I(\tilde{f}^m)$ . If for every $d$ dividing a certain natural number $n$ we have
 $L(f^{n/d})\not= 0$, then one has for that particular $n$,
$$
 \sum_{d\mid n} \mu(d) N(f^{n/d}) \equiv 0 \mod n.
$$
 
\end{theorem}
{\sc Proof}
From the results of Jiang \cite{j} we have that  $N(f^{n/d})=R(f^{n/d})$ for the same particular $n$ and $ \tilde{f}_*$ is eventually commutative. The result now follows from theorem 58.

\begin{corollary}
Let $I(id_{\tilde{X}})=\pi$ and for every $d$ dividing a certain natural number $n$ we have
 $L(f^{n/d})\not= 0$, then theorem 61  applies

\end{corollary}

\begin{corollary}
Suppose that $X$ is aspherical, $f$ is eventually commutative and for every $d$ dividing a certain natural number $n$ we have
 $L(f^{n/d})\not= 0$, then theorem 61  applies
\end{corollary}

\begin{example}
Let $f:T^n\rightarrow T^n$
 be a hyperbolic endomorphism.
Then for every  natural $n$
$$
 \sum_{d\mid n} \mu(d) N(f^{n/d}) \equiv 0 \mod n.
$$
 \end{example}

\begin{example}
Let $g:M\to M$ be an expanding map \cite{sh} of the orientable smooth compact manifold $M$.Then $M$ is aspherical and is a $K(\pi_1(M),1)$, and $\pi_1(M)$ is torsion free \cite{sh}.According to Shub \cite{sh} any lifting $\tilde g$ of $g$ has exactly one fixed point.From this and the covering homotopy theorem it follows that the fixed point of $g$ are pairwise inequivalent.The same is true for all iterates $g^n$.Therefore $N(g^n)=\# \fix(g^n)$ for all $n$.So the sequence of the  Nielsen numbers $N(g^n)$ of an expanding map satisfies the congruences as above.
 \end{example}

\begin{theorem}
Let $X$ be a connected, compact polyhedron with finite fundamental group $\pi$.
Suppose that the action of $\pi$ on the rational homology of the
universal cover $\tilde{X}$ is trival,
i.e. for every covering translation $\alpha\in\pi$,
$\alpha_*=id:H_*(\tilde{X},\bbbq)\rightarrow H_*(\tilde{X},\bbbq)$.
If  for every $d$ dividing a certain natural number $n$ we have
 $L(f^{n/d})\not= 0$, then one has for that particular $n$,
$$
 \sum_{d\mid n} \mu(d) N(f^{n/d}) \equiv 0 \mod n.
$$
 
\end{theorem}
{\sc Proof}
From the results of Jiang \cite{j} we have that  $N(f^{n/d})=R(f^{n/d})$ for the same particular $n$.The result now follows from theorem 58.

\begin{lemma}
Let $X$ be a polyhedron with finite fundamental group $\pi$ and let
$p:\tilde{X}\rightarrow X$ be its universal covering. Then the action
of $\pi$ on the rational homology of $\tilde{X}$ is trivial iff
$H_*(\tilde{X};\bbbq) \cong H_*(X;\bbbq)$.
\end{lemma}

\begin{corollary}
Let $\tilde{X}$ be a compact $1$-connected polyhedron which is a 
 rational homology $n$-sphere, where $n$ is odd.
Let $\pi$ be a finite group acting freely on $\tilde{X}$ and let
$X=\tilde{X}/\pi$.
Then theorem 62 applies.
\end{corollary}

{\sc Proof}
The projection $p:\tilde{X}\rightarrow X= \tilde{X}/\pi$ is a
universal covering space of $X$. For every $\alpha\in\pi$, the degree
of $\alpha:\tilde{X}\rightarrow\tilde{X}$ must be 1, because
$L(\alpha)=0$ ($\alpha$ has no fixed points). Hence
$\alpha_* = id: H_*(\tilde{X};\bbbq)\rightarrow H_*(\tilde{X};\bbbq)$.

\begin{corollary}
If $X$ is a closed 3-manifold with finite $\pi$, then theorem 62
applies.
\end{corollary}

{\sc Proof}
$\tilde{X}$ is an orientable, simply connected manifold, hence a 
 homology 3-sphere. We apply corollary 22.

\begin{example}
Let $X=L(m,q_1,\ldots,q_r)$ be a generalized lens space
and $f:X\to X$ a continuous map with
$f_{1*}(1)=k$ where $\mid k \mid\not= 1$.
Then for every  natural $n$
$$
 \sum_{d\mid n} \mu(d) N(f^{n/d}) \equiv 0 \mod n.
$$

\end{example}

{\sc Proof}

By corollary  we see that theorem 
applies for lens spaces.
Since $\pi_1(X)=\bbbz/m\bbbz$,
the map $f$ is eventually commutative.
A lens space has a structure as a CW complex with one
cell $e_i$ in each dimension $0\leq i\leq 2n+1$.
The boundary map is given by $\partial e_{2k}=m.e_{2k-1}$
for even cells, and $\partial e_{2k+1}=0$ for odd cells.
From this we may calculate the Lefschetz numbers:
$$
 L(f^n) = 1-k^{(l+1)n} \not= 0.
$$
This is true for any $n$ as long as $\mid k\mid\not=1$.
The result now follows from theorem .

\begin{remark}
 It is known that in previous example 
$$
 N(f^n) = R(f^n) = \#\coker(1-f^n_{1*})=hcf(1-k^n,m)
$$
for every $n$.So we obtain pure arithmetical fact: the sequence $n\mapsto hcf(1-k^n,m)$ satisfies 
 congruences above for every natural $n$ if $\mid k \mid\not= 1$.
\end{remark}

\section {Some conjectures for wider classes of \break \hbox{groups}}

 For the case of almost nilpotent groups (ie. groups with polynomial growth, in view of Gromov's theorem \cite{gromov}) we believe that the congruences for the Reidemeister numbers are also true.We intend to prove this conjecture by
 identifying the Reidemeister number on the nilpotent part of the group
 with the number of fixed points in
 the direct sum of the duals of the quotients of successive terms in the central series. We then hope to show that the Reidemeister number of the whole endomorphism is a sum of numbers of orbits of such fixed points under the
 action of the finite quotient group (ie the quotient of the whole group by
 the nilpotent part).
The situation for groups with exponential growth is very different.
There one can expect the Reidemeister number to be infinite as long as the endomorphism is injective.

\chapter{The Reidemeister torsion }
\markboth{\sc  Reidemeister torsion }{\sc  Reidemeister torsion }

  \section{Preliminaries}
 
Like the Euler characteristic, the Reidemeister torsion is algebraically defined.
Roughly speaking, the Euler characteristic is a
graded version of the dimension, extending
the dimension from a single vector space to a complex
of vector spaces.
In a similar way, the Reidemeister torsion
is a graded version of the absolute value of the determinant
of an isomorphism of vector spaces: for this to make sense,
both vector spaces should be equipped with a positive
density.

 Recall that a density $f$ on a complex space $V$ of dimension $n$ is a 
 map $ f: \wedge^nV \rightarrow R$ with $ f(\lambda\cdot x)= |\lambda|\cdot f(x) $
 for all $x\in \wedge ^nV,  \lambda \in \bbbc$. The densities on $V$ clearly form
 a real vector space $|V|$ of dimension one. If $f$ is nonzero and takes values
 in $[o,\infty)$, we say that it is positive . If $V_1$ and $V_2$ are both $n$-dimensional and $A:
 V_1 \rightarrow V_2$ is linear over $\bbbc$, then there is an induced map $ A^*: |V_2| \rightarrow |V_1| $. If each $V_i$ carries a preffered positive density $f_i$ then $A^*f_2=a\cdot f_1$, for some $a\geq 0$, and we write $a=|detA |$.In the case $V_1=V_2, f_1=f_2$ then $a$
 is indeed the absolute value of the determinant of $A$, so this notation is consistent.
 
 If $0\to V_1 \to V_2\to V_3 \to 0 $ is an exact sequence of finite dimensional vector spaces
 over $\bbbc$ , then there is a natural isomorphism $\wedge^{n_1}V_1 \otimes \wedge^{n_3}V_3\rightarrow \wedge^{n_2}V_2,\> n_i=dim V_i$.This induces a natural
 isomorphism $|V_1|\otimes |V_3| \cong |V_2|$. If $V$ is zero, then absolute value is 
 a standard generator for $|V|$ and sets up a natural isomorphism $|V|\cong R$.
 
 Let $d^i : C^i \rightarrow C^{i+1} $ be a cochain complex of finite-dimensional vector spaces over $\bbbc$ with $C^i=0$ for $i<0$ or $i$ large. Let $Z^i=ker d^i, \> B^i= im d^{i-1}$, and
 $H^i=Z^i/ B^i$ be the cocycles, coboundaries, and cohomology of $C^* $ respectively.
 If one is given positive densities $\Delta_i$ on $C^i$ and $D_i$ on $H^i$(with the standard choice when $i$ is negative or large), then the Reidemeister torsion $\tau(C^*, \Delta_i, D_i) \in (0,\infty) $ is defined as follows. The short exact sequences
 $ 0\to Z^i\to C^i \to B^{i+1} \to 0 $ and $0\to B^i\to Z^i\to H^i\to o$
 together with the trivializations $|C^i|=R\cdot\Delta_i, |H^i|=R\cdot D_i$ give isomorphisms
 $|Z^i|\otimes |B^{i+1}| \cong R, \> |B^i|\cong |Z^i|$ for each $i$. This gives isomorphisms 
 $|B^i|\otimes |B^{i+1}| \cong R$, hence also $|B^i|\cong |B^{i+2}|$. But for $i\leq 0$ or $i$
 large, $B^i=0$ and so $|B^i|=R$. So if $j$ is large and even, we find $R=|B^j|\cong |B^0|=R$
 and this isomorphism is the multiplication by scalar $\tau\in R^*$. This is the Reidemeister
 torsion $ \tau(C^*, \Delta_i, D_i)$.

If the cohomology $H^i=0$ for all $i$ we say that
$C^*$ is {\it acyclic} and we write  $\tau(C^*, \Delta_i)$ for the torsion when all
the $D_i$ are standard . 

When $C^*$ is acyclic , there is another way to define $\tau$.

\begin{definition}
 Consider a chain contraction $\delta^i:C^i\rightarrow C^{i-1}$,
 i.e. a linear map such that $d\circ\delta + \delta\circ d = id$.
 Then $d+\delta$ determies a map
 
 $ (d+\delta)_+ : C^+:=\oplus C^{2i}
                      \rightarrow C^- :=\oplus C^{2i+1}$
 and a map
 $ (d+\delta)_- : C^- \rightarrow C^+ $.
Since the map
$(d+\delta)^2 = id + \delta^2$ is unipotent,
$(d+\delta)_+$ must be an isomorphism.
One defines $\tau(C^*,\Delta_i):= \mid \det(d+\delta)_+\mid$
(see \cite{fri2}).
\end{definition}

Reidemeister torsion is defined in the following geometric setting.
Suppose $K$ is a finite complex and $E$ is a flat, finite dimensional,
complex vector bundle with base $K$.
We recall that a flat vector bundle over $K$ is essentially the
same thing as a representation of $\pi_1(K)$ when $K$ is
connected.
If $p\in K$ is a basepoint then one may move the fibre at $p$
in a locally constant way around a loop in $K$. This
 defines an action of $\pi_1(K)$ on the fibre $E_p$ of $E$
above $p$. We call this action the holonomy representation
$\rho:\pi\to GL(E_p)$.
Conversely, given a representation $\rho:\pi\to GL(V)$
of $\pi$ on a finite dimensional complex vector space $V$,
one may define a bundle $E=E_\rho=(\tilde{K}\times V) / \pi$.
Here $\tilde{K}$ is the universal cover of $K$, and
$\pi$ acts on $\tilde{K}$ by covering tranformations and on $V$
by $\rho$.
The holonomy of $E_\rho$ is $\rho$, so the two constructions
give an equivalence of flat bundles and representations of $\pi$.
 
If $K$ is not connected then it is simpler to work with
flat bundles. One then defines the holonomy as a
representation of the direct sum of $\pi_1$ of the
components of $K$. In this way, the equivalence of
flat bundles and representations is recovered.
 
Suppose now that one has on each fibre of $E$ a positive density
which is locally constant on $K$.
In terms of $\rho_E$ this assumption just means 
$\mid\det\rho_E\mid=1$.
Let $V$ denote the fibre of $E$.
 Then the cochain complex $C^i(K;E)$ with coefficients in $E$
can be identified with the direct sum of copies
of $V$ associated to each $i$-cell $\sigma$ of $K$.
The identification is achieved by choosing a basepoint in each
component of $K$ and a basepoint from each $i$-cell.
By choosing a flat density on $E$ we obtain a
preferred density $\Delta_i$ on $C^i(K,E)$. Let $ H^i(K;E)$ be the $i$-dimensional
cohomology of $K$ with coefficients in $E$ , i.e. the twisted cohomology of $E$.
Given a density $D_i$ on each $H^i(K;E)$, one defines the
 Reidemeister torsion of $(K;E,D_i)$ to be
$\tau(K;E,D_i)=\tau(C^*(K;E),\Delta_i, D_i)\in(0,\infty)$. A case of particular interest is when  $E$ is an acyclic bundle, meaning that the twisted cohomology of $E$ is zero ($H^i(K;E)=0$), then one can take $D_i$ to be the absolute value map on $\wedge^0(0)=\bbbc $ and the resulting 
Reidemeister torsion is denoted by $\tau(K;E)$. In this case it does not depend on the choice of flat density on $E$.

 The Reidemeister torsion of an acyclic bundle $E$ on $K$ has many nice properties.
 Suppose that $A$ and $B$ are subcomplexes of $K$. Then we have a multiplicative law:
 \begin{equation}
 \tau(A\cup B;E)\cdot \tau(A\cap B;E) =\tau(A;E)\cdot \tau(B;E)           
 \end{equation}
that is interpreted as follows. If three of the bundles $E| A\cup B, \> E |A\cap B, \> E|A,\> E|B$
are acyclic then so is the fourth and the equation (6.1) holds.

    Another property is the simple homotopy invariance of the Reidemeister torsion.
Suppose $K^{\prime}$ is a subcomplex of $K$ obtained by an elementary collapse of an $n$-cell
$\sigma$ in $K$. This means that $K= K^{\prime}\cup \sigma\cup \sigma^{\prime}$ where 
$\sigma^{\prime}$ is an $(n-1)$ cell of $K$ so set up that $\partial \sigma^{\prime}=\sigma^{\prime}\cap K^{\prime}$ and $\sigma^{\prime}\subset  \partial \sigma $,
i.e. $\sigma^{\prime}$ is a free face of $\sigma$. So one can push $\sigma^{\prime}$ through
$\sigma$ into $K^{\prime}$ giving a homotopy equivalence.Then $H^*(K;E)=H^*(K^{\prime};E)$
and
 \begin{equation}
\tau(K;E,D_i)=\tau(K^{\prime};E,D_i)                                           
 \end{equation}
By iterating a sequence of elementary collapses and their inverses, one obtains a homotopy
equivalence of complexes that is called {\it simple} . Plainly one has,by iterating (6.2) , that
the Reidemeister torsion is a simply homotopy invariant. In particular $\tau$ is invariant under
subdivision. This implies that for a smooth manifold, one can unambiguously define $\tau(K;E,D_i)$ to be the torsion of any smooth triangulation of $K$.

In the case $K= S^1$ is a circle, let $A$ be the holonomy of a generator of the fundamental group 
$\pi_1(S^1)$. One has that $E$ is acyclic iff $I-A$ is invertible and then
 \begin{equation}
\tau(S^1;E)= \mid \det(I-A) \mid                                     
 \end{equation}
Note that the choice of generator is irrelevant as $I-A^{-1}= (-A^{-1})(I-A) $ and $|det(-A^{-1}|=1$.

  These three  properties of the Reidemeister torsion are the analogues of the properties of Euler
  characteristic ( cardinality law, homotopy invariance and normalization on a point), but there are 
  differences.Since a point has no acyclic representations ($H^0\not=0$) one cannot normalise 
  $\tau$  on a point as we do for the Euler characteristic, and so one must use $S^1$ instead. 
  The multiplicative cardinality law for the Reidemeister torsion can be made additive just by using $\log \,\tau $ , so the difference here is inessential. More important for some purposes is that 
the Reidemeister torsion is not an invariant under a general homotopy equivalence: as mentioned earlier this is in fact why it was first invented. 

It might be expected that the Reidemeister torsion counts something geometric(like the Euler characteristic). D. Fried  showed that it counts the periodic orbits of a flow and the 
periodic points of a map. We will show that the Reidemeister torsion counts the periodic point classes of a map( fixed point classes of the iterations of the map).

 Some further properties of $\tau$ describe its behavior under bundles. 
 
Let  $p: X\rightarrow B$ be a simplicial bundle with fiber $F$ where $F, B, X$ are finite 
complexes and $p^{-1}$ sends subcomplexes of $B$ to subcomplexes of $X$ . 
over the circle $S^1$.
We assume here that $E$ is a flat, complex vector bundle over $B$ . We form its pullback $p^*E$
over $X$.
Note that the vector spaces $H^i(p^{-1}(b),\bbbc)$ with
$b\in B$ form a flat vector bundle over $B$,
which we denote $H^i F$. The integral lattice in
$H^i(p^{-1}(b),\bbbr)$ determines a flat density by the condition
that the covolume of the lattice is $1$.
We suppose that the bundle $E\otimes H^i F$ is acyclic for all
$i$. Under these conditions D. Fried \cite{fri2} has shown that the bundle
$p^* E$ is acyclic, and
\begin{equation}
 \tau(X;p^* E) = \prod_i \tau(B;E\otimes H^i F)^{(-1)^i}.
\end{equation}
 the opposite extreme is when one has a bundle $E$ on $X$ for which the restriction $E|F$ is
 acyclic. Then, for $B$ connected,
  \begin{equation}
  \tau(X;E)= \tau(F;E|F)^{\chi(B)}               
  \end{equation}
  Suppose in (6.5) that $F=S^1$ i.e. $X$ is a circle bundle . Then (6.5) can be regarded as saing that
  $$\log \, \tau(X;E)= \chi(B)\cdot \log\, \tau(F;E|F)$$ 
   is counting the circle fibers in $X$ in the way that
  $\chi$ counts points in $B$, with a weighting factor of $\log\, \tau(F;E|F)$.

\section{The Reidemeister zeta Function and the Reidemeister
Torsion of the Mapping To\-rus of the dual map.}
 
Let $f:X\rightarrow X$ be a homeomorphism of
a compact polyhedron $X$.  Let $T_f := (X\times I)/(x,0)\sim(f(x),1)$ be the
mapping tori of $f$.
We shall consider the bundle $p:T_f\rightarrow S^1$
over the circle $S^1$.
We assume here that $E$ is a flat, complex vector bundle with 
finite dimensional fibre and base $S^1$. We form its pullback $p^*E$
over $T_f$.
Note that the vector spaces $H^i(p^{-1}(b),c)$ with
$b\in S^1$ form a flat vector bundle over $S^1$,
which we denote $H^i F$. The integral lattice in
$H^i(p^{-1}(b),\bbbr)$ determines a flat density by the condition
that the covolume of the lattice is $1$.
We suppose that the bundle $E\otimes H^i F$ is acyclic for all
$i$. Under these conditions D. Fried \cite{fri2} has shown that the bundle
$p^* E$ is acyclic, and we have
\begin{equation}
 \tau(T_f;p^* E) = \prod_i \tau(S^1;E\otimes H^i F)^{(-1)^i}.
\end{equation}
Let $g$ be the prefered generator of the group
$\pi_1 (S^1)$ and let $A=\rho(g)$ where
$\rho:\pi_1 (S^1)\rightarrow GL(V)$.
Then the holonomy around $g$ of the bundle $E\otimes H^i F$
is $A\otimes f^*_i$.
 
Since $\tau(S^1;E)=\mid\det(I-A)\mid$ it follows from (6.6)
that
\begin{equation}
 \tau(T_f;p^* E) = \prod_i \mid\det(I-A\otimes f^*_i)\mid^{(-1)^i}.
\end{equation}
We now consider the special case in which $E$ is one-dimensional,
so $A$ is just a complex scalar $\lambda$ of modulus one.
Then in terms of the rational function $L_f(z)$ we have \cite{fri2}:
\begin{equation}
 \tau(T_f;p^* E) = \prod_i \mid\det(I-\lambda \cdot f^*_i)\mid^{(-1)^i}
             = \mid L_f(\lambda)\mid^{-1}
\end{equation}

\begin{theorem}
Let $\phi:G\to G$ be an automorphism of $G$,where  $G$ is the direct sum of a finite group with a finitely generated free Abelian group, then
 \begin{equation}
 \tau\left(T_{\hat{\phi}};p^*E\right)
 =
 \mid L_{\hat{\phi}}(\lambda) \mid^{-1}
 =
 \mid R_{\phi}(\sigma\cdot \lambda) \mid^{(-1)^{r+1}},
 \end{equation}
where $\lambda$ is the holonomy of the one-dimensional
flat complex bundle $E$ over $S^1$, $r$ and $\sigma$ are the constants described in theorem 13 .
\end{theorem}
 
{\sc Proof}
We  know from the theorem 52 
that $R(\phi^n)$ is the number of fixed points
of the map $\hat{\phi}^{n}$.
In general it is only necessary to check that
the number of fixed points of $\hat{\phi}^n$ is equal to the
absolute value of its Lefschetz number.
We assume without loss of generality that $n=1$.
We are assuming that $R(\phi)$ is finite, so
the fixed points of $\hat{\phi}$ form a discrete set.
We therefore have
 $$
 L(\hat{\phi})
 =
 \sum_{x\in\fix\hat{\phi}} \ind(\hat{\phi},x).
 $$
Since $\phi$ is a group endomorphism, the trivial representation $x_0\in\hat{G}$ is always fixed.
Let $x$ be any fixed point of $\hat{\phi} $.Since $\hat{G}$ is union of tori $\hat{G}_0,...,\hat{G}_t$ and $\hat{\phi}$ is a linear map, we  can
shift any two fixed points onto one another without altering the map $\hat\phi$.
This gives us for any fixed point $x$ the equality
 $$ \ind(\hat{\phi},x) = \ind(\hat{\phi},x_0) $$
and so all fixed points have the same index.
It is now sufficient to show that $\ind(\hat{\phi},x_0)=\pm 1$.
This follows because the map on the torus
 $$ \hat{\phi}:\hat{G}_0\to\hat{G}_0 $$
lifts to a linear map of the universal cover, which is an euclidean space. The index is then the sign of the determinant of the identity map  minus this lifted map.
This determinant cannot be zero, because $1-\hat{\phi}$ must have finite
kernel by our assumption that the Reidemeister number of $\phi$ is
finite
(if $\det(1-\hat{\phi})=0$ then the kernel of $1-\hat{\phi}$ is a positive dimensional subspace of $\hat{G}$, and therefore infinite).

\begin{corollary}
Let $f:X\to X$ be a homeomorphism of a compact
polyhedron $X$. If $\pi_1(X)$ is the direct sum of a finite group with a free Abelian group, then 
then 
$$
 \tau\left(T_{\widehat{(f_{1*})}};p^*E\right)
 =
 \left|\ L_{\widehat{(f_{1*})}}(\lambda) \ \right|^{-1}
 =
 \Big|\ R_f(\sigma\cdot \lambda) \ \Big|^{(-1)^{r+1}},
 $$
where $r$ and $\sigma$ are  the constants described in theorem 13 .
\end{corollary}

  \section{The connection between the Reidemeister torsion,
eta--invariant, the Rochlin invariant and theta multipliers via the dynamical
zeta functions}

In this section we establish a connection
 between the Reidemeister torsion of a mapping tori,
 the eta-invariant, the Rochlin invariant, and theta multipliers via
 the Lefchetz zeta function and the Bismut-Freed-Witten holonomy theorem.

\subsection{Rochlin invariant}

We begin by recalling the definition of a spin structure on an oriented
 Riemannian manifold $M^m$
 with special attention to the notion of a spin diffeomorphism.
The tangent bundle $TM$ is associated to a principal $GL_+(m)$
 bundle $P$,
 the bundle of oriented tangent frames.
This last group $GL_+(m)$
 has a unique connected two-fold covering group $\widetilde{GL}_+(m)$.
If the tangent bundle $TM$ is associated to a princial $\widetilde{GL}_+(m)$
 bundle $\tilde{P}$,
 then we call $\tilde{P}$ a {\it spin structure} on $M$.
Thus, spin structures are in
 one-to-one correspondence with double coverings 
 $\tilde{P}\rightarrow P$ which are nontrivial on each fiber.
In terms of cohomology,
 such coverings are given by elements $w$ of
 $H^1(P,\bbbz/2\bbbz )$ such that $w\mid\pi^{-1}(x)=0$ for every $x$ in $M$,
 where $\pi: P\rightarrow M$ is the bundle projection and $\pi^{-1}(x)$
 is the fiber over $x$.
We shall refer to the cohomology class $w$
 as a spin structure on $M$, and the pair $(M,w)$ as a spin manifold.

Given an orientation preserving diffeomorphism
 $f:M\rightarrow M$,
 the differential $df$ of $f$ gives us a diffeomorphism
 $df: P\rightarrow P$
 and hence an isomorphism on the cohomology
 $(df)^*: H^1(P,\bbbz/2\bbbz )\rightarrow H^1(P,\bbbz/2\bbbz )$.
We say that an orientation-preserving diffeomorphism
 $f:M\rightarrow M$
 preserves the spin structure $w$
 if $(df)^*(w)=w$ in $H^1(P,\bbbz/2\bbbz )$.
This is equivalent to the existence of a bundle map
 $b:\tilde{P}\rightarrow\tilde{P}$
 making the following diagram commutative 
$$
\begin{array}{ccc}
\tilde{P} & \stackrel{b}{\rightarrow} & \tilde{P} \\
 \downarrow & & \downarrow \\
P & \stackrel{df}{\rightarrow} & P
\end{array}
$$ 
By a spin diffeomorphism $F$ of $(M,w)$ we mean a pair $F=(f,b)$
 consisting not only of a spin preserving diffeomorphism $f$
 but also of a bundle map $b:\tilde{P}\rightarrow\tilde{P}$
 covering $df$.
Given a spin diffeomorphism $F=(f,b)$ of $(M,w)$
 there is a well defined spin structure $w^\prime$
 on the mapping torus $T_f$ (see \cite{lm}).

We shall now define an invariant
 of spin diffeomorphisms $F$ of $(M,w)$
 where $M$ has dimension $8k+2$.
This will actually be defined via $(T_f,w^\prime)$,
 a spin manifold of dimension $8k+3$.
As usual, a spin manifold $(N^{8k+3},w)$
 is a spin boundary if there is
 a compact $8k+4$-dimensional spin manifold  $(X^{8k+4},W)$
 with $\partial{X}=N$ and with $W$ restricting to $w$.
By a result of \cite{an}
 the manifold $N^{8k+3}$ is a spin boundary
 if and only if it is an unoriented boundary.
A necessary and sufficient condition for $N$ to be a boundary
 is that all its Stiefel-Whitney numbers vanish.
In particular, this is always the case when $k=0$ or $k=1$.
If $M^{8k+2}$ has vanishing Stiefel-Whitney classes,
 then all the Stiefel-Whitney numbers of $T_f$ are zero (see \cite{lm}).
For a spin boundary $(N^{8k+3},w)=\partial{(X^{8k+4},W)}$
 the {\it Rochlin invariant} $R(N,w)$ in $\bbbz/16\bbbz$ is defined by
 $$
 R(N,w) \equiv \sigma{(X)} \pmod{16},
 $$
 where $\sigma{(X)}$ denotes
 the signature of the $8k+4$-dimensional manifold $X$.

\subsection[The Bismut-Freed-Witten Holonomy Theorem]{
 Determinant line bundles,
 the Eta-invariant
 and the Bismut-Freed-Witten theorem}

Let $\pi: Z\rightarrow N$ be
 a smooth fibration of manifolds
 with base manifold $N$
 and total manifold $Z$.
The fiber above a point $x\in N$
 is an $8k+2$-dimensional manifold $M_x^{8k+2}$
 which is equipped with a metric and a compatible spin structure.
The latter vary smoothly with respect to the parameter $x$
 in the base manifold;
 in other words the structure group of the fibration
 $\pi: Z\rightarrow N$ is a subgroup of the spin diffeomorphism group.

In this situation,
 along a fiber $M_x$,
 we have a principal $Spin(8k+2)$-bundle $P(TM_x)$.
Since the dimension $8k+2$ is even
 there are two half-spin representation $S_\pm$ of $Spin(8k+2)$
 and associated to them two vector bundles
 $E_x^\pm=P(TM_x)\otimes S^\pm$.
On the space $C^\infty(E_x^\pm)$
 of $C^\infty$-sections of these bundles,
 there is a Dirac operator
 $\partial_x: C^\infty(E_x^\pm)\rightarrow C^\infty(E_x^\mp)$
 which is a first order, elliptic, differential operator \cite{eich}.
If we replace the $C^\infty$-sections of $E_x^\pm$
 by square-integrable sections,
 the Dirac operator can be extended to an operator
 $\partial_x: L^2(E_x^\pm)\rightarrow L^2(E_x^\mp)$
 of Hilbert spaces.
As we vary $x$ over $N$,
 these Hilbert spaces $L^2(E_x^\pm)$
 form Hilbert bundles $L^2(E^\pm)$
 and the operators $\partial_x$
 form a continuous family of operators
 $\partial: L^2(E^\pm)\rightarrow L^2(E^\mp)$
 on these Hilbert bundles.

From the work Atiyah-Singer, Bismut-Freed \cite{bf} and Quillen \cite{q}
 it follows that there exists a well-defined complex line bundle
 $\det\partial$ over $N$.
Over a point $x$ in $N$,
 the fiber of this line bundle $(\det\partial)_x$
 is isomorphic to \\
 $(\Lambda^{max}ker\partial_x)^*\otimes(\Lambda^{max}coker\partial_x)$.
Bismut and Freed \cite{bf} undertook an extensive study
 of the geometry of this determinant line bundle $\det\partial$.
They showed that $\det\partial$ admits a Bismut-Freed connection
 $\bigtriangledown$,
 and proved a formula for the curvature
 associated to this connection.
One of the basic results in \cite{bf}
 is the holonomy formula for the Bismut-Freed connection
 $\bigtriangledown$ of the determinant line bundle $\det\partial$
 around an immersed circle $\gamma :S^1\rightarrow N $
 in the base manifold $N$.
We now describe this formula.
Pulling back by $\gamma$,
 there is an $8k+3$-dimensional manifold
 which is diffeomorphic to a mapping torus $T_{\phi}$,
 with the diffeomorphism $\phi$ specified by $\gamma$.
Choosing an arbitrary metric $g_{S^1}$ on $S^1$,
 and using the projection
 $\Phi : \tau(T_{\phi})\rightarrow \tau_{fiber}(T_{\phi})$
 of tangent bundles,
 we obtain a Riemannian structure on $T_{\phi}$.
Since the structure group of the fibration
 $\pi : Z \rightarrow N$
 is a subgroup of the spin diffeomorphism group,
 it follows that $\phi$ is covered
 by a canonical spin diffeomorphism
 and the mapping torus $T_{\phi}$ has a natural spin structure.
From this spin structure on $T_{\phi}$
 we obtain a spin bundle over $T_{\phi}$
 with structure group $Spin(8k+3)$
 and Dirac operator $\partial$
 on the space of $C^\infty$-sections of this bundle.

Following Atiyah, Patodi, Singer \cite{ap}
 we define the function $\eta(s,\partial)$
 in terms of the eigenvalues $\lambda$ of $\partial$
 by
 $$
 \eta(s,\partial)
 =
 \sum_{\lambda\ne 0} \frac{sign\lambda} {\mid\lambda\mid^s}.
 $$
This function is holomorphic for $Re(s)>0$
 and its value at $s=0$ is the $\eta$-invariant of $\partial$:
 $\eta(\partial)=\eta(0,\partial)$.
We denote by $h(\partial)$ the dimension of
 the kernel of the operator $\partial$.
Notice that these invariants depend on
 the choice of the metric $g_{S^1}$ on the base circle $S^1$.
In order to be free of this choice,
 we scale the metric $g_{S^1}$ by a factor $\frac{1}{{\epsilon}^2}$,
 and with respect to this new metric $\frac{g_{S^1}}{\epsilon^2}$
 we have a Dirac operator ${\partial}_{\epsilon}$ on $T_{\phi}$
 and a corresponding $\eta$-invariant $\eta(\partial_{\epsilon})$
 and $h(\partial_{\epsilon})$.
As the parameter $\epsilon$ tends to zero,
 the invariant
 $\frac{\eta(\partial_{\epsilon})+h(\partial_{\epsilon})}{2}$
 tends to a fixed limit.

\begin {theorem}{\cite{bf}}
The holonomy of the Bismut-Freed connection
 $\bigtriangledown$
 of the determinant line bundle
 $\det\partial$ around $\gamma$ is given by
 \begin{equation}
hol(\gamma;\det\partial,\bigtriangledown)
=
\lim_{{\epsilon}\to0}
\exp
\left(
-2\pi i \frac{\eta(\partial_\epsilon)+h(\partial_\epsilon)}{2}
\right).
 \end{equation}
\end {theorem}

Now, suppose that $\pi : Z\rightarrow N$
 is a fibration of manifolds $M^{8k+2}$
 with two prefered spin structures
 $w_1$ and $w_2$.
Corresponding to these two spin structures,
 there are families of Dirac operators
 $\partial_{w_1}$ and $\partial_{w_2}$,
 and determinant line bundles
 $\det\partial_{w_1}$ and $\det\partial_{w_2}$.
Notice that from the curvature formula of Bismut-Freed \cite{bf}
 these two complex line bundles
 $\det\partial_{w_1}$ and $\det\partial_{w_2}$
 have the same curvature 2-form.
Hence if we form the bundle
 $\det\partial_{w_1}/\det\partial_{w_2}
 =\det\partial_{w_1}\otimes(\det\partial_{w_2})^*$,
 the result is a flat complex line bundle.
In the language  of contemporary physicists,
 this is known as the cancelation of local anomalies.
For some of their models,
 it is important to investigate
 the holonomies of the flat line bundle
 $\det\partial_{w_1}\otimes(\det\partial_{w_2})^*$ -the
 global anomalies.

Let $w_1^\prime$ and $w_2^\prime$ be the spin structures on $T_\phi$
 induced by $w_1$ and $w_2$,
 where the diffeomorphism $\phi$ is specified by $\gamma$.
Lee, Miller and Weintraub proved the following

\begin{theorem}[\cite{lm}]
The holonomy of the flat complex line bundle\\
 $\det\partial_{w_1}/\det\partial_{w_2}$
 around a loop $\gamma$ is

$$ hol( \gamma ; \det\partial_{w_1} / \det\partial_{w_2} )=$$
$$ =
 \lim_{\epsilon\to 0}\exp\bigg( -2 \pi i\bigg[\frac{ \eta(\partial_{w_1^\prime \epsilon}) + h (\partial_{w_1^\prime \epsilon})}{2}- \frac{\eta(\partial_{w_2^\prime \epsilon})+ h(\partial_{w_2^\prime \epsilon})}{2} \bigg] \bigg)
$$
If the fiber $M^{8k+2}$ is a Riemann surface,
 then
$$
 hol(\gamma; \det\partial_{w_1} / \det\partial_{w_2})
 =
 \exp \left[
 -2 \pi i
 \frac{ R(T_\phi,w_1^\prime) - R(T_\phi,w_2^\prime) }{8}
 \right].
$$
\end{theorem}

\subsection {Connection with Reidemeister torsion}

Let us now suppose that $F=(f,b_i)$, $i=1,2$
 are two spin diffeomorphisms
 of spin manifolds $(M^{8k+2},w_i)$, $i=1,2$.
Then there are well-defined spin structures $w_i^\prime$, $i=1,2$
 on the mapping torus $T_f$.
We may consider $T_f$ as a bundle $p:T_f \rightarrow S^1$
 over the circle $S^1$ with fiber $M^{8k+2}$.
As above we have two Dirac operators
 $\partial_{w_i^\prime\epsilon}$, $i=1,2$
 on the space of $C^\infty$-sections
 of the spin bundles over $T_f$
 and corresponding eta-invariants
 $\eta(\partial_{w_i^\prime\epsilon})$
 and $h(\partial_{w_i^\prime\epsilon})$.
We also have a flat complex determinant line bundle
 $\det\partial_{w_1}/\det\partial_{w_2}$
 over the base manifold $S^1$
 and its pullback $p^*(\det\partial_{w_1}/\det\partial_{w_2})$
 over $T_f$.
Let $\gamma$ be the preferred generator
 of the fundamental group $\pi_1(S^1)$.

\begin{theorem}
 $$
 \tau( T_f;p^*(\det\partial_{w_1}/\det\partial_{w_2}))
 =
 \mid L_f(\lambda)\mid^{-1},
 $$
 where
 $$
 \lambda
 =
 hol( \gamma ; \det\partial_{w_1} / \det\partial_{w_2} )=
$$
$$ =
 \lim_{\epsilon\to 0}
 \exp \left(
 - 2 \pi i
 \left[
 \frac{\eta(\partial_{w_1^\prime\epsilon})
       + h(\partial_{w_1^\prime\epsilon})}{2}
 -
 \frac{\eta(\partial_{w_2^\prime\epsilon})
 +
 h(\partial_{w_2^\prime\epsilon})}{2}
 \right]
 \right)
 $$
If the fiber $M^{8k+2}$ is a Riemann surface, then
$$
 \lambda
 =
 hol( \gamma ; \det\partial_{w_1} / \det\partial_{w_2} )
 =
 \exp\left[
 -2 \pi i
 \frac{R(T_f,w_1^\prime) - R(T_f,w_2^\prime)}{8}
 \right]
$$
\end{theorem}
{\sc Proof}
The connection between the Reidemeister torsion of the mapping torus
 and the Lefschetz zeta function
 follows from formula (6.8):
$$
 \tau( T_f;p^*(\det\partial_{w_1}/\det\partial_{w_2}))
 =
 \mid L_f(\lambda)\mid^{-1},
$$
where
$$
 \lambda = hol(g;\det\partial_{w_1}/\det\partial_{w_2})
$$
The result now follows from the previous theorem 65.

\subsection{The Reidemeister torsion and theta-multipliers}

We recall some well known properties of theta-functions (see \cite{ig}).
Fix an integer $g \geq 1$.
A characteristic ${\rm\bf m}$ is a row vector
 ${\rm\bf m}=(m_1^*,...,m_g^*,m_1^{**},...,m_g^{**})$
 each of whose entries is zero or one.
The parity $e({\rm\bf m})=m_1^*\cdot m_1^{**} +...+m_g^*\cdot m_g^{**} \in \bbbz/2\bbbz $
 of the characteristic ${\rm\bf m}$ is said to be even (odd)
 when $e({\rm\bf m})=0$ ($=1$).

Let $S_g$ denote the Siegel space of degree $g$.
We shall write elements $T\in Sp_{2g}(\bbbz)$ as block matrices:
$$
 T = \left(\begin{array}{cc} A & B \\ C & D \end{array}\right).
$$
The theta function with characteristic ${\rm\bf m}$ is a function
 $\theta_{\rm\bf m} :S_g\times \bbbc^g \rightarrow \bbbc$
 satisfying the following transformations law \cite{ig} : 
$$
 \theta_{T\cdot {\rm\bf m}}(z(C\tau +D)^{-1},(A\tau + B)(C\tau + D)^{-1})=
$$
$$ =
 \gamma_{\rm\bf m}(T)
 \cdot \det(C\tau + D)^{1/2}
 \cdot \exp\Big( \pi i z (C\tau + D)^{-1}\;{ }^tz \Big)
 \cdot \theta_{\rm\bf m}(\tau,z) 
$$
 where
$$
T\cdot {\rm\bf m} = {\rm\bf m}\cdot T^{-1} + ((D \;{ }^tC)_0(B \;{ }^tA)_0)  \pmod{2}
$$
 and $(\  )_0$ denotes the row vector
 obtained by taking the diagonal elements of the matrix.
The number $\gamma_{\rm\bf m}(T)$ is an eighth root of unity,
 and are known as the theta multiplier of $T$ for the characteristic ${\rm\bf m}$.
The action of the symplectic group $Sp_{2g}(\bbbz )$
 on the characteristic (denoted above ${\rm\bf m} \mapsto T.{\rm\bf m}$)
 preserves parity,
 and is transitive on characteristics of a given parity.
We let $\Gamma({\rm\bf m}) = \{ T \in Sp_{2g}(\bbbz ) \mid T\cdot {\rm\bf m}={\rm\bf m} \}$.
If $\Gamma_2$ is the principal congruence subgroup of level 2 in
 $Sp_{2g}(\bbbz)$, i.e.
 $$
 \Gamma_2=\{T\in Sp_{2g}(\bbbz) \mid T\equiv I \pmod{2} \},
 $$
 then $\Gamma_2 \subset \Gamma({\rm\bf m})$ for every ${\rm\bf m}$.

For a diffeomorphism $f: V^2 \rightarrow V^2$
 of a Riemann surface $V^2$ of genus $g$,
 the induced map on the homology
 $T=f_*: H_1(V^2,\bbbz)  \rightarrow H_1(V^2,\bbbz )$
 is an element in the integral symplectic group $Sp_{2g}(\bbbz)$.
There is a bijection between the set of Spin structures on $V^2$
 and the set of characteristics ${\rm\bf m}$.
We shall let $w_{\rm\bf m}$ denote the spin structure
 corresponding to the characteristic ${\rm\bf m}$.
Thus a diffeomorphism $f: V^2\rightarrow V^2$
 preserves the spin structure $w_{\rm\bf m}$
 iff $f_* : H_1(V,\bbbz )\rightarrow H_1(V,\bbbz )$ is in $\Gamma({\rm\bf m})$.
If $f_*$  is in $\Gamma_2$
 then $f$ preserves each of the spin structures on $ V^2 $.
Let $T$ be any element of $\Gamma_2 \subset Sp_{2g}(\bbbz )$
 and $f: V^2\rightarrow V^2$ be a diffeomorphism
 with $f_*=T: H_1(V^2,\bbbz ))\rightarrow H_1(V^2,\bbbz )$.
Consider again the bundle
 $p: T_f\rightarrow S^1$ over the circle $S^1$
 with the fiber $V^2$.
Let $w_{\rm\bf m}$ and $w_{\rm\bf n}$ be two spin structures on $V^2$
 corresponding to any two {\it even} characteristic ${\rm\bf m}$ and ${\rm\bf n}$.
We have a flat complex determinant line bundle
 $\det\partial_{w_{\rm m}}/\det\partial_{w_{\rm n}}$
 over the base $S^1$ and its pullback
 $p^*(\det\partial_{w_{\rm m}}/\det\partial_{w_{\rm n}})$
 over the mapping tori $T_f$.

\begin{theorem}
$$
\tau(T_f;p^*(\det\partial_{w_{\rm m}}/\det\partial_{w_{\rm n}})
=
\mid L_f(\gamma_{\rm\bf m}(T)/\gamma_{\rm\bf n}(T))\mid^{-1}.
$$
\end{theorem}
{\sc Proof }
Lee,Miller and Weintraub \cite{lm} proved that
$$
 \gamma_{\rm\bf m}(T)/\gamma_{\rm\bf n}(T)
 =
 \exp\left( -2 \pi i \frac{R(T_f,w_{\rm m}) - R(T_f,w_{\rm n})}{8} \right)
$$
Now the statement of the theorem it follows from theorem 66.

  \section{ Topology of an attraction domain and the Reidemeister torsion}

  \subsection{Introduction}
Assume that on a smooth compact manifold $M$ of dimension $n$ there is given
a tangential vector field $X$ of class $C^1$, and consider the corresponding system
of differential equations
 \begin{eqnarray}
\frac{dx}{dt} &=& X(x), 
\end{eqnarray}
Let $\phi(t,x)$ is the trajectory of (1) passing through the point $x$ for $t=0$.
We shall say that the set $I$ is the attractor  or the asymptotically stable compact invariant set for  system $(1)$ if for any neighborhood $U$ of $I$ there is the neighborhood $W$, 
\qquad $I\subset W\subset U$ such that \\
    \vspace{0.2cm}        1) for any $x\in W \quad \phi(t,x) \in U,  {\rm if} \> t\in [0,+\infty),$\\
     \vspace{0.2cm}       2) for any $x\in W \quad \phi(t,x)\longrightarrow I, {\rm when}\>  t\longrightarrow +\infty$.\\
     
     By a Lyapunov function $V(x)$ for  attractor $I$ we mean a function that 
     satisfies following conditions\\
    \vspace{0.2cm}        1)$ V(x)\in C^1(U - I),  \quad V(x)\in C(U)$,\\
    \vspace{0.2cm}       2)$ V(x) > 0, x \in U - I;\quad V(x)=0, x\in I $, \\
    \vspace{0.2cm}        3) The derivative by virtue of the system (1) \
    $\frac {dV(x)}{dt} > 0 \ {\rm in } \  U - I $.\\
 Such Lyapunov function $V(x)$  for $I$  always exist \cite{wi}.
Suppose that $S$ is a level surface of Lyapunov function $V(x)$ in $U$.The conditions 3)
and the Implicit Function Theorem imply that the level surface $S$ is a compact smooth 
$n-1$-dimensional manifold transverse to the trajectories of (1) and trajectories of (1)
intersect $S$ on the descending side of the  Lyapunov function $V(x)$. Any two level
surfaces of the  Lyapunov function $V(x)$ are diffeomorphic.Note that  manifold $S$ is determined up to diffeomorphism by the behavior of trajectories  of the system (1) in $U - I$ and does not depend on the choice of the Lyapunov function $V(x)$ and its level.
 Let $ N \supset I, \dim N =n, $
be  a compact smooth manifold with the boundary $\partial N=S$.\\
 In this article we will study the dependence of the topology of the attraction domain
$$ D=\left\{ x \in M - I: \quad \phi(t,x) \longrightarrow I, {\rm when}\>  t\longrightarrow +\infty \right\}$$ of the attractor $I$ and of the level surface $S$  of the Lyapunov function $V(x)$ on the dynamical properties of the system (1) on the attractor. 
The investigation of the topological structure of the level surfaces of the Lyapunov function was initiated by Wilson \cite{wi}. Note that the attraction domain $D$ is diffeomorphic to 
$S\times R^1$,since each trajectory of system (1) in the invariant  attraction domain $D$ intersects  $n-1$-dimensional manifold $S$ exactly once.Hence it follows that the homology groups of $D$ and $S$ are isomorphic.

\subsection{Morse-Smale systems}
We assume in this section that system (1) is given in $R^n$  and is a Morse-Smale system on manifold $N$, i.e.
the following conditions are satisfied:\\
 1) A set  of nonwandering trajectories $ \Omega $ of system (1) is the union of a finite
 number of hyperbolic stationary points and hyperbolic  closed trajectories,\\
 2) Stable  and unstable manifolds of stationary points and closed trajectories
 intersect transversally.
 
 A stable or unstable manifold of a stationary point or a closed trajectory 
 $p$  is denoted by  $ W^s(p)$  and   $W^u(p)$. Let $a^k$ be the number
 of stationary points $p$  of system (1) in $I$  such that $\dim W^u(p)=k$, $b_k$ is 
 the number closed orbits $q$ of system (1) in $I$ such that $\dim W^u(q)=k$,
 $M_k= a_k + b_k + b_{k+1} $,\\
  $ B_k = \dim H_k(D;Q)= \dim H_k(S;Q)$,
 $\chi (D)=\chi (S) $ is the Euler characteristic of $D$ and $S$.
 
\begin{theorem}
The numbers $B_k$ and $ M_k$ satisfy the following inequalities:
$$
B_0  \leq   M_0 + M_{n -1 } - M_n  ,\\
$$
$$
B_1 - B_0  \leq   M_1 - M_0 + M_{n-2} - M_{n-1} + M_n,\\
$$
 \begin{equation}
B_2 - B_1 + B_0  \leq    M_2 -  M_1 + M_0 + M_{n-3} - M_{n-2} + M_{n-1} - M_n ,\\
 \end{equation}
$$
....................................................................................... , \\
$$
$$
\sum\limits_{i=0}^{n-1} (-1)^i \cdot B_i   =   \chi(S)   =   \chi(D)   			  			         =   (1 + (-1)^{n-1}) \sum\limits_{i=0}^{n} (-1)^i \cdot M_i .\\
$$
\end{theorem}

To prove the theorem we need several preliminary definitions and lemmas.

\begin{lemma}
 \begin{equation}
B_r= B_r(N) + B_{n-1-r}(N), 
 \end{equation}
 where $B_r(N)=\dim H_r(N;Q)$.
\end{lemma}

{\sc Proof}
It is possible to assume that $S$ lies on the $n$-dimensional sphere $S^n$.Suppose 
$C=Cl( S^n - N)$  is the closure of the complement of $N$. Then $S= C\cap N$.
The manifolds $C$ and $N$ are compact and intersect only on the boundary.
Consequently, we have the exact reduced Maier-Vietoris sequence \cite{d}:
$$
 .... \rightarrow \tilde H_{r+1}(S^n;Q) \rightarrow \tilde H_{r}(S;Q) \rightarrow \tilde H_{r}(N;Q) 
 \oplus \tilde H_{r}(C;Q)\rightarrow \tilde H_{r}(S^n;Q) \rightarrow ...
  $$

Set $ 1 \leq r <n-1 $.Then $ \tilde H_r(S^n;Q)=\tilde H_{r+}(S^n;Q) =0$, and therefore
$$
 \tilde H_{r}(S;Q) = \tilde H_{r}(N;Q) \oplus \tilde H_{r}(C;Q).
$$
Hence $\tilde B_r(S)=\tilde B_r(N) +\tilde B_r(C)$, where $\tilde B_r$ denotes the dimension of
 $\tilde H_r $.From Alexander duality it follows that $\tilde B_r(C)= \tilde B_r(N -\partial N)$
It is easy to show that $N$ and $N -\partial N$ are homotopically equivalent( for example, by means of the collar theorem). Therefore$ \tilde B_r(N -\partial N)=\tilde B_r(N)$, hence
$$
\tilde B_r(S)=\tilde B_r(N)+ \tilde B_{n-1-r}(N) \ \rm{and} \ B_r(S)=B_r(N) + B_{n-1-r}(N).
$$
The manifold $S$ is closed and orientable ; therefore in case $r=n-1$ we
obtain the following from Poincare duality:
$$
 B_0(S)=B_0(N) + B_{n-1}(N).
$$

Thus the lemma is proved for all $ r=0,1,2,..., n-1$.
We make the following definition:
\begin{definition}
The stationary point $p$ of the vector field $X$ with $\dim W^u(p)=k $ has standard form if 
there exist local coordinates $x_1, x_2, ..., x_k; y_1, y_2, ..,y_{n-k}$ in some neighborhood
of the point $p$ such that
$$
X= x_1\frac{\partial}{\partial x_1} + ...... +x_k\frac{\partial}{\partial x_k} - y_1\frac{\partial}{\partial y_1} - .... - y_{n-k}\frac{\partial}{\partial y_{n-k}}
$$ 
in this neighborhood.
\end{definition}

The standard form for a closed trajectory is defined analogously \cite {fran}.
\begin{lemma} \cite{fran}
If $X_0$ is a Morse-Smale vector field on the manifold $N$,
then there exists a path in the space of smooth vector fields on $N, \  X_t, \  t \in [0,1],$
such that:\\
 1) $X_t $ is  a Morse-Smale vector field for all $ t \in [0,1] $;\\
 2) the stationary points and closed trajectories of the field $X_1$ are all in standard form.\\
  By using this result we replace the original vector field by a vector field for which all stationary points and closed trajectories have standard form.
\end{lemma}

\begin{lemma}  \cite{fran}
Suppose $X$ is a Morse-Smale vector field on an orientable manifold, $\gamma$ is a closed
trajectory in standard form, $\dim W^u(\gamma)= k +1 $, $U$ is a sufficiently small neighborhood of $\gamma$. There exists a Morse-Smale vector field $ Y$ which coincides with $x$ outside of
$U$, in $U$ has stationary points $p$ and $q$, and has no other  stationary points or closed 
orbits.Moreover $\dim W^u(p)=k, \dim W^u(q)= k+1$.
\end{lemma}

By means of this lemma we replace a vector field on the manifold $N$ by the field $Y$
having no closed trajectories.In this connection the number of stationary points of the 
field $Y$  with $ \dim W^u(p)=k$ will be equal to $a_k + b_k + b_{k+1}$

\begin{lemma} \cite{sm}
Suppose $Y$ is a smooth Morse-Smale vector field without closed trajectories on
a compact manifold  $N$ with boundary, the stationary points of $Y$ have standard form, 
and the vector field on $\partial N$ is directed inward. Then there exists a Morse function 
$f$ on $N$ such that:\\
            1) the critical points of the function $f$ coincide with stationary points of the field $Y$,   
	    the  index of a critical point of the function $f$ coincides with the dimension of the
            unstable manifold of this point ;\\
            2) if $p \in N$ is a critical point of $f$ then $ f(p) = \dim W^u(p)$;\\
            3) $f( \partial N) = \frac {n+1}{2}.$ \\
\end{lemma}

\begin{lemma} \cite{sm}
Suppose $f$ is a Morse function on $N$ satisfying \\
conditions 1) - 3).
Then: \\
$$
B_0(N)  \leq   M_0 ,\\
$$
$$
B_1(N) - B_0(N)  \leq   M_1 - M_0 ,\\
$$
 \begin{equation}
B_2(N) - B_1(N) + B_0(N)  \leq    M_2 -  M_1 + M_0  ,\\
 \end{equation}
$$
....................................................................................... , \\
$$
$$
\sum\limits_{i=0}^{n} (-1)^i \cdot B_i(N)      =   \chi(N)   			  			         =    \sum\limits_{i=0}^{n} (-1)^i \cdot M_i .\\
$$
\end{lemma}

\begin{corollary} 
The following inequalities hold:
$$
B_{n-1}(N)  \leq   M_{n -1 } - M_n  ,\\
$$
$$
B_{n-2}(N) - B_{n-1}(N)  \leq M_{n-2} - M_{n-1} + M_n,\\
$$
 \begin{equation}
B_{n-3}(N) - B_{n-2}(N) + B_{n-1}(N)  \leq M_{n-3} - M_{n-2} + M_{n-1} - M_n ,\\
 \end{equation}
$$
....................................................................................... , \\
$$
$$
\sum\limits_{i=0}^{n-1} (-1)^i \cdot B_i (N)=   \chi(N)   			  			         =   (1 + (-1)^{n-1}) \sum\limits_{i=0}^{n-1} (-1)^i \cdot M_i .\\
$$

\end{corollary}
{\sc Proof of the corollary.}
From the fact that $H_n( N)=0$ it follows that \\
\begin{equation}
- B_{n-1}(N) + B_{n-2}(N)  ... + (-1)^n B_{0}(N)= M_{n} - M_{n-1}  ...+ (-1)^n M_0  
\end{equation}
Adding (6.16) to the inequalities (6.14) and subtracting (6.16) from the inequalities (6.14) we obtain system 
of the inequalities (6.15).

{\sc Proof of Theorem 68} 
From lemma 1, inequalities (6.14) and (6.15) it follows that\\
$$
B_k - B_{k-1} + B_{k-2} - ... + (-1)^k B_{0} = B_{k}(N) - B_{k-1}(N) +  ... + (-1)^k B_{0}(N) +\\
 $$                                     
 $$
 + B_{n-1-k}(N) - B_{n-k}(N) + B_{n-k+1}(N) +....+(-1)^k B_{n-1}(N) \leq \\
$$				        
$$	
\leq M_{k} - M_{k-1} + ......+ (-1)^k M_0 +M_{n-1-k} - M_{n-k} + ......+ (-1)^k M_{n-1} - (-1)^k M_n ; \\
$$ 

$$
 \chi(S)= \sum\limits_{k=0}^{n-1} (-1)^k \cdot B_k =   \sum\limits_{k=0}^{n-1} (-1)^k( B_k(N) + B_{n-1-k}(N)).
$$
Using the fact that $H_n(N)=0$ and changing the index of summation we obtain\\

$$
\chi(S) =(1 + (-1)^{n-1}) \sum\limits_{i=0}^{n} (-1)^i \cdot B_i(N)  			  			   =   (1 + (-1)^{n-1}) \sum\limits_{i=0}^{n} (-1)^i \cdot M_i 
$$
The theorem is proved.

\subsection{ A formula for the Euler characteristic}

The last identity in Theorem 68 is also true in more general situation. Namely, assume that 
system (6.11) is an autonomous system of differential equations having a finite number of stationary points in attractor $I$. Denote  by $ \ind (p)$ the indices of the vector field $X$ at stationary point $p$.
\begin{theorem}
 \begin{equation}
\chi(D)= \chi(S)=((-1)^n - 1)\cdot \sum_{p \in I} \ind (p).
 \end{equation}
\end{theorem}

{\sc Proof}
The vector field $X$ is directed on $\partial N$ into $N$ . Therefore from the Poincare-Hopf
theorem \cite {mi}, by replacing $X$ by $-X$ we obtain  $$\chi(N)=(-1)^n \cdot \sum_{p \in I} ind(p).$$ It is known that $\chi(\partial N)= (1 +(-1)^{n-1})\cdot \chi(N)$. Hence 
$$
\chi(D)= \chi(S)=((-1)^n - 1)\cdot \sum_{p \in I} \ind(p).
$$
\begin{corollary}
Suppose the stationary points on $I$ are hyperbolic , $a_k$ is the number of stationary points of $I$  with $\dim W^u(p)=k$.Since for a hyperbolic stationary point $p$ with $\dim W^u(p)=k$ the
index of the vector field at it is equal to $(-1)^k$, we obtain the following formula:\\
 \begin{equation}
\chi(D)= \chi(S)=((-1)^n - 1)\cdot \sum\limits_{k=0}^{n} (-1)^k \cdot a_k. 
 \end{equation}
\end{corollary}
For $n=3$ $S$ is union of finite number of spheres with handles.Suppose $m$ is the number of connected components, $p$ is the total number of handles of the manifold $S$.
Then $\chi(S)= 2m - 2p $. Hence we obtain 
\begin{corollary}
 \begin{equation}
m - p= - \sum_{p \in I} \ind(p).
 \end{equation} 
In the case, when stationary points are hyperbolic
$$
m  - p =a_o - a_1 + a_2 - a_3.
$$
\end{corollary}

 \subsection{The Reidemeister torsion  of the level surface of a Lyapunov function and of the
attraction domain of the attractor}
 In  this section  we consider the flow (6.11) with circular chain recurrent set $ R \subset I$ .The Reidemeister torsion of the attraction domain $D$ and of the level surface $S$ is the relevant topological invariant  of $D$ and $S$ which is calculated in theorem 70   and  corollary 28    via closed orbits of flow (6.11)  in the attractor $I$.\\
  The point  $x \in M$ is called chain-recurrent for flow (6.11) if for any $ \varepsilon >0 $ there exist points $ x_1=x, x_2, ..., x_n=x $ and real numbers $t(i) \geq 1$ such that $ \rho(\phi(t(i), x_i), x_{i+1}) < \varepsilon $ for $ 1 \leq i <n $ . Let  $ R \subset I $ be a set  of chain-recurrent points of equation (6.11) on the manifold $N$ defined above. We assume in this section  that $R$ is circular
 , i.e. there is a smooth map $ \theta : U \rightarrow R^1 / Z, U $ a neighborhood of $R$ in $N$,
 on which $ \frac {d}{dt}(\theta \circ \phi(t,x)) >0$. In other words, there is a cross-section of the 
 flow (6.11) on $R$ , namely , a level set of $\theta $ on $int(U) $ . For instance, if $R$ is finite i.e.,
 consists of finitely many closed orbits, then $R$ is circular. More generally, if $\phi$ on $R$ has 
 no stationary points and the topological dimension of $R$ is 1, then $R$ is circular. For example 
 ,if $ \phi$  is a nonsingular Smale flow , so that $R$ is hyperbolic and 1-dimensional, then $R$ is circular. If $U \in N$ is such that $ \cap _{t\in R^1} \phi_t(U) = J$ is compact and $J \in int U$ , then we say that $U$ is an isolating neighborhood of the isolated invariant set $J$. According to Conley \cite{co}, there is a continuous function $G: \rightarrow R^1$ such that $G$ is decreasing
 on $ N - R$  and $G(R)$ is nowhere dense in $R^1$. Taking an open neighborhood $W$ of $G(R)$ and $ U=G^{-1}(W)$, we see that $U$ is an isolating neighborhood for some 
 isolating invariant set $J$ and that $J \rightarrow R$ as $W \rightarrow G(R) $ \cite {fri}.This proves that the chain recurrent set $R$ can be approximated by the isolated invariant set $J$ .In particular we can make $J$ circular.
 Further, there are finitely many points $ x_i < x_{i+1}$ in $ R^1 - G(R) $ such that 
 $ G^{-1}[x_i, x_{i+1}] $ isolates an invariant set $J_i $ so that $ J=\cup J_i $ is as closes as we like to $R$ . In particular, we can make $J$ circular. For the sequel we need the isolating blocks \cite {co}. A compact isolating neighborhood $B_i$ of $J_i$ is said to be isolating block if: \\
 1)$ B_i $ is smooth manifold with corners,\\
 2) $ \partial B_i= b_i^+ \cup b_i^- \cup b_i^0  $ where each term is compact manifold with 
 boundary,\\
 3) the trajectories $ \phi(t,x) $ is tangent to $b_i^0  $ and  $ \partial b_i^0=( b_i^+ \cup b_i^- ) \cap b_i^0  $ , \\
 4) the trajectories $ \phi(t,x) $ is transverse to $b_i^+$ and $b_i^-$, enters $ B_i$ on $b_i^+$ and
 exists on $b_i^-$.\\
   Since $J_i$ is circular there is a smooth map $ \theta_i : B_i \rightarrow R^1 / Z $ such that
 $ \frac {d}{dt}(\theta_i \circ \phi(t,x)) >0$ on $B_i$.By perturbing  $ \theta_i  $, we can make
 $ \theta_i  $ transverse to $0 \in R^1 / Z $ on $B_i, b_i^+, b_i^-, b_i^0,  \partial b_i^+, \partial b_i^-, \partial b_i^0 $. Now let $ Y_i=\theta_i^{-1}(0) \cup b_i^-, Z_i=b_i^- $. Then $(Y_i, Z_i )$ is a simplicial pair . We define a continuous map $r_i: Y_i \rightarrow Y_i $ as follows. If $y_i \not \in Z_i$ then $r_i(y)=\phi(\tau,y)$ where $\tau=\tau(y)   >0 $ is the smallest positive time $t$ 
 for which $\phi(t,y) \in Y_i $. Since $ \phi(t,x) $ is transverse out on $ b_i^-0 Z_i $, we see
 that $ \tau(y)  $ is near 0 for $y$ near $Z_i$.Thus $\tau$ extends continuously to $Z_i$ if we set
 $ \tau / Z_i =0$. Now let $ E$ be a flat complex vector bundle of finite dimension on $B_i$.
 There is a bundle map $ \alpha_i : r_i^*(E|Y_i) \rightarrow  E|Y_i $ defined by pulling back along trajectory from $y$ to $r_i(y)$, using the flat connection on $E$.This determines an endomorphism 
  \begin{equation}
  (\alpha_i)_* : H^*(Y_i, Z_i; r_i^*E) \rightarrow  H^*(Y_i, Z_i; E).
  \end{equation} 
Since there is a natural induced map $ r_i^* : H^*(Y_i, Z_i;E) \rightarrow H^*(Y_i, Z_i; r_i^*E) $ we obtain the endomorphism 
 \begin{equation} 
\beta_i= (\alpha_i)_*\cdot r_i^* : H^*(Y_i, Z_i; E) \rightarrow  H^*(Y_i, Z_i; E). 
 \end{equation}
 So the relative Lefschetz number 
  \begin{equation}
 L(\beta_i)= \sum \limits_{k=o}^{n-1} (-1)^k\cdot \tr (\beta_i)_k 
  \end{equation} 
 is defined. According to Atiyah and Bott \cite {ab} the numbers $L(\beta_i)$ 
 can be computed from the fixed point set of $r_i$ in $Y_i  - Z_i$.If $ \fix(r_i)  -  Z_i$ is a finite set of points $p$ with the Lefschetz index $\ind_L(r_i,p) $ and $(\alpha_i)_p: E_p 
 \rightarrow E_p $ is the endomorphism of the fiber at $p$, then one has the relative Lefschetz formula 
 \begin{equation} 
 L(\beta_i)= \sum_p \ind_L(r_i,p) \cdot \tr (\alpha_i)_p.
 \end{equation}
We see that $L(\beta_i^n), n\geq 1$ , counts the periodic points of period $n$ for $\beta_i$
which are not in $Z_i$, i.e. the closed orbits of system (6.11) that wrap $n$ times around $R^1/Z$ under $\theta _i$, with a weight coming from the holonomy of $E$ around these closed orbits.
Now , consider the twisted Lefschetz zeta function \cite{fri} for $E$ and $ (B_i,b_i^-)$:
 \begin{equation}
L_i(z)\equiv L_{\beta_i}(z):= \exp\left(\sum_{n=1}^\infty \frac{L(\beta_i^n)}{n} z^n \right)
 \end{equation}
we now turn to the $R$-torsion of pairs. Suppose that $L$ is a CW-subcomplex of $K$ and consider  the relative cochain complex
$$ 
C^*(K,L;E)= ker(C^*(K;E) \rightarrow C^*(L;E|_L)).
$$
 Then one has a natural isomorphism $ |C^*(K,L;E)|\cong \otimes_j |V| $, where $j$ runs
 over the $i$-cells in $K\setminus L$. So our flat density on $E$ gives a density $\Delta_i$ on 
 the relative $i$-cochains in $E$. Thus we again have an R-torsion denoted 
 $ \tau (K, L; E, D_i) $
 for some choice of positive densities $D_i$ on $H^i(K,L;E)$. If $H^i(K,L;E)=0$, we say that $E$
 is acyclic for $(K,L)$ and we simply write $\tau(K,L;E)$, when the $D_i$ are chosen standard.
 Let $\rho_E: \pi_1(N,p) \rightarrow GL(E_p) $ is holonomy representation for acyclic bundle $E$
 on orientable manifold $N, \dim N=n, \rho_E^*$ is the cotragredient representation of  $\rho_E$ and $E^*$ is a flat complex vector bundle with the holonomy $\rho_E^*$. We suppose that
 $\det \rho_E=1$. Let $L_i^*(z)$ is the twisted Lefschetz zeta function for $E^*$ and $(B_i,b_i^-)$,
 and 
  \begin{equation}
  L^*(z)= \prod_i L_i^*(z), \qquad L(z)=\prod_i L_i(z).
 \end{equation}

\begin{theorem}
 \begin{equation}
\tau(D;E)=\tau(S;E)= |L(1)|^{-1} \cdot |L^*(1)|^{\varepsilon(n)},
 \end{equation}
where $\varepsilon(n)=(-1)^n$.

\end{theorem}

{\sc Proof}
Consider the function $G$. Smoothing the level set $G^{-1}(x_i)$ by sliding it along the flow,
one obtains a smooth region $N_i \subset N$ with 
$$
 G^{-1}((- \infty , x_i -\varepsilon)) \subset
G^{-1}((- \infty , x_i +\varepsilon))
$$
, such that the trajectories $\phi(t,x)$ transverse to $\partial N_i $, for large $i$ we have $N_i=N$ and $\partial N^-=\emptyset$. If $\varepsilon$ is small one has that $N_{i+1}  -  N_i$ isolates $J_i$.Then by properties of the Reidemeister torsion \cite{fri} one finds:
 \begin{equation}
\tau(N;E)= \prod_i \tau(N_{i+1},N_i;E)= \prod_i\tau(B_i,b_i;E)
 \end{equation}
D.Fried proved \cite{fri} that $E$ is acyclic for $(B_i, b_i^-)$ iff $I-\beta_i$ is invertible and 
then
 \begin{equation}
\tau(B_i,b_i;E)=|L_i(z)|^{-1}|_{z=1}
 \end{equation}
So we have
 \begin{equation}
\tau(N;E)= \prod_i |L_i(1)|^{-1}=|L(1)|^{-1}              
 \end{equation}
From the multiplicative law (6.1) for the Reidemeister torsion it follows:
 \begin{equation}
\tau(N;E)= \tau(N, \partial N=S;E)\cdot \tau(\partial N=S;E)    
 \end{equation}
Using Milnor's duality theorem for the Reidemeister torsion \cite {mi} we have:
 \begin{equation}
\tau(N, \partial N=S;E)=\tau(N;E^*)^{(-1)^n}  
 \end{equation}
From formula (6.29) it follows that
 \begin{equation}
\tau(N;E^*)= \prod_i |L_i^*(1)|^{-1}=|L^*(1)|^{-1}    
 \end{equation}
Since the attraction domain $D$ is diffeomorphic to $S\times R^1 $ then
the Reidemeister torsion $\tau(D;E)=\tau(S;E) $ by the simple homotopy invariance of the
Reidemeister torsion.
Now from (6.29), (6.30), (6.31), (6.32) we have:
$$
\tau(D;E)=\tau(S=\partial N;E)=\tau(N;E)\cdot \tau^{-1}(N, \partial N=S;E)=
$$
$$ 
= \tau(N;E) \cdot \tau(N;E^*)^{(-1)^{n+1}}=|L(1)|^{-1}\cdot |L^*(1)|^{(-1)^n}
$$

 Suppose now that the system(6.11) on the manifold $N$ is a nonsingular almost Morse-Smale
 system. This means that (6.11) has finitely many hyperbolic prime periodic orbits $\gamma$
 and no other chain-recurrent points. Over the orbit $\gamma$ lies a strong unstable bundle
 $ E^u(\gamma)$ of some dimension $u(\gamma)$. Let $\delta(\gamma)$ be $+1$ if $E^u$
 is orientable and $-1$ if it is not. Let $\varepsilon(\gamma)=(-1)^{u(\gamma)}$
 
 \begin{corollary}
 $$
 \tau(D;E)=\tau(S=\partial N;E)=
 $$
 $$
 =\prod_{\gamma} |\det(I-\delta(\gamma) \cdot \rho_E(\gamma)) |^{\varepsilon(\gamma)}
 \times (\prod_{\gamma} |\det(I-\delta(\gamma) \cdot \rho^*_E(\gamma)) |^{\varepsilon(\gamma)})^{(-1)^{n+1}}
 $$
 \end{corollary}

{\sc Proof}
According to D.Fried \cite{fri} if $ J_i $ is a prime hyperbolic closed orbit $\gamma $
then 
$$ 
|L_i(1)|^{-1} =|\det(I-\delta(\gamma) \cdot \rho_E(\gamma)) |^{\varepsilon(\gamma)}
$$
Now, the statement it follows from theorem 70.

  \section{Integrable Hamiltonian systems and  the Reidemeister torsion}

Let $M$ be a four-dimensional smooth symplectic manifold and the system (6.11) be 
a Hamiltonian system with a smooth Hamiltonian $H$. In the Darboux coordinates
such system has the form:

\begin{equation}
\frac{dp_i}{dt} =\frac{\partial H}{\partial q_i}
\end{equation}
$$
\frac{dq_i}{dt} =\frac{\partial H}{\partial p_i}.
$$
 As the Hamiltonian $H$ is the integral of the system (6.33), then three-dimensional
 level surface $Q=[H=const ] $  is invariant for the system (6.33). The surface $Q$ is called the isoenergetic surface or the constant -energy surface. Since $M$ is orientable( as a symplectic
 manifold), the surface $Q$ is automatically orientable in all cases. Suppose that the system
 (6.33) is complete integrable (in Liouville's sense) on the surface $Q$. This means, that there is the smooth function $f$(the second integral), which is independent with $H$ and for the Poisson bracket $[H,f]=0 $in the neighborhood of $Q$. We shall call the integral $f$ a Bott integral on the 
 isoenergetic surface $Q$, if its critical points form critical nondegenerate smooth submanifolds
 in $Q$.This means that Hessian $d^2f$ of the function $f$ is nondegenerate on the planes
 normal to the critical submanifolds of the function $f$. A.T. Fomenko \cite{fo} proved that a
 Bott integral on compact nonsingular isoenergetic surface $Q$ can have only three types of
 critical submanifolds: circles, tori, Klein bottles. The investigation of the concrete systems shows \cite{fo} that it is a typical situation when the integral on $Q$ is a Bott integral.
 In the classical integrable cases of the solid body movement( cases of the Kovalevskaya, Goryachev-Chaplygin, Clebsch, Manakov) the Bott integrals are a round Morse functions
  on the isoenergetic surfaces. The round Morse function is a Bott function all whose critical
  manifolds are circles. Note that critical circles of $f$ is a periodic solution of the system(6.33)
   and the number of this circles is finite.
    Suppose now that the Bott integral $f$ is a round Morse function on the closed isoenergetic
    surface $Q$. Let us recall the concept of the  separatrix diagram of the critical circle
    $\gamma$  for a Bott function $f$. Let $x\in \gamma $ be an arbitrary point and $N_X(\gamma)$ be a disc of small radius normal to $\gamma$ at $x$.
    The restriction of $f$ to the $N_X(\gamma)$ is a normal Morse function with the critical
    point $x$ having a certain index $\lambda=o,1,2.$ The separatrix of the critical point $x$
    is the integral trajectory of the field grad$f$, which is entering or leaving $x$. The union of all
    the separatrices entering the point $x$ gives a disc of dimension $\lambda$ and is called the incoming separatrix diagram(disc). The union of outgoing separatrices gives a disc of additional
    dimension and is called the outgoing separatrix diagram(disc). Varing the point $x$ and constructing the incoming  and outgoing separatrix discs for each point $x$, we obtain the incoming and outgoing separatrix diagrams of circle$\gamma$. Let $u(\gamma)$ be the dimensi
    on of the outgoing separatrix diagram of $\gamma$, and $\delta(\gamma)$ be +1 if this 
    separatrix is orientable , and -1 if it is not. Let $\varepsilon(\gamma)=(-1)^{u(\gamma)}$.
 Suppose that $\rho_E: \pi_1(Q,p) \rightarrow GL(E_p) $ is holonomy representation for acyclic bundle $E$ over $Q$; $E_p $ is a fiber at point $p$.

\begin{theorem}
 \begin{equation}
\tau(Q;E) =\prod_{\gamma} |\det(I-\delta(\gamma) \cdot \rho_E(\gamma)) |^{\varepsilon(\gamma)}
 \end{equation}
\end{theorem}

 {\sc Proof}
As a Bott integral $f$ on $Q$ is a round Morse function then according to Thurston \cite{thu}
$Q$ has a round handle decomposition whose core circles are critical circles of $f$.
According to Asimov \cite{asi} if $Q$ has a round handle decomposition then $Q$ has a 
nonsingular Morse-Smale flow whose closed orbits are exactly the core circles of the
round handles, i.e. critical circles $\gamma$ of $f$. Consider this nonsingular 
Morse-Smale flow. For a closed orbit $\gamma$ of such flow the dimension $u(\gamma)$ of
unstable manifold $W^u(\gamma)$ is exactly the dimension of the outgoing
separatrix diagram of $\gamma$ as critical circle of $f$. One can choose
as isolating block $B$ for $\gamma$ to be a bundle over $S^1$ with fiber $F$ a 
simplicial disc so that $F \cong I^u\times I^s, u=u(\gamma), s=4-u(\gamma)$, and
$F\cap b^- \cong \partial I^u\times I^s \subset F$. By collapsing $I^s$ to a point
one can produce a simple homotopy equivalence of $(B,b^-)$ to $(X_{\gamma},\partial X_{\gamma})$ where $X_{\gamma}$ is the unstable disc bundle over $\gamma$.
Give $Q$  its Smale filtration by compact submanifolds $Q_i$ of top dimension
so that $ Q_i \subset intQ_{i+1}, Q_0=\emptyset , Q_i=Q$ for large $i$ , flow
is transverse inward on $\partial Q_i$ and $Q_{i+1} -  Q_i$ is an isolating neighborhood for a hyperbolic closed orbit $\gamma$.Then $(Q_{i+1}, Q_i)$ has the same homotopy type as
$(X_{\gamma},\partial X_{\gamma})$ and by properties of the Reidemeister torsion \cite{fri}
one finds:
$$
\tau(Q;E) =\prod_i \tau(Q_{i+1}, Q_i;E)=\prod_{\gamma}\tau(X_{\gamma},\partial X_{\gamma};E)=
$$
$$
=\prod_{\gamma} |\det(I-\delta(\gamma) \cdot \rho_E(\gamma)) |^{\varepsilon(\gamma)}.
$$

\end{document}